\documentclass[12pt,english,floatfix,nofootinbib,superscriptaddress,aps,prd,preprint]{revtex4}
\usepackage[a4paper, margin=1.6cm]{geometry}
\usepackage[utf8]{inputenc}       
\usepackage[english]{babel}       
\usepackage[utf8]{inputenc}
\usepackage{textcomp}
\usepackage{amsmath,amsthm,amssymb,amsfonts,mathrsfs,amsbsy} 
\usepackage{tensor}               
\usepackage{slashed}  
\usepackage{bbm}
\usepackage{esint}                
\usepackage{cancel}               
\usepackage{graphicx}             
\usepackage{natbib}
\usepackage{float}                
\usepackage{subfig}               
\usepackage[font=small,labelfont=bf]{caption} 
\usepackage{multirow}             
\usepackage{array}                
\usepackage{tikz}                 
\usetikzlibrary{quotes,angles,arrows,decorations.markings} 
\usepackage{makecell} 

\usepackage{lipsum}

\usepackage{xcolor}               
\usepackage{color}                
\usepackage{textcomp}             
\usepackage{units}                

\newcommand{\be}{\begin{equation}}
\newcommand{\ee}{\end{equation}}
\newcommand{\Be}{\begin{eqnarray}}
\newcommand{\Ee}{\end{eqnarray}}

\newcommand{\mincir}{\raise
-3.truept\hbox{\rlap{\hbox{$\sim$}}\raise4.truept\hbox{$<$}\ }}
\newcommand{\magcir}{\raise
-3.truept\hbox{\rlap{\hbox{$\sim$}}\raise4.truept\hbox{$>$}\ }}

\newcolumntype{Y}{>{\centering\arraybackslash}X}
\providecommand{\U}[1]

\usepackage[dvips]{epsfig}        
\usepackage[dvips]{graphicx}      
\usepackage{hyperref}             
\hypersetup{
    colorlinks=true,              
    breaklinks=true,              
    citecolor=blue,               
    linkcolor=[rgb]{0,0.5,0.9},   
    urlcolor=red,                 
    filecolor=green               
}

\newcommand{\ie}{\begin{equation}}
\newcommand{\fe}{\end{equation}}
\newcommand{\se}{\begin{eqnarray}}
\newcommand{\ff}{\end{eqnarray}}

\begin{document}

\title{Perturbative dynamics and relativistic effects of a dyonic Kalb--Ramond black hole}


\author{A. A. Ara\'{u}jo Filho}
\email{dilto@fisica.ufc.br}
\affiliation{Departamento de Física, Universidade Federal da Paraíba, Caixa Postal 5008, 58051--970, João Pessoa, Paraíba,  Brazil.}
\affiliation{Departamento de Física, Universidade Federal de Campina Grande Caixa Postal 10071, 58429-900 Campina Grande, Paraíba, Brazil.}
\affiliation{Center for Theoretical Physics, Khazar University, 41 Mehseti Street, Baku, AZ-1096, Azerbaijan.}


\date{\today}

\begin{abstract}

We investigate perturbative dynamics, tidal effects, and relativistic frequency shifts in a dyonic Kalb--Ramond black hole generated by a Lorentz--violating antisymmetric tensor background. The geometry is controlled by the mass $M$, the electric charge $Q$, the magnetic charge $p$, and the Lorentz--violating parameter $\ell$, with the dyonic sector entering through the effective combination $P_{\ell}^{2}=Q^{2}/(1-\ell)^{2}+p^{2}/(1-2\ell)$. First, we analyze the gravitational Doppler effect for radial signal exchange between freely falling and static observers, showing how the dyonic charges weaken the redshift by shifting the frequency ratio toward unity. We then compute the radial and angular tidal forces in a freely falling frame and determine the characteristic radii at which the usual stretching and compression patterns are reversed. The gravitational time delay is also evaluated for null trajectories, showing that the electric and magnetic sectors reduce the delay relative to the reference configuration. In the perturbative sector, {we study massless scalar, vector, and spinor test fields together with an effective axial metric model, computing sixth--order WKB spectra and independently checking selected modes using the asymptotic iteration method. At fixed mass and charges, increasing the Lorentz--violating parameter raises the tabulated oscillation frequencies and damping rates in the original time coordinate; their interpretation accounts for time normalization and changes in effective parameters. The independent comparisons support the tested higher--multipole fundamental modes. Time--domain evolutions exhibit damped ringdown followed by power--law tails, whose exponents are estimated through fits incorporating finite time corrections and numerical sensitivity checks.}

\end{abstract}


\maketitle

\tableofcontents


\section{Introduction }

At energies where quantum--gravitational effects may leave residual traces in the effective spacetime description, local Lorentz invariance need not remain an exact symmetry. This possibility has motivated extensive work on effective frameworks in which tiny deviations from Lorentz symmetry are incorporated without abandoning the low--energy field--theoretic setting \cite{kostelecky1999constraints,colladay1997cpt,kostelecky2004gravity,kostelecky1989spontaneous,kostelecky2011data}. In many such scenarios, the symmetry is broken spontaneously: a dynamical field acquires a nonzero vacuum configuration and thereby selects privileged spacetime directions. The bumblebee model realizes this mechanism in one of its simplest forms. In this approach, a vector field develops a vacuum expectation value, and its coupling to gravity produces well--defined modifications of the general relativistic dynamics \cite{bluhm2008spontaneous,bluhm2005spontaneous,Bluhm:2023kph,Maluf:2013nva,Maluf:2014dpa,Bluhm:2019ato}.

Bumblebee scenarios offer a direct realization of spontaneous Lorentz breaking in gravitational physics. Instead of assuming Lorentz invariance as an exact microscopic symmetry, these models allow it to arise only as an effective low--energy limit. Their formulation was influenced by string--motivated mechanisms in which tensor fields acquire nonzero vacuum values \cite{kostelecky1989spontaneous,kostelecky1991photon}, by vector--tensor modifications of Einstein gravity \cite{jacobson2004einstein}, and by effective field--theory descriptions that incorporate fixed background configurations associated with symmetry breaking \cite{kostelecky2004gravity,bluhm2005spontaneous}. In the bumblebee case, the dynamics of the vector field $B_{\mu}$ are governed by a potential $V(B_{\mu}B^{\mu}\mp b^{2})$, which forces the field toward a vacuum state with $B_{\mu}B^{\mu}=\pm b^{2}$. Once this vacuum configuration is reached, spacetime contains a distinguished direction, and local Lorentz symmetry is spontaneously violated \cite{bluhm2005spontaneous,bluhm2008spontaneous}. The excitations around this vacuum split into massless Nambu--Goldstone modes, which may behave as photon--like degrees of freedom \cite{bluhm2005spontaneous}, and massive modes, which measure deviations away from the fixed--norm vacuum condition \cite{bluhm2008spontaneous}.

Bumblebee gravity has also been extended to curved spacetimes, where the nonzero vacuum value of the vector field participates directly in the gravitational dynamics and changes the resulting geometry \cite{Bertolami:2005bh}. Black hole spacetimes constitute one of the main arenas in which these effects have been examined. The solution proposed in Ref.~\cite{Casana:2017jkc} became a reference point for many later analyses, ranging from quantum processes near the horizon \cite{Liu:2024wpa,AraujoFilho:2025hkm} and Lorentz--violating corrections to gravitational wave propagation \cite{Liang:2022hxd} to quasinormal spectra \cite{Oliveira:2021abg} and even--parity perturbations \cite{Liu:2026cxs}. Further extensions have incorporated topological-defect--like structures \cite{Gullu:2020qzu}, non--commutative effects \cite{KumarJha:2020ivj,AraujoFilho:2025rvn}, cosmological constant contributions \cite{Maluf:2020kgf}, cosmological applications \cite{Gonzalez-Espinoza:2025fmi}, nonlinear electrodynamics \cite{Li:2026tae}, and approximate rotating geometries \cite{Ding:2019mal,Liu:2019mls}. Light propagation in these spacetimes has likewise been analyzed through weak- and strong-deflection methods, including Gauss--Bonnet techniques, direct geodesic treatments, perturbative rotation schemes, and Newman--Janis-based constructions \cite{Ovgun:2018ran,Li:2020wvn,Sekhmani:2025zen,Deng:2025uvp}. More recently, mass--ratio inspirals in bumblebee gravity have also been used to extract phenomenological constraints on Lorentz--violating parameters \cite{Long:2026dcb}.

A different route to Lorentz--symmetry breaking can be built from an antisymmetric tensor field instead of a vector degree of freedom. The Kalb--Ramond field, introduced as a rank--two tensor in bosonic string theory \cite{kalb}, provides such a framework when it is coupled nonminimally to gravity. Under suitable conditions, this tensor sector develops a nonzero vacuum configuration, so that spacetime no longer remains directionally neutral \cite{maluf2019antisymmetric,brett}. In this case, the preferred background (similarly to the bumblebee case) is generated by the vacuum structure of the antisymmetric field, and Lorentz symmetry is broken spontaneously.

Tensor--based extensions of Lorentz--violating gravity have led to what is commonly referred to as Kalb--Ramond gravity, in which the antisymmetric sector replaces the vector order parameter used in bumblebee models. In this context, black hole configurations have formed a central class of exact backgrounds. The static and spherically symmetric geometry obtained in Ref.~\cite{Yang:2023wtu} provided the first explicit realization (as far as we know), after which the motion of particles and fields around this spacetime was investigated in Ref.~\cite{Atamurotov}. This solution has since been used to study quasinormal modes \cite{Filho:2023ycx}, greybody factors and transmission bounds \cite{Guo:2023nkd}, weak-- and strong--field lensing signatures \cite{junior2024gravitational}, phenomenological bounds on the symmetry-breaking scale \cite{junior2024spontaneous}, and gravitational wave emission from periodic orbits \cite{Junior:2024tmi}. Related analyses have also addressed circular motion and quasi--periodic oscillations \cite{jumaniyozov2024circular}, accretion modeled by Vlasov gases \cite{jiang2024accretion}, particle production effects \cite{AraujoFilho:2024ctw,AraujoFilho:2025hkm}, and neutrino propagation \cite{Shi:2025xkd,Shi:2025rfq}.

On the other hand, charged Kalb--Ramond black holes were later obtained in Ref.~\cite{Duan:2023gng}, opening the way to analyses of their thermodynamic properties, emission mechanisms, geometric observables and energy extraction \cite{al-Badawi:2024pdx,Zahid:2024ohn,araujo2025antisymmetric,araujo2025impact,yao2026energy,chen2024thermal}. A non--commutative version of the Kalb--Ramond solution was subsequently proposed in Ref.~\cite{AraujoFilho:2025jcu}. The class of available geometries has also been enlarged by solutions surrounded by anisotropic fluids \cite{Sekhmani:2026gup}, ModMax sectors \cite{Sekhmani:2025epe,Sekhmani:2025jbl,Sucu:2026nkw}, clouds of strings \cite{Ahmed:2026vjw}, perfect fluid dark matter \cite{Ahmed:2026doo}, dyonic ModMax fields \cite{Ahmed:2026vbh}, and other configurations \cite{Lessa:2025kln,Liu:2024oas}. Rotating extensions have provided another direction of development. In the slow--rotation regime, the corresponding shadow profiles were investigated in Ref.~\cite{Liu:2024lve}, while the scalar, vector, and tensor quasinormal spectra were computed in Ref.~\cite{Deng:2025atg}. Kalb--Ramond black holes carrying a global--monopole contribution have also been derived, further broadening the range of Lorentz--violating compact objects described by this antisymmetric tensor sector \cite{Belchior:2025xam}.

More recently, the Kalb--Ramond black hole sector was extended through the construction of a charged dyonic configuration \cite{Lin:2026ewo}. Despite this progress, the dynamical and observational properties of this spacetime remain largely unexplored. In particular, its quasinormal spectrum, time--domain response, tidal behavior, and related relativistic effects have not yet been analyzed. The present work addressed this gap by developing an initial study of the gravitational properties associated with the dyonic Kalb--Ramond black hole.

Black holes now serve as remarkable laboratories for probing deviations from general relativity in the strong--field regime. This role has become even more prominent after the direct detection of gravitational waves by the LIGO and VIRGO collaborations, which made it possible to test compact-object dynamics through interferometric observations \cite{LIGOScientific:2016aoc}. At the same time, the Event Horizon Telescope has resolved horizon--scale structures associated with supermassive compact objects, which provides independent observational access to the geometry of the near--horizon region \cite{Akiyama2022,Akiyama2019}.

Black hole perturbations provide a direct set of probes for the geometry of compact objects. In the post--merger stage, the distorted remnant approaches equilibrium through a ringdown signal governed by complex quasinormal frequencies \cite{Konoplya:2011qq,Konoplya:2013rxa,karmakar2022quasinormal,Konoplya:2019hlu,karmakar2024quasinormal,Konoplya:2007zx,Kokkotas:2010zd,berti2009quasinormal,cardoso2009geodesic}. The real part fixes the oscillation scale, whereas the imaginary part controls the damping time. Since both quantities are determined by the background metric, the perturbing sector, and the boundary conditions, any Lorentz--violating modification of the black hole spacetime can shift the ringdown spectrum. Quasinormal modes therefore offer a useful diagnostic for distinguishing modified geometries from their general--relativistic counterparts. Their role is also connected with other observables, since relations have been explored between the eikonal quasinormal spectrum and shadow properties \cite{Jusufi:2020dhz}, as well as between perturbative dynamics and greybody emission \cite{Konoplya:2024lir,Konoplya:2024vuj,Sakalli:2022xrb}. Although the identification of individual modes in present gravitational wave data remains debated \cite{Franchini:2023eda}.

This study examines the dynamics of perturbations, tidal responses, and relativistic frequency shifts in a dyonic Kalb--Ramond black hole that arises from a Lorentz--violating antisymmetric tensor background. The spacetime is specified by the mass $M$, the electric charge $Q$, the magnetic charge $p$, and the Lorentz--violating parameter $\ell$. The charge contribution enters the metric through the effective combination $P_{\ell}^{2}=Q^{2}/(1-\ell)^{2}+p^{2}/(1-2\ell)$, which allows the electric and magnetic sectors to be treated within a single radial structure. We first consider the gravitational Doppler shift associated with radial signal propagation between a freely falling emitter and a static receiver. The analysis shows that the dyonic charges drive the frequency ratio closer to unity, thereby reducing the redshift effect relative to less charged configurations. We then evaluate the tidal-force components in the freely falling frame. This calculation identifies the radii where the standard radial stretching and angular compression can change character due to the combined influence of $Q$, $p$, and $\ell$. The null propagation sector is addressed through the gravitational time delay, for which the electric and magnetic charges lower the delay with respect to the reference case. {The perturbative analysis considers massless scalar, vector, and spinor test fields, together with an effective axial metric perturbation model whose source closure is specified in Appendix~A.} For each sector, we obtain the corresponding effective potential and compute the quasinormal frequencies with the sixth--order WKB approximation. The numerical results show that the Lorentz--violating parameter produces the strongest changes in the spectra, increasing the real oscillation frequencies and altering the damping rates. By contrast, the electric and magnetic charges generate weaker corrections. The time--domain evolution supports the frequency--domain analysis, displaying an intermediate stage in which damped quasinormal ringing dominates and a late--time regime in which power--law tails control the signal.


\section{The dyonic Kalb--Ramond black hole }
\label{sec:dyonic_KR_BH}

The starting point of our analysis is a dyonic black hole spacetime arising from a Lorentz--violating theory driven by a Kalb--Ramond background. In this setup, the gravitational dynamics receive contributions from the antisymmetric tensor $B_{\mu\nu}$ through a nonminimal curvature coupling. The action for the theory is written as
\begin{equation}
S= \int \mathrm{d}^{4}x \sqrt{-g} \left[
\frac{1}{2\kappa} \left( R-2\Lambda+\xi B_{\mu}{}^{\rho}B_{\nu\rho}R^{\mu\nu} \right)
-\frac{1}{12}H_{\mu\nu\rho}H^{\mu\nu\rho}
-V(X) +\mathcal{L}_{\mathrm{m}} \right],
\label{action_KR}
\end{equation}
with, in this notation, $\kappa=8\pi G/c^{4}$ fixes the gravitational coupling, $\Lambda$ represents the cosmological constant, and $\xi$ sets the strength of the nonminimal interaction between the Kalb--Ramond tensor and the Ricci curvature. The antisymmetric field $B_{\mu\nu}$ gives rise to the field strength $H_{\mu\nu\rho}=\partial_{\mu}B_{\nu\rho}+\partial_{\nu}B_{\rho\mu}+\partial_{\rho}B_{\mu\nu}$. The self-interaction potential is written as a function of $X=B_{\mu\nu}B^{\mu\nu}\pm b^{2}$, where $b^{2}$ fixes the vacuum scale of the tensor field. Once $V(X)$ reaches its minimum, the Kalb--Ramond field develops a nonzero vacuum expectation value, thereby selecting preferred spacetime directions and inducing spontaneous Lorentz--symmetry breaking.

The electromagnetic contribution is also deformed by the Kalb--Ramond background. In this sector, the gauge field $A_{\mu}$ enters through the Maxwell tensor $F_{\mu\nu}=\partial_{\mu}A_{\nu}-\partial_{\nu}A_{\mu}$, and its dynamics are supplemented by two curvature--independent interactions involving $B_{\mu\nu}$ and $F_{\mu\nu}$. With these contributions included, the matter Lagrangian takes the form
\begin{equation}
\mathcal{L}_{\mathrm{m}} = -\frac{1}{2\kappa}
\left( F_{\mu\nu}F^{\mu\nu} +
\gamma_{1}B^{\mu\nu}B^{\rho\sigma}F_{\mu\nu}F_{\rho\sigma} + \gamma_{2}B^{\mu\nu}B_{\mu\nu}F_{\rho\sigma}F^{\rho\sigma} \right),
\label{matter_lagrangian_KR}
\end{equation}
with the constants $\gamma_{1}$ and $\gamma_{2}$ determine the strength of the two electromagnetic nonminimal interactions. Varying the full action with respect to $g_{\mu\nu}$, $A_{\mu}$, and $B_{\mu\nu}$ yields the corresponding gravitational, gauge field, and Kalb--Ramond equations, namely
\begin{equation}
G_{\mu\nu}+\Lambda g_{\mu\nu} =
\kappa \left( T_{\mu\nu}^{\mathrm{KR}}
+ T_{\mu\nu}^{\mathrm{EM}} \right),
\label{modified_Einstein_KR}
\end{equation}
\begin{equation}
\nabla_{\nu} \left[ \left( 1+\gamma_{2}B_{\alpha\beta}B^{\alpha\beta} \right) F^{\mu\nu} + \gamma_{1}B_{\alpha\beta}F^{\alpha\beta}B^{\mu\nu} \right] =0,
\label{modified_Maxwell_KR}
\end{equation}
and
\begin{equation}
\nabla_{\alpha}H^{\mu\nu\alpha} -
4B^{\mu\nu}V'(X) - \frac{2\gamma_{1}}{\kappa}
B_{\alpha\beta}F^{\alpha\beta}F^{\mu\nu}
- \frac{2\gamma_{2}}{\kappa} F_{\alpha\beta}F^{\alpha\beta}B^{\mu\nu}
- \frac{\xi}{\kappa} \left( R^{\mu}{}_{\alpha}B^{\nu\alpha}
- R^{\nu}{}_{\alpha}B^{\mu\alpha} \right) =0.
\label{KR_eom}
\end{equation}
The resulting field equations make the coupling structure explicit: the Maxwell and Kalb--Ramond contributions interact through the parameters $\gamma_{1}$ and $\gamma_{2}$, while their backreaction on the metric provides an additional gravitational channel connecting both sectors.

We consider a static, spherically symmetric geometry described by the line element
\begin{equation}
\mathrm{d}s^{2} = -F(r)\mathrm{d}t^{2}
+ \frac{\mathrm{d}r^{2}}{F(r)} + r^{2} \left( \mathrm{d}\theta^{2} + \sin^{2}\theta\,\mathrm{d}\phi^{2} \right).
\label{metric_ansatz_KR}
\end{equation}
A pseudo-electric ansatz is adopted for the Kalb--Ramond background, so that the tensor field carries only a temporal--radial component,
\begin{equation}
B_{\mu\nu} =
\begin{pmatrix}
0 & \dfrac{b}{\sqrt{2}}\sqrt{F(r)G(r)} & 0 & 0\\
-\dfrac{b}{\sqrt{2}}\sqrt{F(r)G(r)} & 0 & 0 & 0\\
0 & 0 & 0 & 0\\
0 & 0 & 0 & 0
\end{pmatrix}.
\label{Bmunu_ansatz_KR}
\end{equation}
Notice that, with this choice, the Kalb--Ramond invariant is fixed by $B_{\mu\nu}B^{\mu\nu}=-b^{2}$, while its field strength vanishes, $H_{\mu\nu\rho}=0$. The dyonic sector is incorporated through the gauge potential $A_{\mu}=(-\Phi(r),0,0,p\cos\theta)$, where $p$ denotes the magnetic charge. This potential produces the electric component $F_{tr}=-F_{rt}=-\Phi'(r)$ and the magnetic component $F_{\theta\phi}=-F_{\phi\theta}=-p\sin\theta$.

We restrict the solution to the asymptotically non--flat branch with $\Lambda=0$ and choose the quadratic potential $V(X)=\lambda X^{2}/2$. The vacuum configuration follows from $X=0$, which also gives $V(X)=0$ and $V'(X)=0$. An exact dyonic black hole geometry arises after fixing the electromagnetic couplings according to $\gamma_{1}+\gamma_{2}=\xi/2$ and $\gamma_{2}=\xi/(2b^{2}\xi-2)$. The Lorentz--violating contribution is encoded in the dimensionless parameter $\ell=\xi b^{2}/2$. After expressing the solution in terms of the physical electric charge $Q$, the electrostatic potential becomes $\Phi(r)=-Q/[(1-\ell)r]$. The corresponding metric function takes the form
\begin{equation}
F(r) = \frac{1}{1-\ell} -
\frac{2M}{r} + \frac{Q^{2}}{(1-\ell)^{2}r^{2}}
+ \frac{p^{2}}{(1-2\ell)r^{2}}.
\label{metric_function_KR_flat}
\end{equation}
The resulting dyonic Kalb--Ramond black hole geometry can therefore be written as \cite{Lin:2026ewo}
\begin{equation}
\begin{split}
\mathrm{d}s^{2} &= - \left[ \frac{1}{1-\ell} - \frac{2M}{r} + \frac{Q^{2}}{(1-\ell)^{2}r^{2}} + \frac{p^{2}}{(1-2\ell)r^{2}} \right]\mathrm{d}t^{2} \\
& + \left[ \frac{1}{1-\ell} - \frac{2M}{r} + \frac{Q^{2}}{(1-\ell)^{2}r^{2}} + \frac{p^{2}}{(1-2\ell)r^{2}} \right]^{-1} \mathrm{d}r^{2} + r^{2}\mathrm{d}\Omega^{2},
\end{split}
\label{dyonic_KR_metric}
\end{equation}
Here, $\mathrm{d}\Omega^{2}=\mathrm{d}\theta^{2}+\sin^{2}\theta\,\mathrm{d}\phi^{2}$ denotes the metric on the unit two--sphere. The constants $M$, $Q$, and $p$ characterize, respectively, the mass, electric charge, and magnetic charge of the black hole, whereas $\ell$ parametrizes the correction generated by the Kalb--Ramond background. When the Lorentz--violating contribution is switched off, $\ell\to0$, the metric function becomes $F(r)\to 1-2M/r+(Q^{2}+p^{2})/r^{2}$, and the usual dyonic Reissner--Nordström spacetime is recovered. The sector $p=0$ gives the electrically charged Kalb--Ramond black hole, while the choice $Q=p=0$ leaves the neutral Lorentz--violating Kalb--Ramond solution.

The subsequent analysis employs the metric \eqref{dyonic_KR_metric} as the fixed background for the study of perturbations, time--domain evolution, tidal behavior, and relativistic effects in the dyonic Kalb--Ramond geometry.


\section{The gravitational Doppler effect }
\label{sec:gravitational_doppler_effect}

We next consider the gravitational Doppler shift produced by radial photon exchange in the dyonic Kalb--Ramond spacetime. The emitting source follows a freely falling trajectory, whereas the receivers are taken to be static observers. Our treatment uses the usual construction of local velocity measurements and frequency ratios in static black hole geometries \cite{Cordeiro:2025krdoppler,Crawford:2002,Radosz:2009,Augousti:2018,Radosz:2019}. It also generalizes the charged Kalb--Ramond case studied in Ref.~\cite{Cordeiro:2025krdoppler} by including both electric and magnetic charges. Since the metric function has already been specified in Eq.~\eqref{metric_function_KR_flat}, we only introduce here the effective dyonic charge parameter
\begin{equation}
\mathcal{P}_{\ell}^{2} \equiv \frac{Q^{2}}{(1-\ell)^{2}} + \frac{p^{2}}{1-2\ell}.
\label{effective_charge_doppler}
\end{equation}
In this notation, the roots of $F(r)=0$ determine the horizon radii,
\begin{equation}
r_{\pm} = (1-\ell) \left[ M \pm \sqrt{ M^{2}
- \frac{Q^{2}}{(1-\ell)^{3}} - \frac{p^{2}}{(1-\ell)(1-2\ell)}} \right],
\label{horizons_doppler}
\end{equation}
where these radii exist when the radicand remains non--negative. The limiting case in which this quantity vanishes defines the extremal dyonic Kalb--Ramond configuration.

For radial motion along a timelike trajectory, the four velocity normalization $u_{\mu}u^{\mu}=-1$ leads to
\begin{equation}
-F(r)\dot{t}^{2} + \frac{\dot{r}^{2}}{F(r)} = -1.
\label{radial_timelike_norm_doppler}
\end{equation}
Here, a dot indicates differentiation with respect to the proper time $\tau$. The static character of the geometry implies the existence of a conserved energy per unit rest mass, written as
\begin{equation}
E=F(r)\dot{t}.
\label{energy_conservation_doppler}
\end{equation}
In this manner, the radial geodesic equation turns out to be
\begin{equation}
\dot{r}^{2} = E^{2}-F(r).
\label{radial_geodesic_doppler}
\end{equation}
For a particle that starts from rest at $r=b$, the conserved energy satisfies $E^{2}=F(b)$. When the initial position is taken in the asymptotic region, the non-flat behavior of the spacetime at large $r$ instead leads to
\begin{equation}
E^{2}=F(\infty)=\frac{1}{1-\ell}.
\label{energy_infinity_doppler}
\end{equation}

In the exterior domain, a static observer assigns to the infalling particle a local velocity defined by
\begin{equation}
\mathrm{d}\tau^{2} =  F(r)\mathrm{d}t^{2} \left(1-v^{2}\right), \qquad
v^{2} = \frac{1}{F(r)^{2}}
\left( \frac{\mathrm{d}r}{\mathrm{d}t} \right)^{2}.
\label{velocity_definition_doppler}
\end{equation}
Combining Eqs.~\eqref{energy_conservation_doppler} and \eqref{radial_geodesic_doppler}, the velocity becomes
\begin{equation}
v^{2}(r) = \frac{E^{2}-F(r)}{E^{2}}.
\label{velocity_general_doppler}
\end{equation}
For an infalling particle whose initial rest position is placed at spatial infinity, Eq.~\eqref{velocity_general_doppler} simplifies to
\begin{equation}
v^{2}(r) = (1-\ell) \left(
\frac{2M}{r} - \frac{\mathcal{P}_{\ell}^{2}}{r^{2}} \right).
\label{velocity_explicit_doppler}
\end{equation}
At the outer horizon, $r=r_{+}$, the metric function vanishes, $F(r_{+})=0$. Consequently, the local free--fall speed satisfies $v(r_{+})=1$, meaning that a static observer outside the black hole measures the infalling particle as reaching the speed of light at the horizon.

In the interval bounded by the inner and outer horizons, $r_{-}<r<r_{+}$, the radial and temporal coordinates interchange their causal roles. The free--fall motion in this interior domain is therefore characterized by the velocity
\begin{equation}
V^{2}(r) = \frac{E^{2}}{E^{2}-F(r)}.
\label{interior_velocity_general_doppler}
\end{equation}
For $E^{2}=1/(1-\ell)$, this yields
\begin{equation}
V^{2}(r) = \frac{r^{2}}{(1-\ell)\left(2Mr-\mathcal{P}_{\ell}^{2}\right)}.
\label{interior_velocity_explicit_doppler}
\end{equation}
The minimum of $V(r)$ is obtained from $\mathrm{d}V^{2}/\mathrm{d}r=0$, giving
\begin{equation}
r_{\mathrm{m}} = \frac{\mathcal{P}_{\ell}^{2}}{M}
= \frac{1}{M} \left[ \frac{Q^{2}}{(1-\ell)^{2}}
+ \frac{p^{2}}{1-2\ell} \right].
\label{rm_doppler}
\end{equation}
{For non--extremal black holes, this radius lies between the horizons and marks the minimum of the frequency ratio given below.}

We now describe radial photon propagation by introducing the wave vector $k^{\mu}=\mathrm{d}x^{\mu}/\mathrm{d}\sigma$, where $\sigma$ is an affine parameter along the null trajectory. The condition $k_{\mu}k^{\mu}=0$ then gives
\begin{equation}
g_{\mu\nu}k^{\mu}k^{\nu}=0.
\label{null_condition_doppler}
\end{equation}
The conserved photon energy gives
\begin{equation}
k^{t} = \frac{\omega_{\infty}}{F(r)},
\qquad k^{r} = \pm \omega_{\infty},
\label{photon_wave_vector_doppler}
\end{equation}
with the plus sign describes an outgoing radial photon, whereas the minus sign refers to an ingoing one. For an observer with $u^{\mu}$, the measured photon frequency is
\begin{equation}
\omega = -k_{\mu}u^{\mu}.
\label{measured_frequency_doppler}
\end{equation}

We initially consider the photon emitted by the freely falling source and detected by a static observer located at the release radius $r=b$. Using the initial rest condition $E^{2}=F(b)$, the frequency measured at reception relative to the emitted one is
\begin{equation}
\frac{\omega^{(r)}_{b}}{\omega^{(s)}_{\mathrm{ff}}} = 1-v(r).
\label{frequency_ratio_outgoing_doppler}
\end{equation}

When the source approaches the event horizon, $v\rightarrow1$, and we have
\begin{equation}
\frac{\omega^{(r)}_{b}}{\omega^{(s)}_{\mathrm{ff}}} \longrightarrow 0.
\label{infinite_redshift_doppler}
\end{equation}
Consequently, the radiation produced by the freely falling emitter is observed in the exterior frame with an unbounded redshift.

On the other hand, when the photon is sent from the static observer at $r=b$ and detected by the freely falling observer, the corresponding frequency ratio is
\begin{equation}
\frac{\omega^{(r)}_{\mathrm{ff}}}{\omega^{(s)}_{b}} = \frac{1}{1+v(r)}.
\label{frequency_ratio_ingoing_doppler}
\end{equation}
In addition, at the outer horizon, this ratio becomes
\begin{equation}
\frac{\omega^{(r)}_{\mathrm{ff}}}{\omega^{(s)}_{b}} \longrightarrow \frac{1}{2}.
\label{ingoing_horizon_limit_doppler}
\end{equation}
This nonreciprocal behavior is a direct consequence of the horizon causal structure. Once the emitter crosses the outer horizon, photons directed outward no longer return to the exterior region, while photons sent inward from outside the black hole may still intercept the falling observer.

In the interior domain, $r_{-}<r<r_{+}$, an observer placed between the two horizons measures the frequency ratio as
\begin{equation}
\frac{\omega^{(r)}_{r_{0}}}{\omega^{(s)}_{\mathrm{ff}}}
= \left(1+V^{-1}\right)^{-1}.
\label{frequency_ratio_inside_general_doppler}
\end{equation}
Also, taking into account Eq.~\eqref{interior_velocity_explicit_doppler}, we obtain
\begin{equation}
\frac{\omega^{(r)}_{r_{0}}}{\omega^{(s)}_{\mathrm{ff}}}
= \left[ 1+ \sqrt{ (1-\ell) \left(
\frac{2M}{r} - \frac{\mathcal{P}_{\ell}^{2}}{r^{2}} \right) } \right]^{-1}.
\label{frequency_ratio_inside_explicit_alt_doppler}
\end{equation}
Since the metric function vanishes at the two horizons, $F(r_{\pm})=0$, the interior velocity satisfies $V(r_{\pm})=1$. In this manner,
\begin{equation}
\frac{\omega^{(r)}_{r_{0}}}{\omega^{(s)}_{\mathrm{ff}}} \longrightarrow
\frac{1}{2}, \qquad r\rightarrow r_{\pm}.
\label{inside_horizon_limits_doppler}
\end{equation}
{The frequency ratio reaches its minimum at $r = r_m$. Along an inward trajectory, it decreases for $r_m < r <r_+$ and increases for $r_-< r < r_m$. Nevertheless, Eq.~(\ref{frequency_ratio_inside_explicit_alt_doppler}) remains below unity throughout this region, approaching $1/2$ at either horizon. In this view, such an increasing represents a weakening redshift relative to the emitted frequency.}

When the Lorentz--violating parameter is removed, $\ell\rightarrow0$, Eq.~\eqref{frequency_ratio_inside_explicit_alt_doppler} recovers the dyonic Reissner--Nordstr\"om expression,
\begin{equation}
\frac{\omega^{(r)}_{r_{0}}}{\omega^{(s)}_{\mathrm{ff}}} \longrightarrow
\left[ 1+ \sqrt{ \frac{2M}{r} - \frac{Q^{2}+p^{2}}{r^{2}}}\right]^{-1}.
\label{RN_limit_doppler}
\end{equation}
{Setting $p=0$ gives the purely electric Kalb--Ramond case. When $Q=p=0$, the inner horizon disappears and there is no finite positive radius $r_m$ at which the frequency ratio reverses its radial trend.}

\begin{figure}
    \centering
    \includegraphics[scale=0.55]{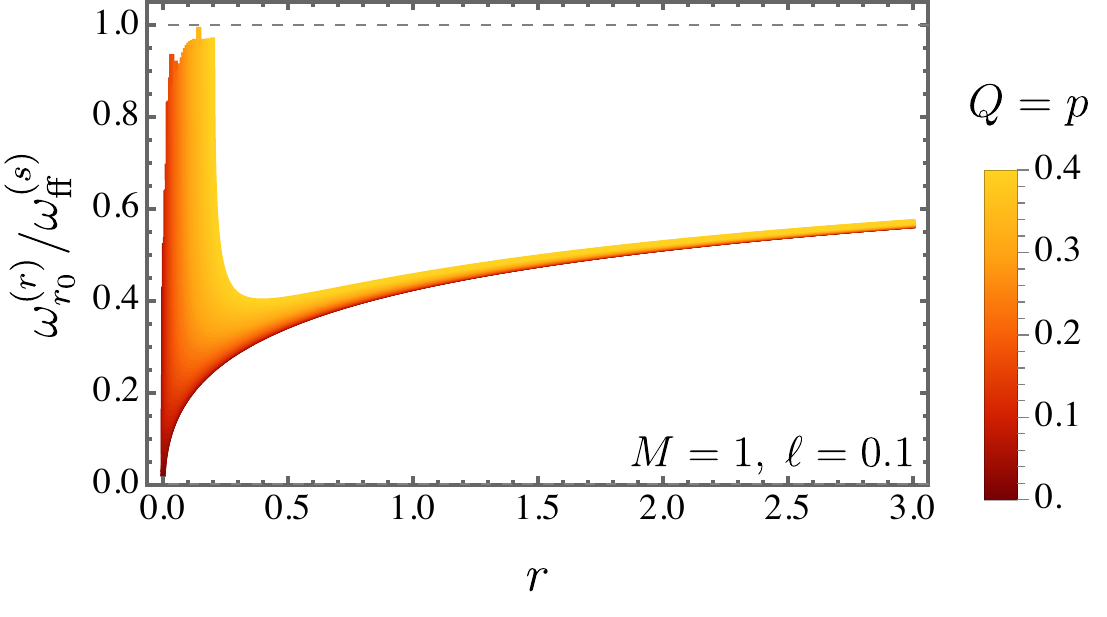}
    \caption{Radial behavior of the gravitational Doppler frequency ratio 
$\omega^{(r)}_{r_{0}}/\omega^{(s)}_{\mathrm{ff}}$ for different values of the dyonic charge parameter, assuming $Q=p$, with $M=1$ and $\ell=0.1$. }
    \label{Dopplereffect}
\end{figure}

Fig.~\ref{Dopplereffect} displays the radial profile of the frequency ratio $\omega^{(r)}_{r_{0}}/\omega^{(s)}_{\rm ff}$ obtained from Eq.~\eqref{frequency_ratio_inside_explicit_alt_doppler} for several dyonic configurations with $Q=p$, fixing $M=1$ and $\ell=0.1$. The ratio remains below unity in the plotted branch, indicating a redshift of the received signal with respect to the emitted one. The electric and magnetic charges enter through $P_{\ell}^{2}$ and reduce the square root contribution. Larger values of $Q=p$ shift the curves upward and bring the ratio closer to unity, corresponding to a weaker Doppler redshift.  {For nonzero charges, the minimum occurs at $r_m=\mathcal{P}_\ell^2/M$, separating the decreasing and increasing cases of the frequency ratio along the inward trajectory.}

\section{Tidal forces }
\label{sec:tidal_forces}

We study the tidal response of a neutral test body falling radially into the dyonic Kalb--Ramond black hole. In this setting, tidal forces are obtained from the relative acceleration of nearby geodesics and give a local characterization of the curvature acting on an extended body \cite{Misner:1973prb,Chandrasekhar:1983,Wald:1984rg,Poisson:2004}. In charged black hole spacetimes, the standard Schwarzschild behavior -- radial stretching together with angular compression -- may be altered, since charge terms can change the sign of the tidal components near the inner region \cite{Crispino:2016pnv,Cordeiro:2025krdoppler}. For the dyonic Kalb--Ramond solution, the electric and magnetic charges contribute with distinct Lorentz--violating factors. As a result, the radii at which the radial and angular tidal sectors change sign depend explicitly on $\ell$.


\subsection{Tidal forces for static and spherically symmetric spacetimes }
\label{subsec:tidal_general}

The analysis is based on the static, spherically symmetric metric given in Eq.~\eqref{dyonic_KR_metric}. To keep the tidal expressions compact, we define the effective dyonic charge parameter
\begin{equation}
P_{\ell}^{2} \equiv \frac{Q^{2}}{(1-\ell)^{2}} + \frac{p^{2}}{1-2\ell},
\label{eq:Pell_tidal}
\end{equation}
and with this definition, the metric function in Eq.~\eqref{metric_function_KR_flat} assumes the compact form
\begin{equation}
F(r) = \frac{1}{1-\ell} - \frac{2M}{r} + \frac{P_{\ell}^{2}}{r^{2}}.
\label{eq:F_tidal_compact}
\end{equation}
When the Lorentz--violating parameter is removed, $\ell\rightarrow0$, the effective charge reduces to $P_{\ell}^{2}\rightarrow Q^{2}+p^{2}$, and the metric returns to the usual dyonic Reissner--Nordstr\"om form. The sector with $p=0$ instead gives the electrically charged Kalb--Ramond solution.

The tidal acceleration is governed by the geodesic deviation equation,
\begin{equation}
\frac{D^{2}\eta^{\mu}}{D\tau^{2}} + R^{\mu}_{\ \nu\alpha\beta} u^{\nu}\eta^{\alpha}u^{\beta} = 0,
\label{eq:geodesic_deviation_tidal}
\end{equation}
where, in this expression, $\eta^{\mu}$ denotes the separation vector between neighboring geodesics, while $u^{\mu}$ represents the four--velocity of the radially infalling observer. For a timelike radial trajectory, the condition $u_{\mu}u^{\mu}=-1$ yields
\begin{equation}
-F(r)\dot{t}^{2} + \frac{\dot{r}^{2}}{F(r)} = -1,
\label{eq:normalization_tidal}
\end{equation}
where the dot denotes differentiation with respect to the proper time $\tau$. The time translation symmetry of the spacetime leads to a conserved energy per unit rest mass, given by
\begin{equation}
E = F(r)\dot{t}.
\label{eq:energy_tidal}
\end{equation}
Now, let us substitute Eq.~\eqref{eq:energy_tidal} into Eq.~\eqref{eq:normalization_tidal}, so that
\begin{equation}
\dot{r}^{2} = E^{2} - F(r).
\label{eq:radial_motion_tidal}
\end{equation}
For inward radial motion, the radial component satisfies $\dot{r}<0$. In the freely falling frame, one may introduce the following comoving orthonormal tetrad:
\begin{align}
e_{\hat{0}}^{\ \mu} &= \left( \frac{E}{F(r)},
-\sqrt{E^{2}-F(r)}, 0, 0 \right),
\\ e_{\hat{1}}^{\ \mu}
&= \left( -\frac{\sqrt{E^{2}-F(r)}}{F(r)}, E, 0, 0 \right),
\\ e_{\hat{2}}^{\ \mu} &= \left( 0, 0, \frac{1}{r}, 0 \right), \qquad e_{\hat{3}}^{\ \mu} =
\left( 0, 0, 0, \frac{1}{r\sin\theta} \right).
\label{eq:tetrad_tidal}
\end{align}
This local frame satisfies
\begin{equation}
g_{\mu\nu} e_{\hat{a}}^{\ \mu} e_{\hat{b}}^{\ \nu} = \eta_{\hat{a}\hat{b}}, \qquad \eta_{\hat{a}\hat{b}} = \mathrm{diag}(-1,1,1,1).
\label{eq:tetrad_orthonormality}
\end{equation}
It is worth mentioning that the tetrad is normalized so that $e_{\hat{0}}^{\ \mu}=u^{\mu}$. After Eq.~\eqref{eq:geodesic_deviation_tidal} is projected onto this locally inertial frame, the radial component of the tidal acceleration takes the form
\begin{equation}
\frac{D^{2}\eta^{\hat{1}}}{D\tau^{2}} = -\frac{F''(r)}{2}\eta^{\hat{1}},
\label{eq:radial_tidal_general}
\end{equation}
whereas the angular components satisfy
\begin{equation}
\frac{D^{2}\eta^{\hat{i}}}{D\tau^{2}} =
-\frac{F'(r)}{2r}\eta^{\hat{i}}, \qquad \hat{i}=\hat{2},\hat{3}.
\label{eq:angular_tidal_general}
\end{equation}
For the metric function considered here,
\eqref{eq:F_tidal_compact}, the derivatives are
\begin{equation}
F'(r) = \frac{2M}{r^{2}} - \frac{2P_{\ell}^{2}}{r^{3}}, \qquad F''(r) = -\frac{4M}{r^{3}} + \frac{6P_{\ell}^{2}}{r^{4}}.
\label{eq:F_derivatives_tidal}
\end{equation}
In other words, the projected equations give the radial stretching/compression and the transverse tidal response measured by the freely falling observer. 

\subsection{Radial tidal force }
\label{subsec:radial_tidal}

Substituting the derivatives in Eq.~\eqref{eq:F_derivatives_tidal} into the general radial relation \eqref{eq:radial_tidal_general}, one obtains the radial tidal acceleration for the dyonic Kalb--Ramond geometry as
\begin{equation}
\frac{D^{2}\eta^{\hat{1}}}{D\tau^{2}} =
\left( \frac{2M}{r^{3}} - \frac{3P_{\ell}^{2}}{r^{4}} \right) \eta^{\hat{1}}.
\label{eq:radial_tidal_dyonic_compact}
\end{equation}
Equivalently, it can be written as
\begin{equation}
\frac{D^{2}\eta^{\hat{1}}}{D\tau^{2}}
= \frac{1}{2}
\left[ \frac{4M}{r^{3}} -
\frac{6}{r^{4}} \left( \frac{Q^{2}}{(1-\ell)^{2}}
+ \frac{p^{2}}{1-2\ell} \right) \right] \eta^{\hat{1}}.
\label{eq:radial_tidal_dyonic}
\end{equation}
The term $2M/r^{3}$ represents the Schwarzschild part and gives rise to radial stretching. The charge dependent term, $-3\mathcal{P}_{\ell}^{2}/r^{4}$, has the opposite sign and becomes dominant sufficiently close to the central region. Then, in direct analogy with the Reissner--Nordstr\"om spacetime, the radial tidal component can pass from stretching to compression.

\begin{figure}
    \centering
    \includegraphics[scale=0.65]{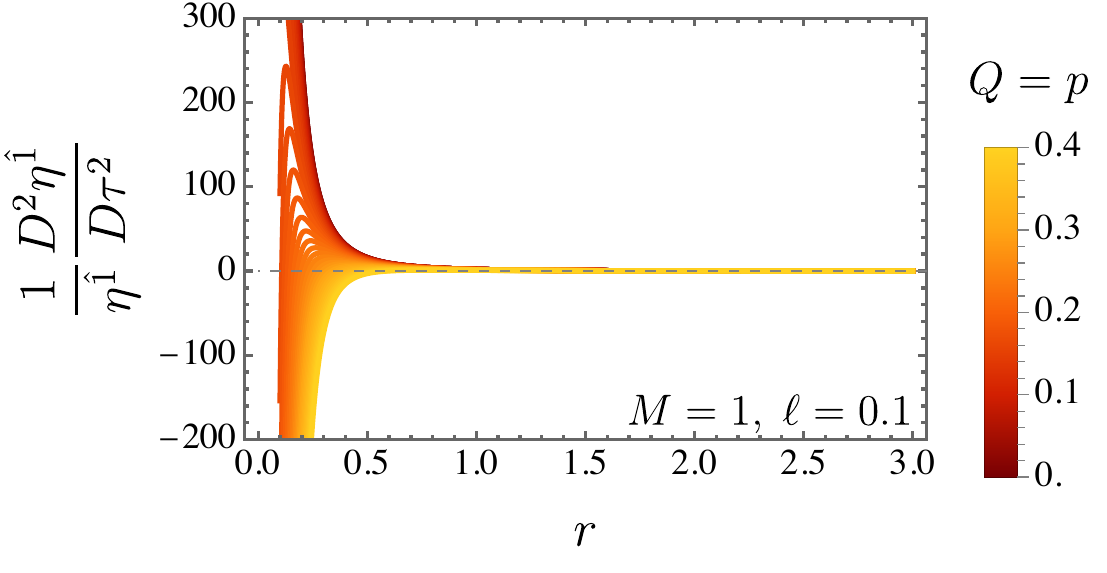}
    \caption{Radial tidal force for different values of the dyonic charges with $Q=p$, fixing $M=1$ and $\ell=0.1$. The dashed horizontal line separates radial stretching from radial compression.}
    \label{radialtidal}
\end{figure}

Figure~\ref{radialtidal} shows the radial tidal acceleration per unit separation, $(D^{2}\eta^{\hat{1}}/D\tau^{2})/\eta^{\hat{1}}$, as a function of $r$ for several dyonic configurations with $Q=p$, keeping $\ell=0.1$ fixed. Positive values describe radial stretching, while negative values correspond to radial compression. At large distances, the Schwarzschild--like contribution $2M/r^{3}$ controls the behavior, so the radial tidal force remains positive. Moving toward smaller radii, the dyonic term $-3\mathcal{P}_{\ell}^{2}/r^{4}$ grows faster and eventually reverses the sign of the force. The transition occurs at $R_{0}^{\mathrm{rad}}=3\mathcal{P}_{\ell}^{2}/(2M)$. Larger values of the common charge $Q=p$ increase $\mathcal{P}_{\ell}^{2}$, displacing the radial maximum and the zero crossing radius toward larger values of $r$.

Here, notice that the radial tidal force vanishes when
\begin{equation}
\frac{2M}{r^{3}} - \frac{3P_{\ell}^{2}}{r^{4}}
= 0,
\end{equation}
which gives us 
\begin{equation}
R_{0}^{\mathrm{rad}} = \frac{3P_{\ell}^{2}}{2M}
= \frac{3}{2M} \left[ \frac{Q^{2}}{(1-\ell)^{2}}
+ \frac{p^{2}}{1-2\ell} \right].
\label{eq:R0_radial_tidal}
\end{equation}
The sign of the radial tidal component changes at $R_{0}^{\mathrm{rad}}$. For $r>R_{0}^{\mathrm{rad}}$, the acceleration is positive, and an extended body is pulled apart along the radial direction. For $r<R_{0}^{\mathrm{rad}}$, the same component becomes negative, so the radial separation between neighboring parts of the body decreases. The stationary point of Eq.~\eqref{eq:radial_tidal_dyonic_compact} is obtained by imposing $\mathrm{d}\!\left[(D^{2}\eta^{\hat{1}}/D\tau^{2})/\eta^{\hat{1}}\right]/\mathrm{d}r=0$, which gives
\begin{equation}
R_{\mathrm{max}}^{\mathrm{rad}} =
\frac{2P_{\ell}^{2}}{M} = \frac{2}{M} \left[
\frac{Q^{2}}{(1-\ell)^{2}} + \frac{p^{2}}{1-2\ell} \right].
\label{eq:Rmax_radial_tidal}
\end{equation}
The radial tidal component attains its maximum at this point. For $P_{\ell}^{2}>0$, larger values of $Q$ or $p$ move both $R_{0}^{\mathrm{rad}}$ and $R_{\mathrm{max}}^{\mathrm{rad}}$ outward. The parameter $\ell$ modifies these characteristic radii through the weights $(1-\ell)^{-2}$ and $(1-2\ell)^{-1}$. For $0<\ell<1/2$, the magnetic sector is amplified by the factor $(1-2\ell)^{-1}$, whereas the electric sector is scaled by $(1-\ell)^{-2}$. Around $\ell=0$, the two charge sectors acquire the same leading-order relative correction. However, as $\ell$ approaches $1/2$, the magnetic contribution increases more strongly.

If we take into account the limit $\ell\rightarrow 0$, Eq.~\eqref{eq:radial_tidal_dyonic_compact} becomes
\begin{equation}
\frac{D^{2}\eta^{\hat{1}}}{D\tau^{2}} \rightarrow
\left[ \frac{2M}{r^{3}} - \frac{3(Q^{2}+p^{2})}{r^{4}} \right] \eta^{\hat{1}},
\label{eq:radial_tidal_RN_limit}
\end{equation}
which coincides with the dyonic Reissner--Nordstr\"om expression. In the neutral limit, $Q=p=0$, the result reduces to the Schwarzschild form,
\begin{equation}
\frac{D^{2}\eta^{\hat{1}}}{D\tau^{2}} = \frac{2M}{r^{3}} \eta^{\hat{1}}.
\label{eq:radial_tidal_schwarzschild_limit}
\end{equation}
In the neutral configuration, the local radial tidal component is independent of $\ell$. Nevertheless, the Lorentz--violating parameter still affects the spacetime through the constant contribution appearing in $F(r)$, and this modifies the position of the horizon.


\subsection{Angular tidal forces }
\label{subsec:angular_tidal}

The transverse tidal sector follows from Eq.~\eqref{eq:angular_tidal_general}. After inserting the derivatives listed in Eq.~\eqref{eq:F_derivatives_tidal}, the angular components take the form
\begin{equation}
\frac{D^{2}\eta^{\hat{i}}}{D\tau^{2}} =
\left( \frac{P_{\ell}^{2}}{r^{4}} - \frac{M}{r^{3}} \right) \eta^{\hat{i}},
\qquad \hat{i}=\hat{2},\hat{3}.
\label{eq:angular_tidal_dyonic_compact}
\end{equation}
Equivalently,
\begin{equation}
\frac{D^{2}\eta^{\hat{i}}}{D\tau^{2}} = \frac{1}{r^{4}} \left[ \frac{Q^{2}}{(1-\ell)^{2}} +
\frac{p^{2}}{1-2\ell} - Mr \right] \eta^{\hat{i}}, \qquad \hat{i}=\hat{2},\hat{3}.
\label{eq:angular_tidal_dyonic}
\end{equation}
Notice that the angular tidal force vanishes at
\begin{equation}
R_{0}^{\mathrm{ang}} = \frac{P_{\ell}^{2}}{M} =
\frac{1}{M} \left[ \frac{Q^{2}}{(1-\ell)^{2}}
+ \frac{p^{2}}{1-2\ell} \right],
\label{eq:R0_angular_tidal}
\end{equation}
and its corresponding minimum occurs at
\begin{equation}
R_{\mathrm{min}}^{\mathrm{ang}} = \frac{4P_{\ell}^{2}}{3M} = \frac{4}{3M}
\left[ \frac{Q^{2}}{(1-\ell)^{2}} + \frac{p^{2}}{1-2\ell} \right].
\label{eq:Rmin_angular_tidal}
\end{equation}
The angular tidal component changes sign at $R_{0}^{\mathrm{ang}}$. Outside this radius, $r>R_{0}^{\mathrm{ang}}$, the transverse acceleration is negative, so the body undergoes compression along the angular directions. Inside this radius, $r<R_{0}^{\mathrm{ang}}$, the angular component becomes positive. Therefore, sufficiently close to the inner region, the dyonic charge contribution overturns the Schwarzschild transverse compression and produces angular stretching.

For a particle starting from rest in the asymptotic region, the conserved energy satisfies
\begin{equation}
E^{2} = F(\infty) = \frac{1}{1-\ell}.
\label{eq:E_infty_tidal}
\end{equation}
The radial turning point is fixed by the condition $E^{2}=F(r)$, or equivalently by solving
\begin{equation}
R_{\mathrm{stop}} = \frac{P_{\ell}^{2}}{2M}
= \frac{1}{2M} \left[ \frac{Q^{2}}{(1-\ell)^{2}}
+ \frac{p^{2}}{1-2\ell} \right].
\label{eq:Rstop_tidal}
\end{equation}
For $P_{\ell}^{2}>0$, the corresponding radii read
\begin{equation}
\nonumber
R_{\mathrm{stop}} < R_{0}^{\mathrm{ang}} <
R_{\mathrm{min}}^{\mathrm{ang}} < R_{0}^{\mathrm{rad}} <
R_{\mathrm{max}}^{\mathrm{rad}}.
\label{eq:ordering_tidal}
\end{equation}
Because the falling body evolves from larger toward smaller radii, this hierarchy shows that the radial sector changes sign before the angular one. Along the trajectory, the radial tidal component first attains its maximum at $R_{\mathrm{max}}^{\mathrm{rad}}$ and then crosses zero at $R_{0}^{\mathrm{rad}}$. The transverse component follows a different sequence: it reaches its minimum at $R_{\mathrm{min}}^{\mathrm{ang}}$ and becomes zero only afterward, at $R_{0}^{\mathrm{ang}}$.

\begin{figure}
    \centering
    \includegraphics[scale=0.65]{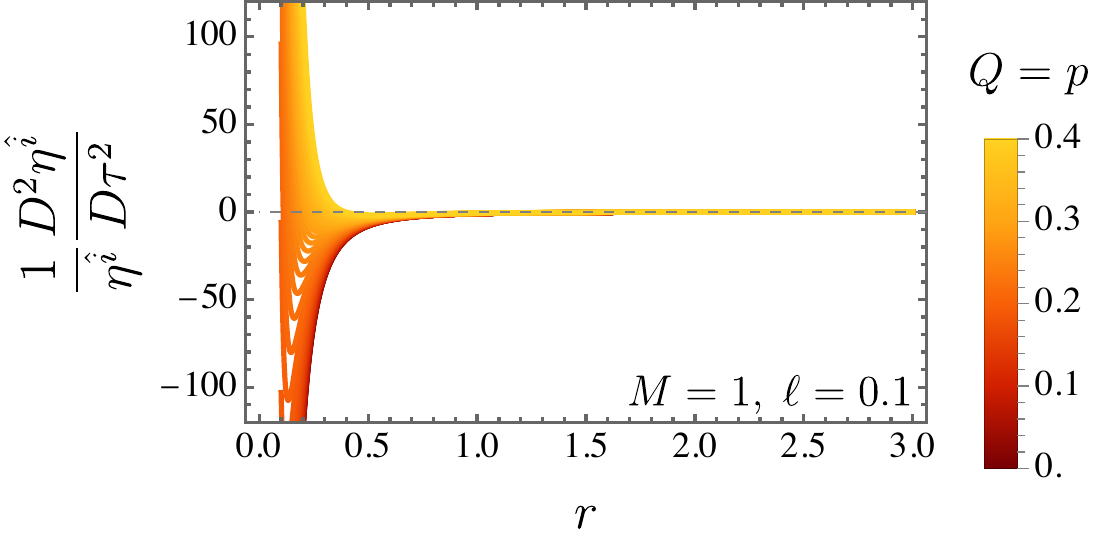}
    \caption{Angular tidal force for different values of the dyonic charges with $Q=p$, fixing $M=1$ and $\ell=0.1$. The dashed horizontal line separates angular compression from angular stretching.}
    \label{angulartidal}
\end{figure}

Figure~\ref{angulartidal} presents the angular tidal acceleration per unit separation, $(D^{2}\eta^{\hat{i}}/D\tau^{2})/\eta^{\hat{i}}$, as a function of $r$ for several dyonic configurations with $Q=p$, while keeping $\ell=0.1$. Unlike the radial component, the angular sector is negative in the large--distance region, which corresponds to transverse compression. This behavior is controlled by the Schwarzschild--like contribution $-M/r^{3}$. At smaller radii, the dyonic term $\mathcal{P}_{\ell}^{2}/r^{4}$ grows faster and may overcome the Schwarzschild part, turning the transverse compression into angular stretching. The zero of the angular component occurs at $R_{0}^{\mathrm{ang}}=\mathcal{P}_{\ell}^{2}/M$, whereas the maximum compression is reached at $R_{\mathrm{min}}^{\mathrm{ang}}=4\mathcal{P}_{\ell}^{2}/(3M)$. As the common charge $Q=p$ increases, $\mathcal{P}_{\ell}^{2}$ also increases, pushing both characteristic radii outward.

When the Lorentz--violating parameter is switched off, $\ell\rightarrow0$, Eq.~\eqref{eq:angular_tidal_dyonic_compact} takes the dyonic Reissner--Nordstr\"om form,
\begin{equation}
\frac{D^{2}\eta^{\hat{i}}}{D\tau^{2}} \rightarrow
\left[ \frac{Q^{2}+p^{2}}{r^{4}} - \frac{M}{r^{3}} \right] \eta^{\hat{i}}, \qquad \hat{i}=\hat{2},\hat{3}.
\label{eq:angular_tidal_RN_limit}
\end{equation}
In the neutral limit, $Q=p=0$, the angular sector reduces to the Schwarzschild transverse compression,
\begin{equation}
\frac{D^{2}\eta^{\hat{i}}}{D\tau^{2}} = -\frac{M}{r^{3}} \eta^{\hat{i}}, \qquad \hat{i}=\hat{2},\hat{3}.
\label{eq:angular_tidal_schwarzschild_limit}
\end{equation}

\begin{figure}
    \centering
    \includegraphics[scale=0.55]{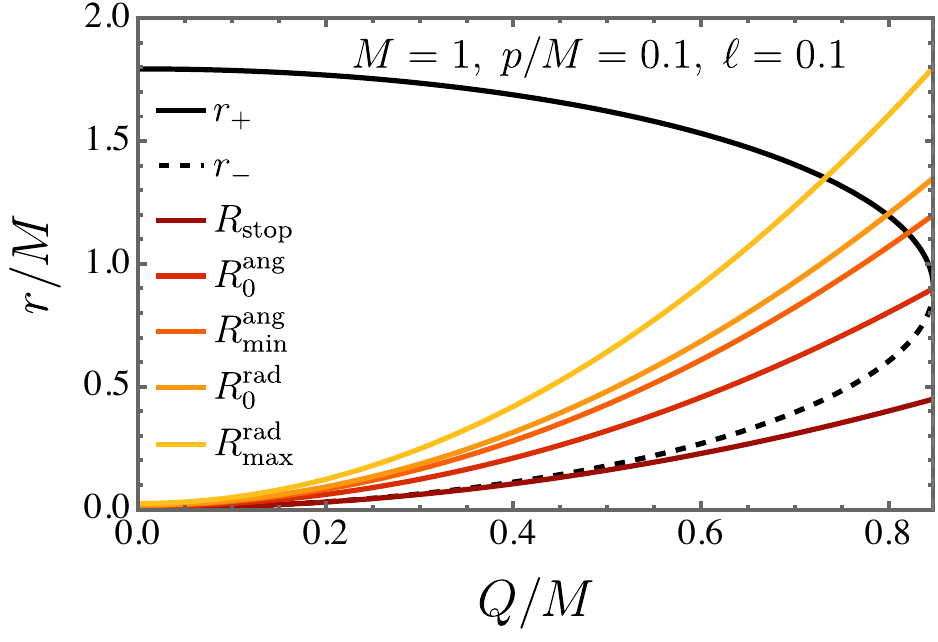}
    \includegraphics[scale=0.55]{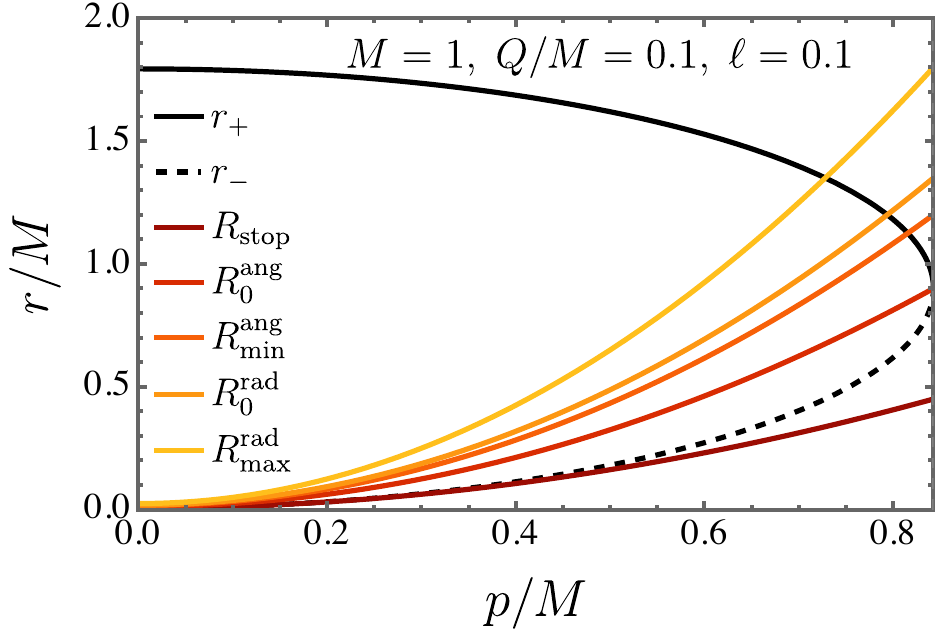}
    \caption{Horizon radii and characteristic tidal radii as functions of the charge parameters. On the left panel, the variation with $Q/M$ for fixed $p/M=0.1$. On the right one, the variation with $p/M$ for fixed $Q/M=0.1$. In both panels we set $M=1$ and $\ell=0.1$, while the vertical dashed line marks the extremal limit where $r_{+}=r_{-}$.}
    \label{rrrssss}
\end{figure}

Figure~\ref{rrrssss} compares the horizon positions with the main tidal radii in the electric and magnetic sectors. The left panel varies $Q/M$ at fixed $p/M=0.1$, while the right panel varies $p/M$ with $Q/M=0.1$. In both panels, a larger charge increases the effective combination $\mathcal{P}_{\ell}^{2}$, moving $R_{\mathrm{stop}}$, $R_{0}^{\mathrm{ang}}$, $R_{\mathrm{min}}^{\mathrm{ang}}$, $R_{0}^{\mathrm{rad}}$, and $R_{\mathrm{max}}^{\mathrm{rad}}$ to larger values of $r$. The horizons exhibit the opposite trend: $r_{+}$ moves inward and $r_{-}$ moves outward, so the two roots approach each other as the extremal limit is approached. The vertical dashed line identifies the critical charge at which the horizons coincide. Throughout the allowed interval, the hierarchy $R_{\mathrm{stop}}<R_{0}^{\mathrm{ang}}<R_{\mathrm{min}}^{\mathrm{ang}}<R_{0}^{\mathrm{rad}}<R_{\mathrm{max}}^{\mathrm{rad}}$ remains unchanged, indicating that an infalling body encounters the radial sign reversal before the angular one.

\section{Time delay }

The present discussion follows the approach of Ref.~\cite{Qiao:2024ehj}. We begin by formulating photon motion in a generic static and spherically symmetric spacetime, instead of introducing the time delay expression directly. Since light rays follow null geodesics, their trajectories are fixed by the geodesic equations together with the condition $\mathrm{d}s^{2}=0$. If the null curve is parameterized by $\lambda$, the spacetime symmetries generate conserved quantities along the photon path. In particular, the Killing field associated with time translations defines the conserved photon energy,
\ie
\label{seffdf}
E = F(r)\,\frac{\mathrm{d}t}{\mathrm{d}\lambda},
\fe
where the spherical symmetry provides a second conserved quantity, namely the photon angular momentum. Restricting the trajectory to the equatorial plane, $\theta=\pi/2$, this constant may be written as
\ie
{J} = r^{2}\sin^{2}\theta\,\frac{\mathrm{d}\phi}{\mathrm{d}\lambda}.
\fe

The photon trajectory is completed by the null constraint imposed on the four velocity. This condition follows from the metric Lagrangian $\mathcal{L}$ and relates the radial motion to the conserved quantities $E$ and {$J$}. Once written in this form, the geodesic system contains all the ingredients needed for the later computation of the gravitational time delay. Using the Lagrangian formulation, we get
\ie
\mathcal{L} = g_{\mu\nu}\mathrm{d}x^{\mu}\mathrm{d}x^{\nu} = F(r) \bigg( \frac{\mathrm{d}t}{\mathrm{d}\lambda} \bigg)^2
- \frac{1}{F(r)} \bigg( \frac{\mathrm{d}r}{\mathrm{d}\lambda} \bigg)^{2}
- r^{2} \bigg( \frac{\mathrm{d}\theta}{\mathrm{d}\lambda} \bigg)^{2} 
- r^{2}\sin^{2}\theta \bigg( \frac{\mathrm{d}\phi}{\mathrm{d}\lambda} \bigg)^{2}.
\fe

Owing to spherical symmetry, the motion can be restricted without loss of generality to the equatorial plane, $\theta=\pi/2$. With this choice, the geodesic system is reduced to the variables $r$ and $\phi$, together with the constants $E$ and {$J$}, which determine the radial and angular evolution of the light ray
\ie
\frac{1}{2} \bigg( \frac{\mathrm{d}r}{\mathrm{d}\lambda} \bigg)^{2} + \frac{1}{2} F(r) \bigg[ \frac{{J^{2}}}{r^{2}} + \mathcal{L} \bigg]
= \frac{1}{2} \bigg( \frac{\mathrm{d}r}{\mathrm{d}\lambda} \bigg)^{2} + V(r)
= \frac{1}{2}E^{2}.
\fe

In a static and spherically symmetric geometry, the radial part of the null geodesic equations can be written in the form of an effective one dimensional motion. After the constants of motion are introduced, the radial evolution is governed by a single function whose zeros determine the possible turning points of the photon trajectory. This function plays the role of an effective potential and, for the geometry under consideration, is given by
\ie
V(r) = \frac{F(r)}{2} \left[ \frac{{J^2}}{r^2} + \mathcal{L} \right].
\fe 
The impact parameter is introduced as the length scale fixed by the conserved quantities, $b=|{J}/E|$. For photons, the null condition sets the norm of the four velocity to zero, so the Lagrangian contribution $\mathcal{L}$ drops out of the radial constraint. After this restriction is applied to the lightlike trajectory, the radial equation assumes the reduced form
\ie
\frac{\mathrm{d}r}{\mathrm{d}t} = \frac{\mathrm{d}r}{\mathrm{d}\lambda}  \frac{\mathrm{d}\lambda}{\mathrm{d}t}
= \pm F(r) \sqrt{1 - b^{2}\frac{F(r)}{r^{2}}}.
\fe
For lightlike motion, the constraint $\mathcal{L}=0$ eliminates the norm contribution from the geodesic system, while the conserved energy enters through Eq.~\eqref{seffdf}. The resulting radial equation contains two branches, distinguished by the sign $\pm$. These branches describe the two portions of the photon trajectory: one in which the photon approaches the compact object and another in which it moves away from it. During the incoming part, $r$ decreases until the orbit reaches the closest-approach radius $r_{0}$. At this point, the radial derivative changes sign, and the photon continues along the outgoing branch. With this interpretation, the trajectory is described by the two relations
\ie
\frac{\mathrm{d}r}{\mathrm{d}t} = - F(r) \sqrt{1 - b^{2} \frac{F(r)}{r^{2}}} < 0.
\fe

The null orbit separates into an ingoing and an outgoing portion. Along the ingoing segment, the radial coordinate decreases until the photon reaches the closest approach radius $r_{0}$, where the radial motion turns around. After this turning point, the trajectory follows the outgoing branch, and $r$ increases as the photon moves away from the gravitational source. In this split description, the radial geodesic equations are written as
\ie
\frac{\mathrm{d}r}{\mathrm{d}t} = F(r) \sqrt{1 - b^{2} \frac{F(r)}{r^{2}}} > 0,  
\fe
After crossing the point of closest approach $r_{0}$, the photon follows the outgoing branch, and the radial coordinate increases until the ray reaches the observer at $r=r_{\text{O}}$; also, $b= r_{0}/\sqrt{F(r_{0}) }$ is the impact parameter. The source is located at $r=r_{S}$, while the detector is placed at $r=r_{\text{O}}$. The coordinate time collected over the ingoing and outgoing parts of the trajectory contains the delay produced by the gravitational field. Using the prescription of Ref.~\cite{Qiao:2024ehj}, this time delay is expressed as
\ie
\begin{split}
\label{asdasdddd}
 \Delta T & = T - T_{0} \\
 & = -\int_{r_{S}}^{r_{0}} \frac{\mathrm{d}r}{F(r)\sqrt{1-\frac{b^{2} F(r)}{r^{2}}}}
	      + \int_{r_{0}}^{r_{\text{O}}} \frac{\mathrm{d}r}{F(r)\sqrt{1-\frac{b^{2} F(r)}{r^{2}}}}
	      - T_{0}
	      \\
	& = \int_{r_{0}}^{r_{S}} \frac{\mathrm{d}r}{F(r)\sqrt{1-\frac{b^{2} F(r)}{r^{2}}}}
          + \int_{r_{0}}^{r_{\text{O}}} \frac{\mathrm{d}r}{F(r)\sqrt{1-\frac{b^{2} F(r)}{r^{2}}}}
          - \sqrt{r_{S}^{2}-r_{0}^{2}} - \sqrt{r_{\text{O}}^{2}-r_{0}^{2}}.
\end{split}
\fe
{For comparison, the Minkowski travel time between the prescribed radial endpoints, with closest approach radius $r_0$, is}
\ie
T_{0}= \sqrt{r_{S}^{2}-r_{0}^{2}} \; + \; \sqrt{r_{\text{O}}^{2}-r_{0}^{2}} .
\fe

\begin{figure}
    \centering
    \includegraphics[scale=0.5]{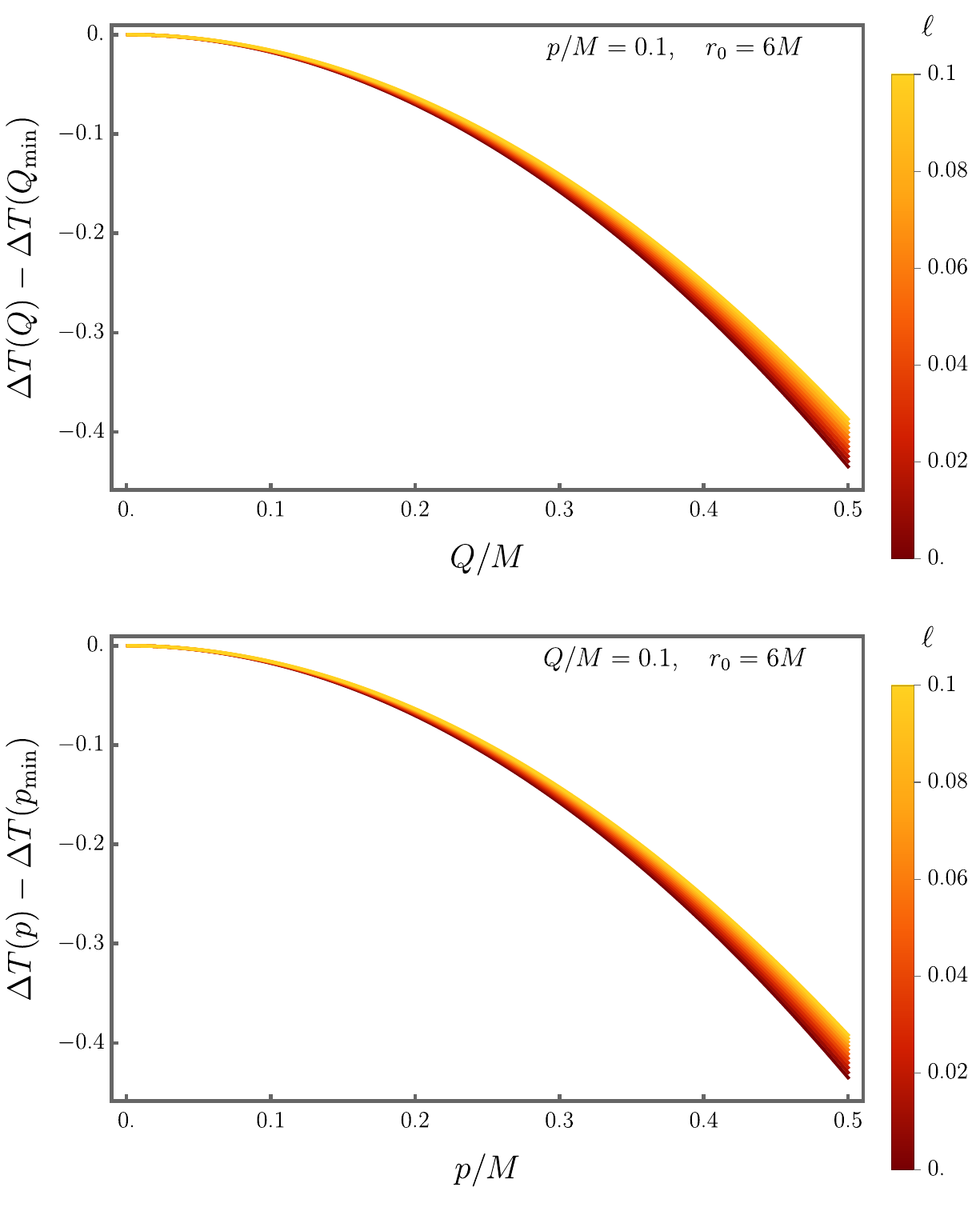}
    \caption{Shifted time delay as a function of the normalized electric and magnetic charges. Upper panel: $\Delta T(Q)-\Delta T(Q_{\min})$ versus $Q/M$ for fixed $p/M=0.1$. Bottom panel: $\Delta T(p)-\Delta T(p_{\min})$ versus $p/M$ for fixed $Q/M=0.1$. In both panels, $M=1$, $r_{0}=6M$, and $r_{\mathrm{S}}=r_{\mathrm{O}}=80M$, while the color gradient denotes the variation of the Lorentz--violating parameter $\ell$ in the interval $0\leq \ell\leq 0.1$.}
    \label{delayyy}
\end{figure}

{At fixed $\ell$, the background obtained by setting $M = Q = p = 0$ has $F(r)=1/(1-\ell)$. For the same radial endpoints and turning radius, its coordinate travel time is $T_{\mathrm{bg}} = (1 - \ell)T_0$. The quantity defined in Eq.~(\ref{asdasdddd}) separates as
\begin{equation}
\Delta T = - \ell \, T_0 + \left(T-T_{\mathrm{bg}}\right).
\end{equation}
The first term is independent of the mass and charges and arises from the {difference} between the Minkowski reference and the background at fixed $\ell$. For distant endpoints, it becomes
\begin{equation}
-\ell \,T_0\simeq-\ell(r_S+r_O), \qquad r_S,r_O\gg r_0.
\end{equation}
The remaining contribution, $T-T_{\mathrm{bg}}$, is the delay produced by the mass and charges relative to this background. In normalized asymptotic time, it is $\Delta\widehat T_{\mathrm{cone}}=(T-T_{\mathrm{bg}})/\sqrt{1-\ell}$, as discussed in Section~VII.F. The charge differences plotted in Fig.~\ref{delayyy} cancel the term $-\ell T_0$ because each subtraction is performed at fixed $\ell$, $M$, $r_S$, $r_O$, and $r_0$. In other words, these curves display the change in the delay caused by varying the charge.}

Figure~\ref{delayyy} shows the charge--induced variation of the time delay after subtracting the value computed at the smallest charge in the numerical range. This subtraction removes the common offset and isolates the dependence on the parameters of the dyonic Kalb--Ramond geometry. In the upper panel, the shifted quantity $\Delta T(Q)-\Delta T(Q_{\min})$ is plotted against $Q/M$ for fixed $p/M=0.1$, with $M=1$, $r_{0}=6M$, and $r_{\text{S}}=r_{\text{O}}=80M$. Each curve corresponds to a different value of the Lorentz--violating parameter $\ell$ within $0\leq \ell \leq 0.1$. The lower panel repeats the same procedure for $\Delta T(p)-\Delta T(p_{\min})$ as a function of $p/M$, keeping $Q/M=0.1$ fixed. In both panels, the shifted delay decreases as the varied charge increases, showing that both the electric and magnetic contributions reduce the travel time relative to the reference configuration.


\section{The corresponding effective potentials }

The dynamics of quantum fields on a curved spacetime can be reduced to an effective scattering problem in one spatial dimension. For bosonic and fermionic perturbations, described respectively by the Klein--Gordon and Dirac equations, the angular dependence is separated first. This procedure isolates the radial sector and allows the corresponding wave equation to be written in a Schrödinger--type form, with an effective potential $\mathrm{V}_{s,v,t,\psi}$.

Here, $\mathrm{V}_{s}$, $\mathrm{V}_{v}$, $\mathrm{V}_{t}$, and $\mathrm{V}_{\psi}$ denote the potentials associated with scalar, vector, tensor, and spinor perturbations, respectively. These functions control how the perturbing field propagates across the black hole geometry, fixing the reflection and transmission properties of the radial modes. They also enter directly in the calculation of quasinormal frequencies, absorption probabilities, and greybody factors. We therefore derive the explicit form of $\mathrm{V}_{s,v,t,\psi}$ for each spin sector before analyzing the quasinormal spectra and the time--domain profiles.

Although the metric used in the present work, Eq.~\eqref{metric_ansatz_KR}, has the simplified structure $-g_{tt}=1/g_{rr}$, it is useful to formulate the perturbation equations in a more general static and spherically symmetric background. We therefore consider
\begin{equation}
\nonumber
\mathrm{d}s^{2}=
-A(r)\mathrm{d}t^{2}
+\frac{\mathrm{d}r^{2}}{B(r)}
+r^{2}\left(
\mathrm{d}\theta^{2}
+\sin^{2}\theta\,\mathrm{d}\phi^{2}
\right),
\label{asssasss}
\end{equation}
which allows the effective potentials for the different spin sectors to be obtained in a form that can later be specialized to the dyonic Kalb--Ramond geometry.

\subsection{Bosonic sector  }

Bosonic perturbations with spin $s=0,1,$ and $2$ may be cast into a Schrödinger--like radial equation after the angular variables are separated. The construction is performed separately for each sector, since the scalar, vector, and tensor fields obey different covariant equations and therefore lead to distinct effective potentials. For the scalar case, $s=0$, the starting point is the Klein--Gordon equation
\ie
\label{klein}
\frac{1}{\sqrt{-g}} \partial_\mu \!\left( \sqrt{-g}\, g^{\mu\nu} \partial_\nu \Psi \right) = 0.
\fe
After a separation ansatz is imposed, the scalar field is written as a product of temporal, radial, and angular factors. This decomposition reduces the covariant Klein--Gordon equation to separate equations for each coordinate sector
\ie
\label{ansatz}
    \Psi_{\omega \ell m}(\mathbf{r},t) = \frac{\psi_{\omega \ell}(r)}{r} Y_{\ell m}(\theta, \varphi) e^{-i\omega t},
\fe
After introducing the tortoise coordinate $r^{*}$, the radial sector takes the form
\ie
\label{rdsdtdadr}
\mathrm{d}r^{*} = \frac{\mathrm{d}r}{\sqrt{A(r)B(r)}},
\fe
With this coordinate choice, the radial part of the Klein--Gordon equation is rewritten as a one--dimensional Schrödinger--like equation for the scalar radial mode,
\ie
\label{waves}
\left[\frac{\mathrm{d}^2}{\mathrm{d}r^{*2}} + \bigl(\omega^2 - \mathrm{V}_{s}\bigr)\right] \psi_{\omega \ell}(r) = 0.
\fe

For the vector sector, $s=1$, the perturbation is described by the Proca field equation,
\ie
\label{proca}
\nabla_\nu F^{\mu\nu} + m^2 A^\mu = 0.
\fe
{Throughout this work, we set $m=0$ and treat the vector sector as a minimally coupled Maxwell test field on the fixed background. After angular separation, its radiative modes satisfy the radial equation~(\ref{waves}), with the effective potential given in Eq.~(\ref{vectorrrrpot}).}

For tensor perturbations, $s=2$, the linearized Einstein equations split into axial and polar sectors, described respectively by Regge--Wheeler--type and Zerilli--type equations. Each sector is governed by its own effective potential. In the present analysis, we restrict attention to the axial gravitational sector only.

In this manner, scalar, vector, and axial tensor perturbations in a static and spherically symmetric background can be described by a unified master equation, with the spin dependence encoded in the corresponding effective potential. For the general metric in Eq.~\eqref{asssasss}, the scalar potential follows the form reported in Refs.~\cite{Heidari:2024bkm,AraujoFilho:2024xhm,AraujoFilho:2024lsi,AraujoFilho:2025zaj}
\ie
\mathrm{V}_{s}  = A(r)\left[\frac{{l(l + 1)}}{{{r^2}}} + \frac{1}{{r\sqrt {{A(r)}{B(r)^{ - 1}}} }}\frac{\mathrm{d}}{{\mathrm{d}r}}\sqrt {{A(r)}B(r)}\right]
\label{effssssss}
\fe
vector, on the other hand, reads \cite{Baruah:2025ifh,Filho:2023abd,Filho:2024ilq}
\ie
\mathrm{V}_{v} =  A(r)\left[\frac{{l(l + 1)}}{{{r^2}}} \right]
\label{vectorrrrpot}
\fe
and, moreover, the tensor perturbations are governed by \cite{AraujoFilho:2024xhm,AraujoFilho:2025zzf,AraujoFilho:2025hnf,Baruah:2025ifh,Chen:2019iuo,Bouhmadi-Lopez:2020oia,AraujoFilho:2025zaj,zerilli1970gravitational}
\ie
\mathrm{V}_{t} = A(r) \left[\frac{2 (B(r)-1)}{r^2}+\frac{l(l+1)}{r^2}   - \frac{1}{r \sqrt{A(r)B^{-1}(r)}} \frac{\partial}{\partial r} \left(  \sqrt{A(r) B(r)} \right) \right].
\label{tententen}
\fe

{The axial potential used here follows from the effective source prescription specified in Appendix \ref{dervitensor}, where the independent axial source amplitudes satisfy $\tau_t-p_\perp h_0=0$, $\tau_r-p_\perp h_1=0$, and $\tau_2=0$. These conditions close the metric perturbation equation. The resulting frequencies describe this effective axial model; in addition, a full derivation of the coupled gravitational, electromagnetic, and Kalb--Ramond spectrum from the action is outside the present calculation.}

It is important to mention that the effective potentials determine how each spin sector probes the black hole background. They contain the curvature contributions that enter the radial wave equation and therefore fix the scattering behavior of the perturbing field. Through these functions, we can analyze, for instance, the wave propagation, reflection and transmission coefficients, absorption probabilities, and the quasinormal mode spectrum of the massless bosonic perturbations.


\subsection{Fermionic sector }

The fermionic sector is treated separately from the bosonic perturbations. It is represented by a spin--$\tfrac{1}{2}$ field whose dynamics in the curved black hole background are governed by the covariant Dirac equation
\ie
\gamma^{\alpha}\!\left( \partial_{\alpha} - \omega_{\alpha} \right)\!\Psi = 0,
\fe
where, $\gamma^{\alpha}$ denote the gamma matrices adapted to the curved spacetime, and $\omega_{\alpha}$ is the spin connection constructed from the tetrad frame. Once the spinor field is decomposed into its angular and radial sectors, the Dirac equation reduces to a pair of radial equations. These equations can be decoupled by introducing the functions $\Psi^{\pm}$, which satisfy wave equations of Schrödinger type and define the corresponding spinor effective potentials
\ie
\left[\frac{\mathrm{d}^2}{\mathrm{d}{r^*}^2} + \bigl(\omega^2 - \mathrm{V}_{\psi}^{\pm}\bigr)\right]\Psi^\pm = 0.
\fe
We {highlight} that the signs $\pm$ distinguish the two independent spinor sectors obtained after decoupling the radial Dirac equations. 

{The angular dependence of the Dirac field is described by spinor spherical harmonics with total angular momentum $j=1/2,3/2,5/2,\ldots$. The corresponding Dirac quantum number is $\kappa_{\mathrm D}=\pm(j+1/2)$, where the subscript distinguishes it from the gravitational coupling $\kappa$. Defining $k\equiv|\kappa_{\mathrm D}|=j+1/2$, the effective potentials are \cite{albuquerque2023massless,al2024massless,arbey2021hawking,devi2020quasinormal}
\begin{equation}
\mathrm{V}_{\psi}^{\pm}(r)=\frac{k^2A(r)}{r^2}\pm k\sqrt{A(r)B(r)}\,\frac{\mathrm{d}}{\mathrm{d}r}\left(\frac{\sqrt{A(r)}}{r}\right).
\end{equation}
The signs distinguish the two supersymmetric partner potentials obtained from the radial Dirac equations. We use the positive partner in the calculations,
\begin{equation}
\mathrm{V}_{\psi}(r)\equiv \mathrm{V}_{\psi}^{+}(r)=\frac{k^2A(r)}{r^2}+k\sqrt{A(r)B(r)}\,\frac{\mathrm{d}}{\mathrm{d}r}\left(\frac{\sqrt{A(r)}}{r}\right).
\end{equation}
It is important to mention that, in the spinor sector, the half integer label $l$ used in the tables and figures denotes the total angular momentum $j$. Thereby, $k=|\kappa_{\mathrm D}|=j+1/2=l+1/2$, and the labels $l=1/2,3/2,5/2$ correspond to the Dirac angular eigenvalues $|\kappa_{\mathrm D}|=1,2,3$, respectively.}

After obtaining the potentials $\mathrm{V}_{s}$, $\mathrm{V}_{v}$, $\mathrm{V}_{t}$, and $\mathrm{V}_{\psi}$, we use them as the starting point for the dynamical analysis. The next sections compute the quasinormal spectra and construct the corresponding time--domain profiles {for all four perturbative sectors}.


\section{Quasinormal frequencies }

This section analyzes the role of the Lorentz--violating parameter $\ell$, together with the electric and magnetic charges $Q$ and $p$, in shaping the field perturbations supported by the black hole geometry. {Here, one additional remark is worthy to be commented: unless stated otherwise, the frequencies refer to the original time coordinate $t$ of the background metric. The scans in $\ell$ keep $M$, $Q$, and $p$ fixed. The distinction between these parameter features and comparisons using normalized time and effective parameters is discussed in Section~VII~F.}


\subsection{Scalar perturbations }

Using the metric of Eq.~(\ref{dyonic_KR_metric}) in the scalar perturbation potential introduced in Eq.~(\ref{effssssss}) gives
\ie
\mathrm{V}_{s} = \left(\frac{1}{1-\ell }-\frac{2 M}{r}+\frac{Q^2}{r^2 (1-\ell )^2}+\frac{p^2}{r^2 (1-2 \ell )}\right) \left(\frac{l (l+1)}{r^2}+\frac{2 M}{r^3}-\frac{2 Q^2}{r^4 (1-\ell )^2}-\frac{2 p^2}{r^4 (1-2 \ell )}\right).
\fe
In the simultaneous limit $Q,p,\ell \to 0$, the scalar potential recovers the usual Schwarzschild expression, which provides a direct consistency check of the result.

\begin{figure}
    \centering
    \includegraphics[scale=0.45]{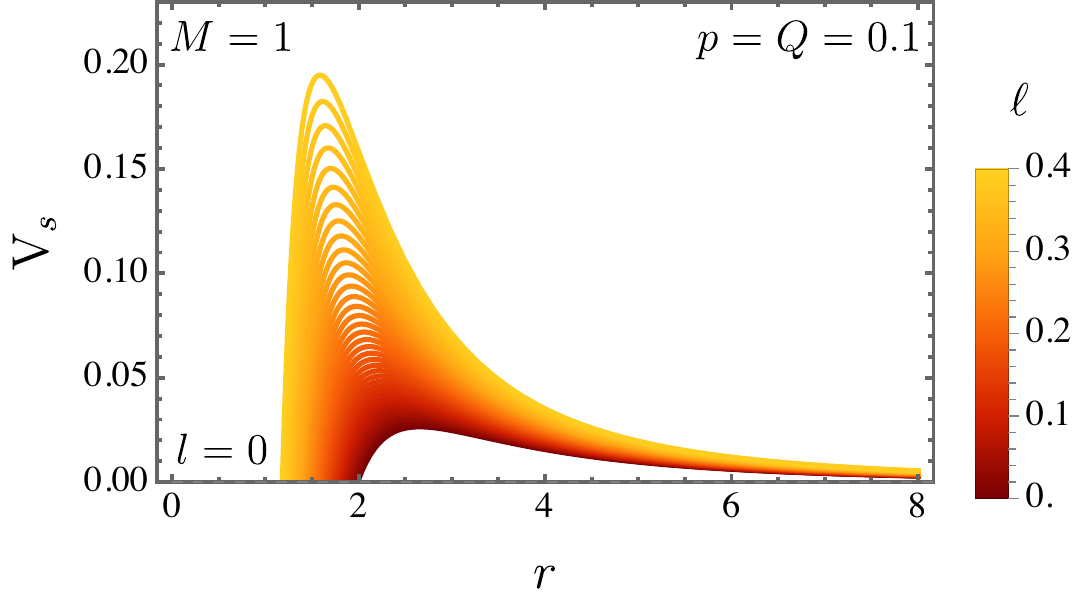}
    \includegraphics[scale=0.45]{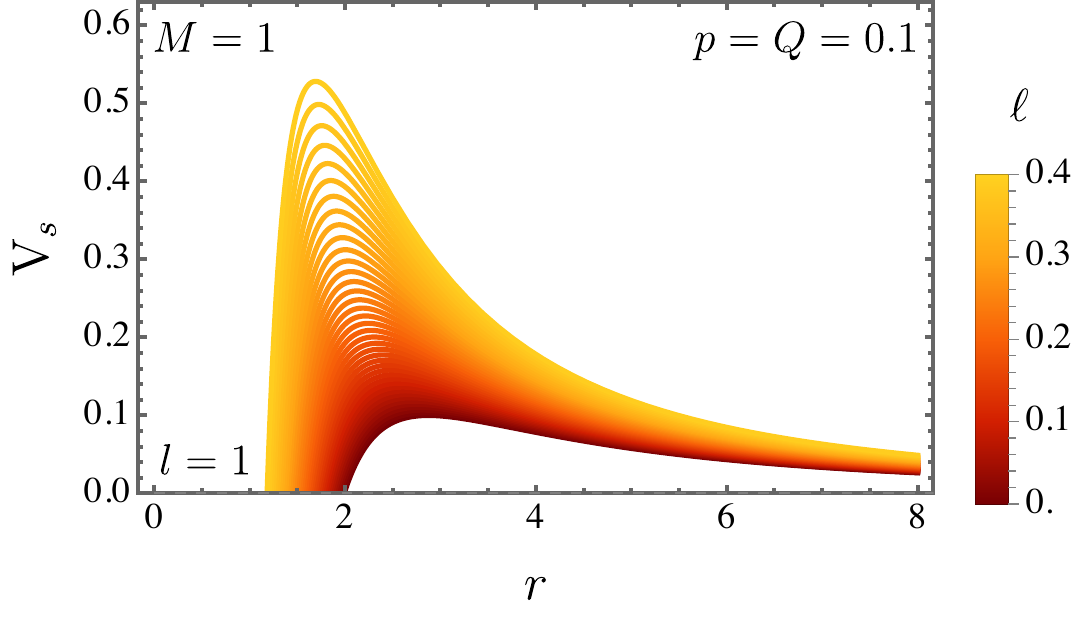}
     \includegraphics[scale=0.45]{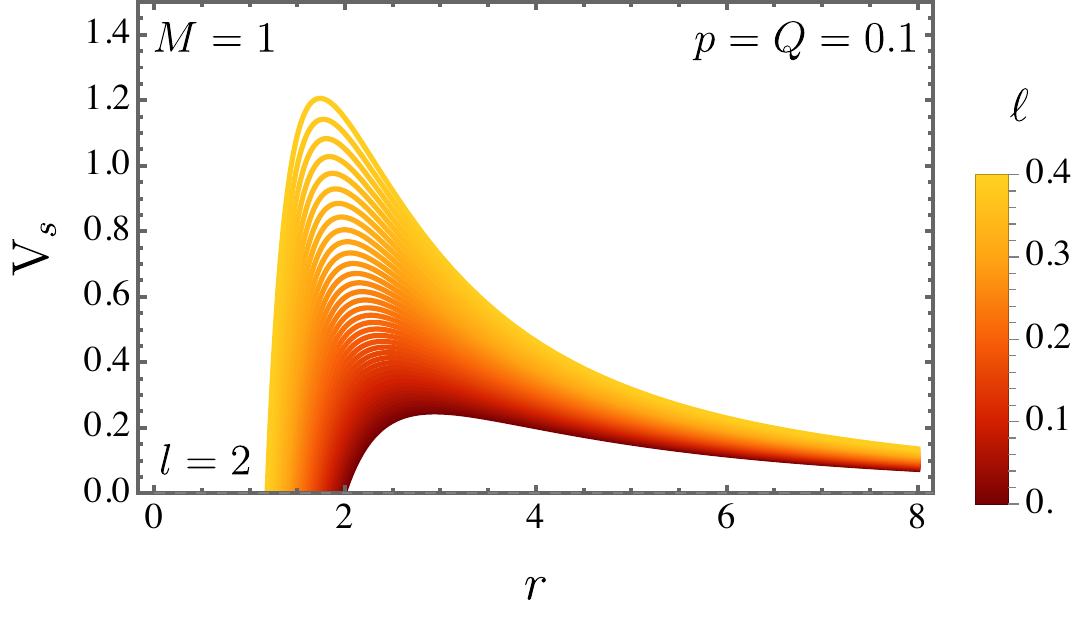}
    \caption{The scalar effective potential $\mathrm{V}_{s}$ as a function of the radial coordinate $r$ for $M=1$ and several values of $\ell$. The panels correspond to different multipole numbers: top--left, $l=0$; top--right, $l=1$; and bottom, $l=2$. }
    \label{spot}
\end{figure}

Figure~\ref{spot} displays the scalar effective potential as a function of the radial coordinate $r$ for different choices of $l$ and $\ell$. The potential barrier becomes higher as the Lorentz--violating parameter grows, showing that $\ell$ strengthens the radial barrier experienced by the scalar modes. A complementary description follows from rewriting $\mathrm{V}_{s}$ in terms of the tortoise coordinate $r^{*}$. Substitution of the metric in Eq.~(\ref{dyonic_KR_metric}) into Eq.~(\ref{rdsdtdadr}) gives
\ie
\begin{split}
& r^{*}  =    r(1-\ell) + M (\ell -1)^2 \ln \bigg[p^2 (\ell -1)^2-(2 \ell -1) \left(Q^2-r (\ell -1) (2 M (\ell -1)+r)\right)\bigg]\\
& +\frac{\sqrt{\ell -1} \left(-2 M^2 (2 \ell -1) (\ell -1)^3+p^2 (\ell -1)^2+Q^2 (1-2 \ell )\right) \tan ^{-1}\left(\frac{\sqrt{\ell -1} \sqrt{2 \ell -1} (M (\ell -1)+r)}{\sqrt{-M^2 (2 \ell -1) (\ell -1)^3+p^2 (\ell -1)^2+Q^2 (1-2 \ell )}}\right)}{\sqrt{2 \ell -1} \sqrt{-M^2 (2 \ell -1) (\ell -1)^3+p^2 (\ell -1)^2+Q^2 (1-2 \ell )}} .
\end{split}
\fe

\begin{figure}
    \centering
    \includegraphics[scale=0.5]{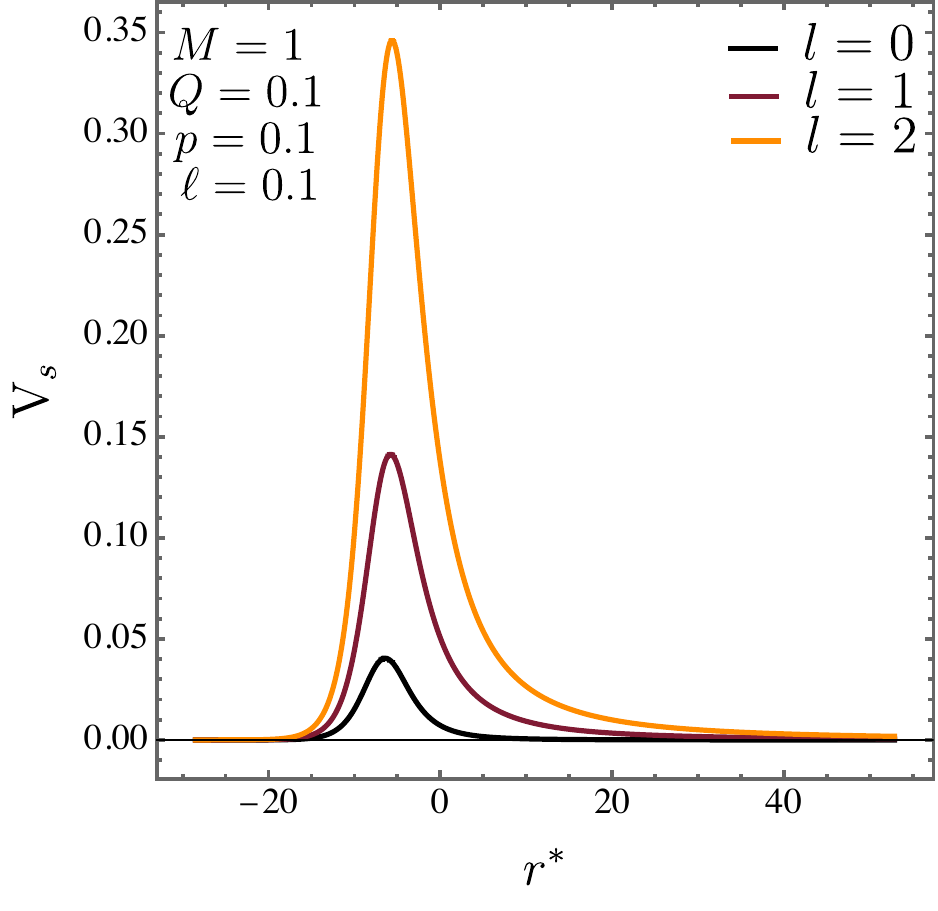}
    \caption{Scalar effective potential $\mathrm{V}_{s}$ as a function of the tortoise coordinate $r^{*}$ for $M=1$ and $\ell=p=Q=0.1$. The curves correspond to the multipole numbers $l=0,1,2$.}
    \label{stortoise}
\end{figure}

Figure~\ref{stortoise} shows $\mathrm{V}_{s}$ in terms of the tortoise coordinate $r^{*}$ for several values of the multipole number $l$, while the remaining parameters are fixed as $M=1$ and $\ell=p=Q=0.1$. {The resulting profile forms a smooth barrier with a single maximum, motivating the use of the WKB approximation. Its accuracy remains mode dependent and is independently addressed for selected modes in Section~VII~E.} Increasing $l$ raises the barrier height, in agreement with the behavior commonly found for scalar perturbations in Schwarzschild, bumblebee, and Kalb--Ramond black hole geometries. For each value of $l$, the potential contains only one peak; therefore, no secondary (and so forth) barrier appears in this sector. This feature indicates that echo--like contributions are not expected, a point that will be checked below through the time--domain evolution.

After characterizing the behavior of $\mathrm{V}_{s}$, we proceed to the scalar quasinormal spectrum computed with the 6th--order WKB method. The numerical implementation follows the publicly available code of Ref.~\cite{Konoplya:2019hlu,iyer1987black}. The frequencies obtained for $l=0$, $l=1$, and $l=2$ are reported in Tables~\ref{qnmsscalarl0}, \ref{qnmsscalarl1}, and \ref{qnmsscalarl2}, respectively. In all cases, the mass is fixed at $M=1$, while the Lorentz--violating parameter $\ell$ and the dyonic charges $Q$ and $p$ are varied.

{For fixed $\ell=0.1$, Tables \ref{qnmsscalarl0}--\ref{qnmsscalarl2} show that increasing the equal charges $Q=p$ raises both $\operatorname{Re}(\omega_0)$ and $|\operatorname{Im}(\omega_0)|$ for $l=0,1,2$. In the scalar monopole sector, increasing $Q=p$ from $0.01$ to $0.25$ changes the fundamental frequency from $0.136386-0.124466i$ to $0.140777-0.125473i$. The corresponding changes are from $0.345686-0.120879i$ to $0.356449-0.121858i$ for $l=1$, and from $0.568048-0.119538i$ to $0.585557-0.120585i$ for $l=2$. In this manner, throughout this charge interval, the tabulated fundamental modes oscillate faster and decay more rapidly as the charges increase.}

The parameter $\ell$ produces a more significant modification of the spectrum. For fixed $Q=p=0.1$, increasing $\ell$ raises both $\mathrm{Re}(\omega_{n})$ and $|\mathrm{Im}(\omega_{n})|$ in all scalar sectors analyzed here. Therefore, the Lorentz--violating contribution makes the scalar modes oscillate at higher frequencies and attenuate more quickly. Rather than reducing dissipation, this correction strengthens the decay of the ringdown signal.

{The tabulated fundamental modes have smaller imaginary magnitudes than the corresponding overtones and decay more slowly during ringdown. For the scalar $l=0$, $n=2$ mode at $M=1$ and $Q=p=\ell=0.1$, the sixth-order WKB approximation gives $\omega_2^{(6)}=0.236232-0.590051i$, whose real part exceeds that of the fundamental mode. As we shall {see}, the accuracy of this estimate is examined in Section~VII.E.}

\begin{table}[!h]
\centering
\caption{Scalar quasinormal spectrum for the $l=0$ sector. The frequencies $\omega_n$ are evaluated using the sixth order WKB approximation for the listed values of $M$, the equal charges $Q=p$, and the Lorentz symmetry breaking parameter $\ell$. {$^{*}$The scalar $l=0$, $n=2$ WKB estimates should not be used
as reliable quasinormal frequencies; see the independent AIM
comparison in Section~VII~E and Table~XV.}}
\label{qnmsscalarl0}
\begin{tabular}{c|c|c|c}
\hline\hline
$M,\; Q=p,\; \ell$ & $\omega_0$ & $\omega_1$ & $\omega_2^{{*}}$ \\[0.2ex]
\hline
$1.0,\; 0.01,\; 0.10$ & $0.136386 - 0.124466\,i$ & $0.109920 - 0.425347\,i$ & $0.236814 - 0.588174\,i$ \\
$1.0,\; 0.10,\; 0.10$ & $0.137045 - 0.124638\,i$ & $0.110637 - 0.425607\,i$ & $0.236232 - 0.590051\,i$ \\
$1.0,\; 0.15,\; 0.10$ & $0.137897 - 0.124847\,i$ & $0.111567 - 0.425903\,i$ & $0.235474 - 0.592423\,i$ \\
$1.0,\; 0.20,\; 0.10$ & $0.139127 - 0.125129\,i$ & $0.112919 - 0.426252\,i$ & $0.234366 - 0.595747\,i$ \\
$1.0,\; 0.25,\; 0.10$ & $0.140777 - 0.125473\,i$ & $0.114751 - 0.426584\,i$ & $0.232853 - 0.600036\,i$ \\
\hline\hline
$1.0,\; 0.10,\; 0.10$ & $0.137045 - 0.124638\,i$ & $0.110637 - 0.425607\,i$ & $0.236232 - 0.590051\,i$ \\
$1.0,\; 0.10,\; 0.15$ & $0.153792 - 0.139770\,i$ & $0.124199 - 0.477207\,i$ & $0.264708 - 0.661933\,i$ \\
$1.0,\; 0.10,\; 0.20$ & $0.173843 - 0.157844\,i$ & $0.140456 - 0.538803\,i$ & $0.298630 - 0.747893\,i$ \\
$1.0,\; 0.10,\; 0.21$ & $0.178327 - 0.161879\,i$ & $0.144095 - 0.552549\,i$ & $0.306189 - 0.767102\,i$ \\
$1.0,\; 0.10,\; 0.22$ & $0.182990 - 0.166071\,i$ & $0.147880 - 0.566829\,i$ & $0.314036 - 0.787068\,i$ \\
\hline\hline
\end{tabular}
\end{table}

\begin{table}[!h]
\centering
\caption{Scalar quasinormal spectrum for the $l=1$ sector. The frequencies $\omega_n$ are evaluated using the sixth order WKB approximation for the listed values of $M$, the equal charges $Q=p$, and the Lorentz symmetry breaking parameter $\ell$.}
\label{qnmsscalarl1}
\begin{tabular}{c|c|c|c}
\hline\hline
$M,\; Q=p,\; \ell$ & $\omega_0$ & $\omega_1$ & $\omega_2$ \\[0.2ex]
\hline
$1.0,\; 0.01,\; 0.10$ & $0.345686 - 0.120879\,i$ & $0.309752 - 0.380215\,i$ & $0.270176 - 0.673692\,i$ \\
$1.0,\; 0.10,\; 0.10$ & $0.347302 - 0.121044\,i$ & $0.311552 - 0.380601\,i$ & $0.272167 - 0.674067\,i$ \\
$1.0,\; 0.15,\; 0.10$ & $0.349389 - 0.121248\,i$ & $0.313883 - 0.381070\,i$ & $0.274748 - 0.674498\,i$ \\
$1.0,\; 0.20,\; 0.10$ & $0.352403 - 0.121524\,i$ & $0.317261 - 0.381686\,i$ & $0.278496 - 0.675011\,i$ \\
$1.0,\; 0.25,\; 0.10$ & $0.356449 - 0.121858\,i$ & $0.321813 - 0.382396\,i$ & $0.283561 - 0.675489\,i$ \\
\hline\hline
$1.0,\; 0.10,\; 0.10$ & $0.347302 - 0.121044\,i$ & $0.311552 - 0.380601\,i$ & $0.272167 - 0.674067\,i$ \\
$1.0,\; 0.10,\; 0.15$ & $0.380427 - 0.135856\,i$ & $0.339923 - 0.427890\,i$ & $0.297045 - 0.758330\,i$ \\
$1.0,\; 0.10,\; 0.20$ & $0.419231 - 0.153563\,i$ & $0.373042 - 0.484520\,i$ & $0.326349 - 0.859152\,i$ \\
$1.0,\; 0.10,\; 0.21$ & $0.427797 - 0.157519\,i$ & $0.380339 - 0.497183\,i$ & $0.332846 - 0.881681\,i$ \\
$1.0,\; 0.10,\; 0.22$ & $0.436663 - 0.161630\,i$ & $0.387888 - 0.510346\,i$ & $0.339582 - 0.905094\,i$ \\
\hline\hline
\end{tabular}
\end{table}

\begin{table}[!h]
\centering
\caption{Scalar quasinormal spectrum for the $l=2$ sector. The frequencies $\omega_n$ are evaluated using the sixth order WKB approximation for the listed values of $M$, the equal charges $Q=p$, and the Lorentz symmetry breaking parameter $\ell$.}
\label{qnmsscalarl2}
\begin{tabular}{c|c|c|c}
\hline\hline
$M,\; Q=p,\; \ell$ & $\omega_0$ & $\omega_1$ & $\omega_2$ \\[0.2ex]
\hline
$1.0,\; 0.01,\; 0.10$ & $0.568048 - 0.119538\,i$ & $0.542564 - 0.365849\,i$ & $0.500313 - 0.631168\,i$ \\
$1.0,\; 0.10,\; 0.10$ & $0.570676 - 0.119713\,i$ & $0.545327 - 0.366334\,i$ & $0.503302 - 0.631842\,i$ \\
$1.0,\; 0.15,\; 0.10$ & $0.574070 - 0.119931\,i$ & $0.548900 - 0.366930\,i$ & $0.507172 - 0.632661\,i$ \\
$1.0,\; 0.20,\; 0.10$ & $0.578973 - 0.120226\,i$ & $0.554070 - 0.367732\,i$ & $0.512780 - 0.633738\,i$ \\
$1.0,\; 0.25,\; 0.10$ & $0.585557 - 0.120585\,i$ & $0.561023 - 0.368695\,i$ & $0.520343 - 0.634982\,i$ \\
\hline\hline
$1.0,\; 0.10,\; 0.10$ & $0.570676 - 0.119713\,i$ & $0.545327 - 0.366334\,i$ & $0.503302 - 0.631842\,i$ \\
$1.0,\; 0.10,\; 0.15$ & $0.623357 - 0.134297\,i$ & $0.594331 - 0.411370\,i$ & $0.546754 - 0.710503\,i$ \\
$1.0,\; 0.10,\; 0.20$ & $0.684799 - 0.151727\,i$ & $0.651310 - 0.465264\,i$ & $0.597119 - 0.804774\,i$ \\
$1.0,\; 0.10,\; 0.21$ & $0.698325 - 0.155619\,i$ & $0.663831 - 0.477311\,i$ & $0.608168 - 0.825865\,i$ \\
$1.0,\; 0.10,\; 0.22$ & $0.712310 - 0.159665\,i$ & $0.676769 - 0.489833\,i$ & $0.619579 - 0.847793\,i$ \\
\hline\hline
\end{tabular}
\end{table}


\subsection{Vector perturbations }

We now apply the same procedure to the vector sector. Inserting the metric of Eq.~(\ref{dyonic_KR_metric}) into the general vector potential in Eq.~(\ref{vectorrrrpot}) yields the following effective potential:
\ie
\mathrm{V}_{v} =  \left(\frac{1}{1-\ell }-\frac{2 M}{r}+\frac{Q^2}{r^2 (1-\ell )^2}+\frac{p^2}{r^2 (1-2 \ell )}\right) \left(\frac{l (l+1)}{r^2}\right).
\fe
As it is straightforward to be verified, in the Schwarzschild limit, obtained by taking $\ell,Q,p \to 0$, this expression recovers the usual effective potential for vector perturbations.

\begin{figure}
    \centering
    \includegraphics[scale=0.45]{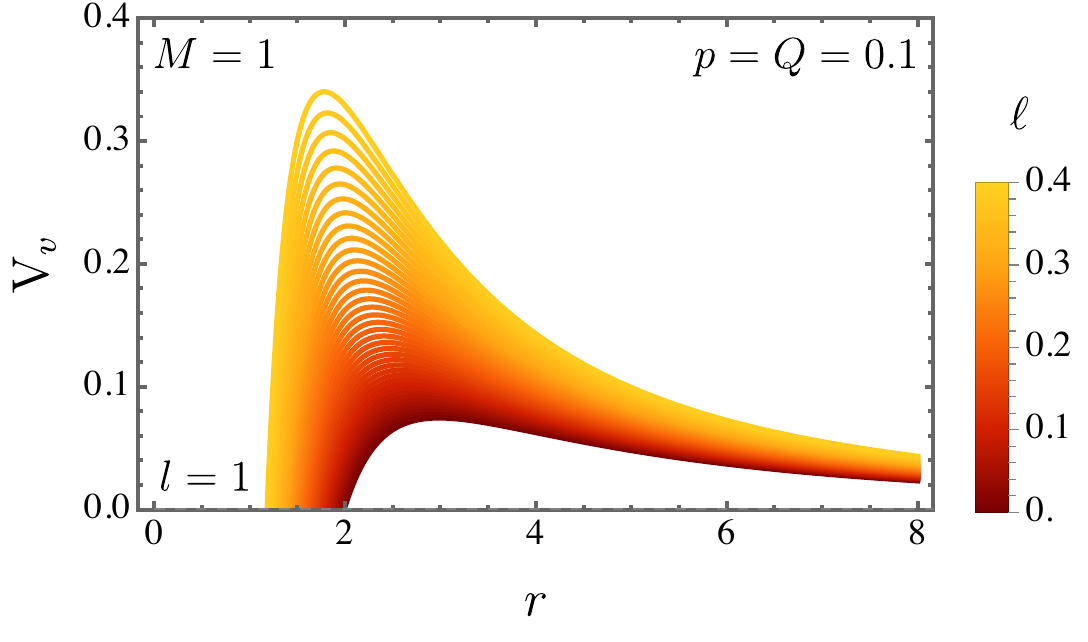}
    \includegraphics[scale=0.45]{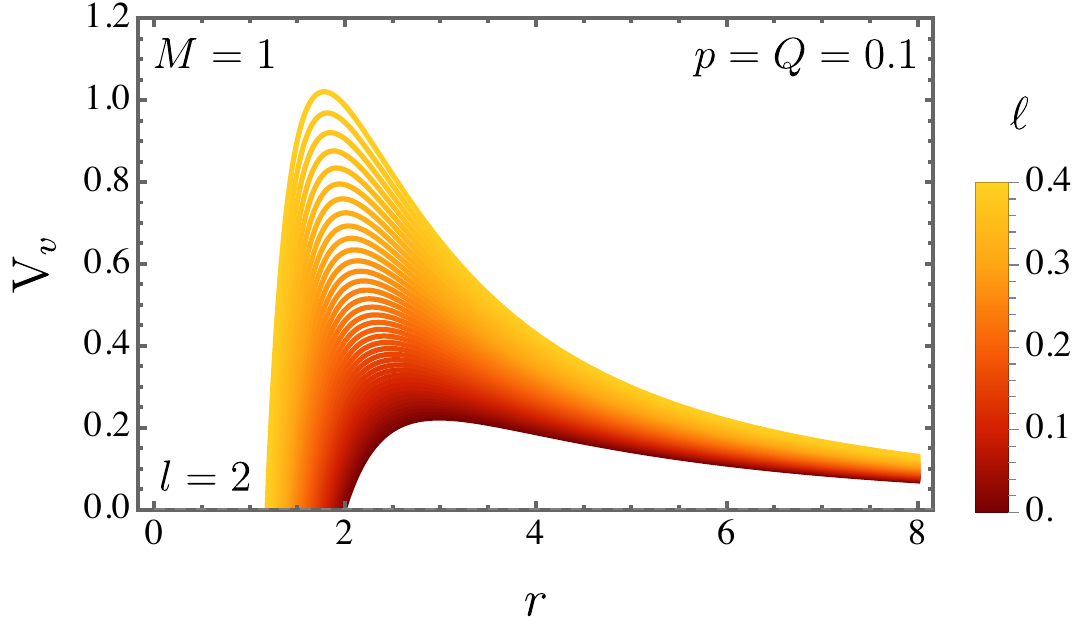}
     \includegraphics[scale=0.45]{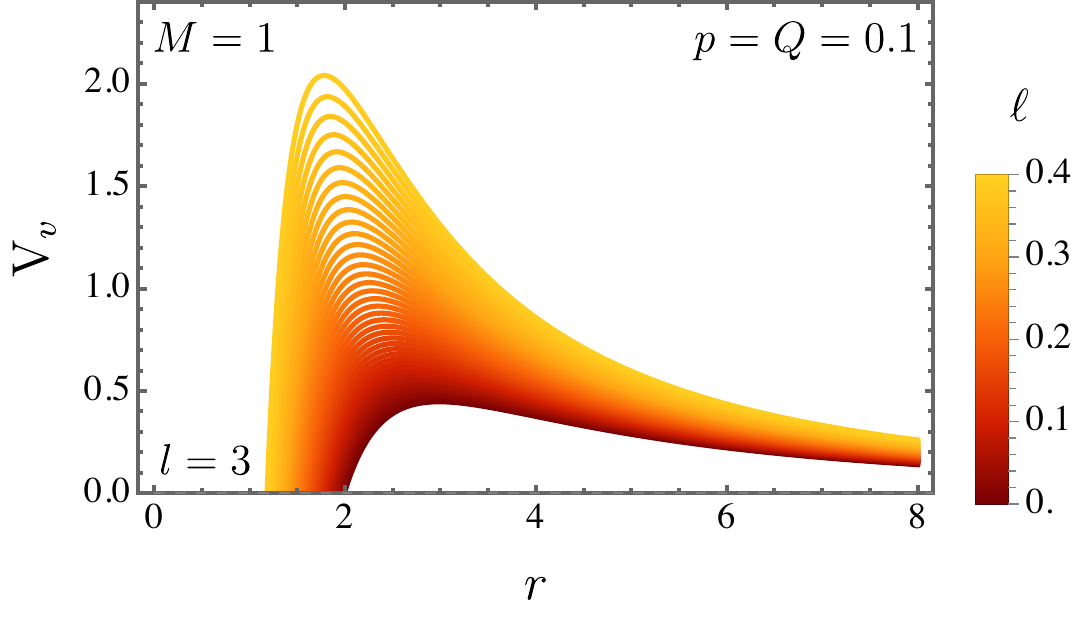}
    \caption{Vector effective potential $\mathrm{V}_{v}$ as a function of the radial coordinate $r$ for $M=1$ and several values of $\ell$. The panels correspond to the multipole sectors $l=1$ top--left, $l=2$ top--right, and $l=3$ bottom. }
    \label{vpot}
\end{figure}

Fig.~\ref{vpot} presents the vector effective potential $\mathrm{V}_{v}$ as a function of the radial coordinate $r$ for several choices of $l$, $\ell$, $Q$, and $p$. The plots show that larger values of the Lorentz--violating parameter raise the height of the potential barrier. As in the scalar sector, the parameter $\ell$ strengthens the radial barrier governing the propagation of the vector modes.

\begin{figure}
    \centering
    \includegraphics[scale=0.5]{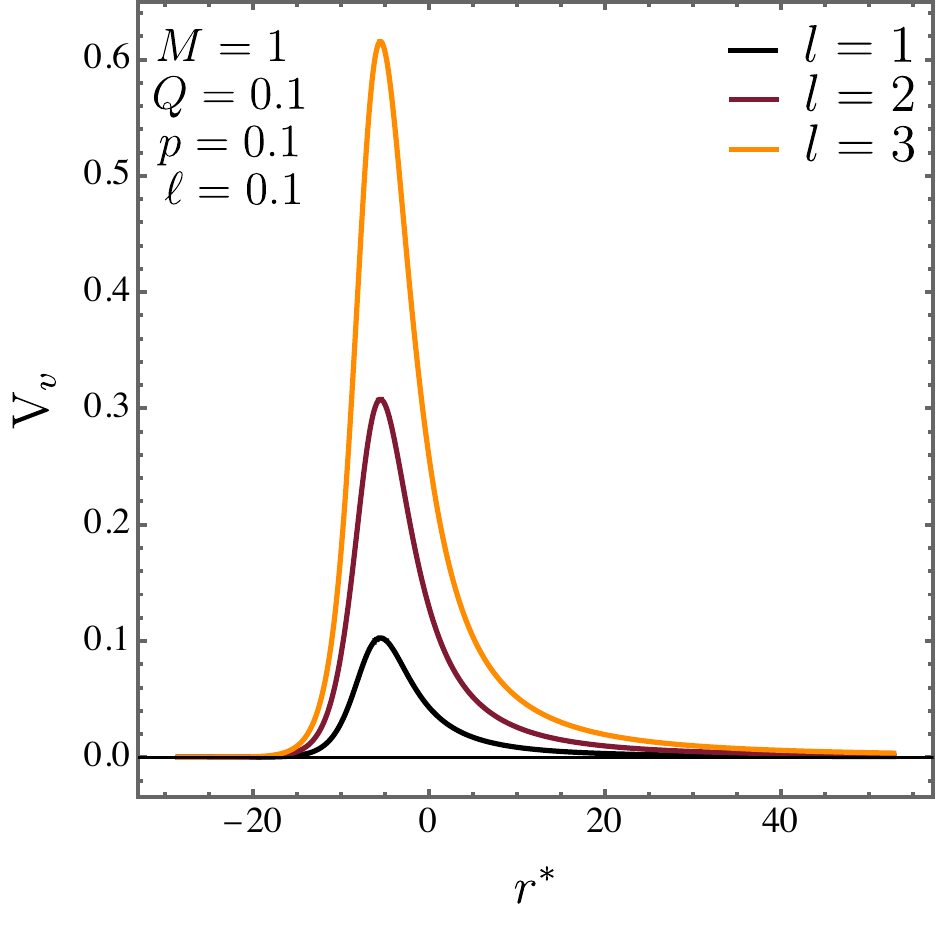}
    \caption{ Vector effective potential $\mathrm{V}_{v}$ expressed in terms of the tortoise coordinate $r^{*}$ for $M=1$ and $\ell=Q=p=0.1$, considering the multipole sectors $l=1,2,3$. }
    \label{vtortoise}
\end{figure}

{Fig.~\ref{vtortoise} displays the vector potential in the tortoise coordinate $r_*$ for $M=1$ and $\ell=Q=p=0.1$, considering the multipoles $l=1,2,3$. Each profile has a single maximum, with no secondary barrier in the displayed region.}

The vector quasinormal frequencies obtained from the 6th--order WKB approximation are collected in Tables~\ref{qnmsvectorl1}--\ref{qnmsvectorl3}. These tables correspond, respectively, to the multipole sectors $l=1$, $l=2$, and $l=3$, with the mass normalized to $M=1$. The remaining parameters are varied so that the separate effects of the dyonic charges and of the Lorentz--violating contribution can be compared.

A first pattern emerges when the Lorentz--violating parameter is fixed at $\ell=0.1$ and the electric and magnetic charges are increased together, $Q=p$. Under this variation, the vector spectrum changes only moderately. The real part of $\omega_n$ grows with the charge value for all multipoles considered, showing that the oscillatory component becomes slightly faster. The imaginary sector follows the same tendency in magnitude, but the change is weaker; then, the damping rate receives only a small enhancement. This charge--induced shift becomes more apparent for larger $l$, with the $l=3$ modes reaching the highest real frequencies among the vector cases listed.

The dependence on $\ell$ is considerably more pronounced. When the charges are held fixed at $Q=p=0.1$, larger values of $\ell$ increase both $\mathrm{Re}(\omega_n)$ and $|\mathrm{Im}(\omega_n)|$ throughout the vector spectrum. Therefore, the Lorentz--violating parameter does not delay the decay of the perturbation. Instead, it drives the ringdown toward higher oscillation frequencies and faster attenuation. This behavior matches the potential profiles discussed above, where increasing $\ell$ raises and narrows the effective barrier.

The ordering with respect to the multipole number also follows the expected structure. For fixed $M$, $Q$, $p$, and $\ell$, the real part of the frequency rises as one goes from $l=1$ to $l=3$. {In this manner}, higher vector multipoles oscillate more rapidly. The damping of the fundamental mode varies less dramatically with $l$, so the main multipolar effect appears in the oscillation frequency rather than in the decay rate. {Within each multipole sector, $\omega_0$ is the least damped tabulated quasinormal mode and dominates the ringdown after the higher overtones have decayed. At sufficiently late times, the power law tail becomes dominant.}

A comparison with the scalar results shows that, for the same parameter choices and the same multipole number, the vector modes generally oscillate at lower frequencies. Their damping rates are also slightly smaller for the lowest modes. Even so, the qualitative response to the parameters is the same in both sectors: the dyonic charges produce relatively mild corrections, whereas the Lorentz--violating parameter $\ell$ gives the dominant enhancement of the oscillation frequency and of the damping rate.

\begin{table}[!h]
\begin{center}
\caption{\label{qnmsvectorl1} Vector quasinormal spectrum in the $l=1$ sector for $M=1$. The frequencies $\omega_n$ are obtained with the 6th--order WKB approximation for several choices of $\ell$, $Q$, and $p$.}
\begin{tabular}{c| c | c | c} 
 \hline\hline\hline 
 \!\!\!\! $M$,  \, $Q=p$, \,\, $\ell$  & $\omega_{0}$ & $\omega_{1}$ & $\omega_{2}$  \\ [0.2ex] 
 \hline 
 \,  1.0, \,  0.01, \, 0.1  & 0.287554 - 0.113896$i$ & 0.243983 - 0.363707$i$ & 0.195321 - 0.658795$i$ \\
 
\,  1.0, \,  0.10, \, 0.1  & 0.289090 - 0.114114$i$ & 0.245756 - 0.364238$i$ & 0.197417 - 0.659249$i$  \\
 
 \, 1.0, \,  0.15, \, 0.1  & 0.291080 - 0.114387$i$ & 0.248064 - 0.364883$i$ & 0.200158 - 0.659730$i$   \\
 
\, 1.0, \,  0.20, \, 0.1  & 0.293960 - 0.114764$i$ & 0.251404 - 0.365777$i$ & 0.204115 - 0.660433$i$ \\
 
\, 1.0, \,  0.25, \, 0.1  & 0.297842 - 0.115235$i$ & 0.255933 - 0.366853$i$ & 0.209495 - 0.661144$i$  \\
   [0.2ex] 
 \hline \hline \hline 
 \,  1.0, \,  0.1, \, 0.10  & 0.289090 - 0.114114$i$ & 0.245756 - 0.364238$i$ & 0.197417 - 0.659249$i$ \\
 
\,  1.0, \,  0.1, \, 0.15  & 0.313295 - 0.127679$i$ & 0.263725 - 0.408841$i$ & 0.210513 - 0.741849$i$  \\
 
 \, 1.0, \,  0.1, \, 0.20  & 0.341132 - 0.143831$i$ & 0.284006 - 0.462193$i$  & 0.225347 - 0.840880$i$ \\
 
\, 1.0, \,  0.1, \, 0.21  & 0.347206 - 0.147430$i$ & 0.288380 - 0.474113$i$ & 0.228563 - 0.863028$i$ \\
 
\, 1.0, \,  0.1, \, 0.22  & 0.353466 - 0.151168$i$ & 0.292869 - 0.486506$i$ & 0.231869 - 0.886073$i$  \\
   [0.2ex] 
 \hline \hline \hline 
\end{tabular}
\end{center}
\end{table}

\begin{table}[!h]
\begin{center}
\caption{\label{qnmsvectorl2} Vector quasinormal spectrum for the $l=2$ multipole with $M=1$. The values of $\omega_n$ are extracted through the 6th--order WKB approximation for different configurations of $\ell$, $Q$, and $p$.}
\begin{tabular}{c| c | c | c} 
 \hline\hline\hline 
 \!\!\!\! $M$,  \, $Q=p$, \,\, $\ell$  & $\omega_{0}$ & $\omega_{1}$ & $\omega_{2}$  \\ [0.2ex] 
 \hline 
 \,  1.0, \,  0.01, \, 0.1  & 0.534168 - 0.117134$i$ & 0.506875 - 0.359192$i$ & 0.461578 - 0.621867$i$ \\
 
\,  1.0, \,  0.10, \, 0.1  & 0.536748 - 0.117322$i$ & 0.509599 - 0.359713$i$ & 0.464548 - 0.62259$i$  \\
 
 \, 1.0, \,  0.15, \, 0.1  & 0.540082 - 0.117556$i$ & 0.513125 - 0.360356$i$ & 0.468401 - 0.623472$i$  \\
 
\, 1.0, \,  0.20, \, 0.1  & 0.544903 - 0.117875$i$ & 0.518231 - 0.361227$i$ & 0.473989 - 0.624642$i$ \\
 
\, 1.0, \,  0.25, \, 0.1  & 0.551385 - 0.118267$i$ & 0.525109 - 0.362283$i$ & 0.481536 - 0.626013$i$  \\
   [0.2ex] 
 \hline \hline \hline 
 \,  1.0, \,  0.1, \, 0.10  & 0.536748 - 0.117322$i$ & 0.509599 - 0.359713$i$ & 0.464548 - 0.622592$i$ \\
 
\,  1.0, \,  0.1, \, 0.15  & 0.584220 - 0.131464$i$ & 0.553009 - 0.403564$i$ & 0.501804 - 0.699711$i$  \\
 
 \, 1.0, \,  0.1, \, 0.20  & 0.639257 - 0.148337$i$ & 0.603089 - 0.455974$i$  & 0.544502 - 0.792089$i$  \\
 
\, 1.0, \,  0.1, \, 0.21  & 0.651327 - 0.152101$i$ & 0.614040 - 0.467679$i$ & 0.553806 - 0.812744$i$ \\
 
\, 1.0, \,  0.1, \, 0.22  & 0.663792 - 0.156011$i$ & 0.625334 - 0.479844$i$ & 0.563385 - 0.834226$i$  \\
   [0.2ex] 
 \hline \hline \hline 
\end{tabular}
\end{center}
\end{table}

\begin{table}[!h]
\begin{center}
\caption{\label{qnmsvectorl3} Vector quasinormal spectrum in the $l=3$ sector for the normalized mass $M=1$. The frequencies $\omega_n$ are computed through the 6th--order WKB approximation while varying $\ell$, $Q$, and $p$.}
\begin{tabular}{c| c | c | c} 
 \hline\hline\hline 
  \!\!\!\! $M$,  \, $Q=p$, \,\, $\ell$  & $\omega_{0}$ & $\omega_{1}$ & $\omega_{2}$  \\ [0.2ex] 
 \hline 
 \,  1.0, \,  0.01, \, 0.1  & 0.768134 - 0.117960$i$ & 0.748443 - 0.357838$i$ & 0.712487 - 0.608899$i$ \\
 
\,  1.0, \,  0.10, \, 0.1  & 0.771754 - 0.118148$i$ & 0.752168 - 0.358366$i$ & 0.716404 - 0.609704$i$  \\
 
 \, 1.0, \,  0.15, \, 0.1  & 0.776431 - 0.118375$i$ & 0.756983 - 0.359019$i$ & 0.721475 - 0.610695$i$  \\
 
\, 1.0, \,  0.20, \, 0.1  & 0.783192 - 0.118685$i$ & 0.763948 - 0.359904$i$ & 0.728818 - 0.612026$i$ \\
 
\, 1.0, \,  0.25, \, 0.1  & 0.792275 - 0.119064$i$ & 0.773317 - 0.360983$i$ & 0.738712 - 0.613620$i$  \\
   [0.2ex] 
 \hline \hline \hline 
 \,  1.0, \,  0.1, \, 0.10  & 0.771754 - 0.118148$i$ & 0.752168 - 0.358366$i$ & 0.716404 - 0.609704$i$ \\
 
\,  1.0, \,  0.1, \, 0.15  & 0.840831 - 0.132442$i$ & 0.818287 - 0.401975$i$ & 0.777353 - 0.684631$i$  \\
 
 \, 1.0, \,  0.1, \, 0.20  & 0.921047 - 0.149507$i$ & 0.894887 - 0.454083$i$ & 0.847685 - 0.774291$i$  \\
 
\, 1.0, \,  0.1, \, 0.21  & 0.938658 - 0.153316$i$ & 0.911680 - 0.465720$i$ & 0.863067 - 0.794332$i$ \\
 
\, 1.0, \,  0.1, \, 0.22  & 0.956850 - 0.157272$i$  & 0.929017 - 0.477811$i$ & 0.878936 - 0.815163$i$  \\
   [0.2ex] 
 \hline \hline \hline 
\end{tabular}
\end{center}
\end{table}


\subsection{Tensor perturbations  }

We close the bosonic part of the analysis with the tensor, or axial, perturbative sector. Following the same construction adopted for the scalar and vector fields, the metric in Eq.~(\ref{dyonic_KR_metric}) is inserted into the general expression of Eq.~(\ref{tententen}). This substitution leads to the tensor effective potential written as follows:
\ie
\begin{split}
\mathrm{V}_{t} = & \left(\frac{1}{1-\ell }-\frac{2 M}{r}+\frac{p^2}{r^2 (1-2 \ell )}+\frac{Q^2}{r^2 (1-\ell )^2}\right) \\
& \times \left(\frac{l (l+1)}{r^2}-\frac{6 M \ell ^2-12 M \ell +6 M}{r^3 (\ell -1)^2}+\frac{4 p^2}{r^4 (1-2 \ell )}+\frac{4 Q^2}{r^4 (\ell -1)^2}+\frac{2 \ell -2 \ell ^2}{r^2 (\ell -1)^2}\right).
\label{tensdas}
\end{split}
\fe
In the simultaneous limit $Q,p,\ell \to 0$, the tensor effective potential reduces to the standard Schwarzschild expression for the corresponding perturbative sector.

\begin{figure}
    \centering
    \includegraphics[scale=0.45]{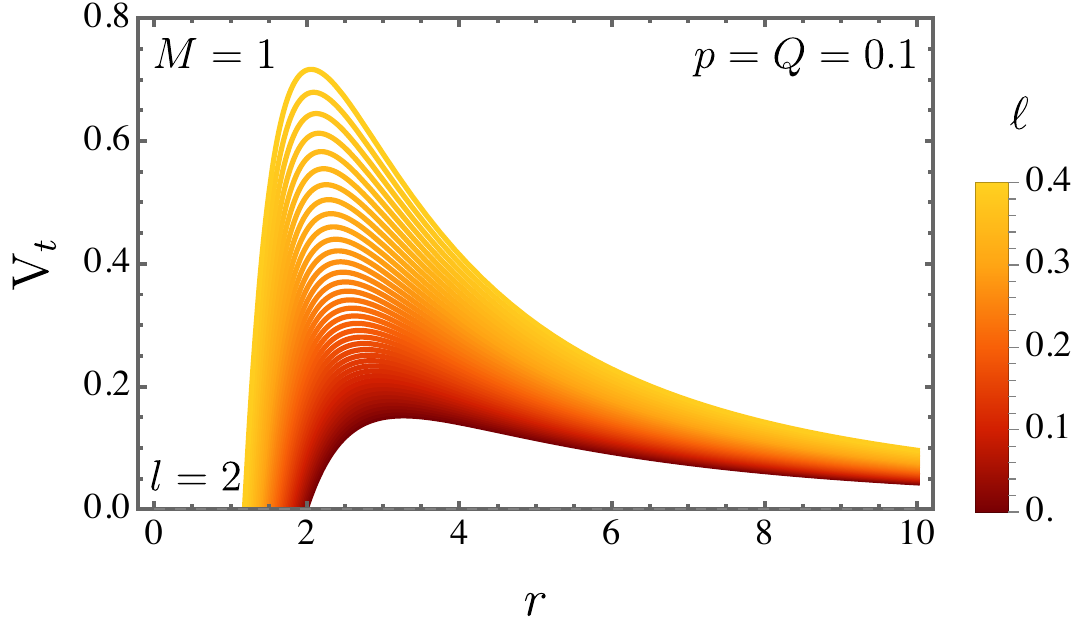}
    \includegraphics[scale=0.45]{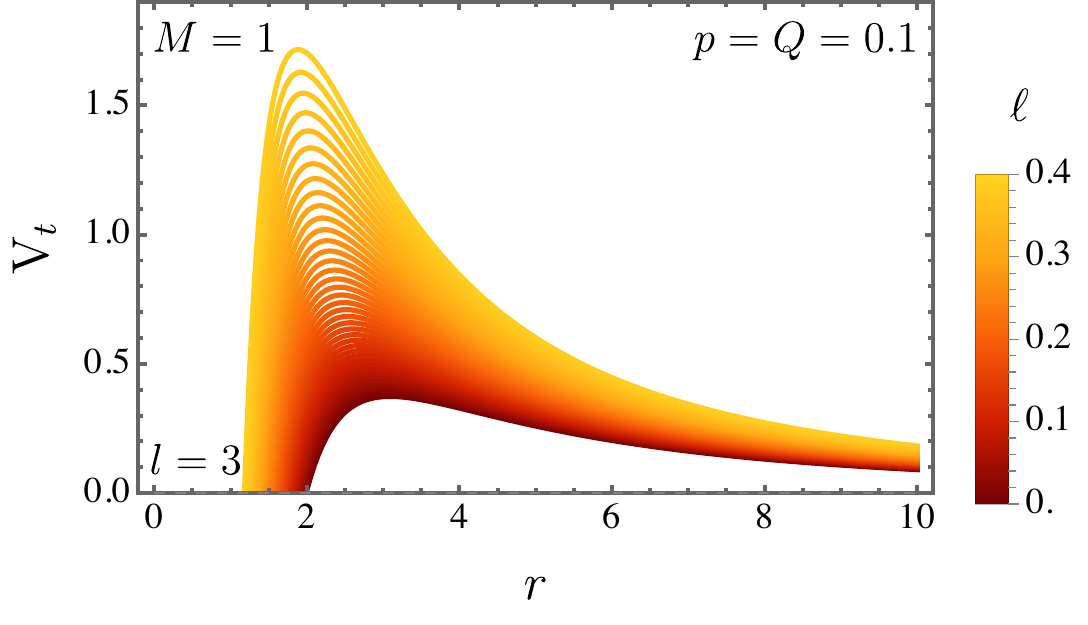}
     \includegraphics[scale=0.45]{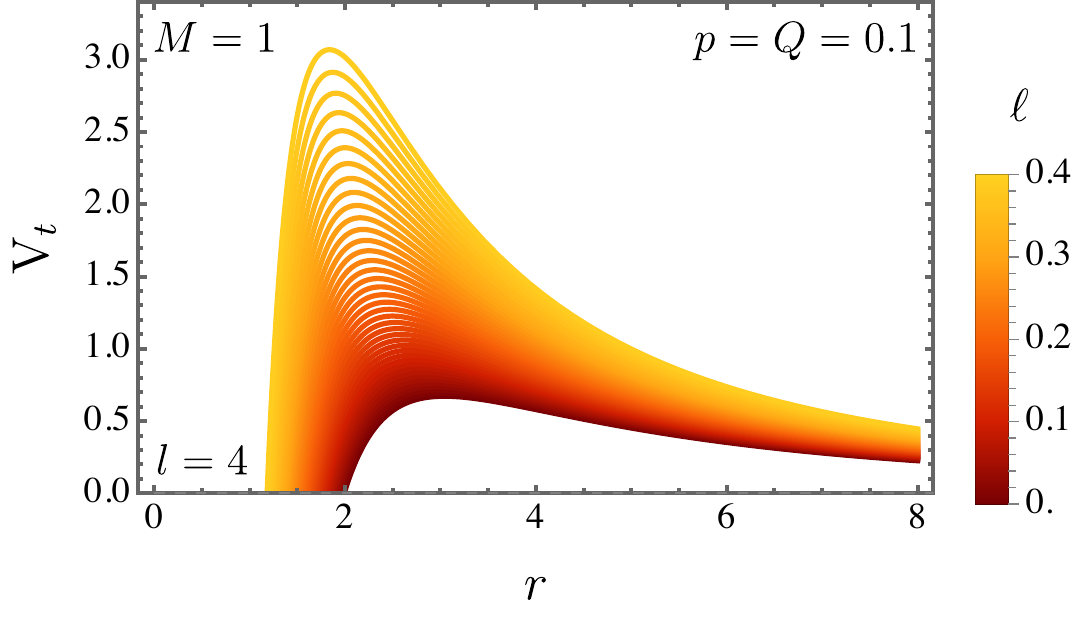}
    \caption{Tensor effective potential $\mathrm{V}_{t}$ as a function of the radial coordinate $r$ for $M=1$ and several choices of the Lorentz--violating parameter $\ell$. The panels represent the multipole sectors $l=2$ top--left, $l=3$ top--right, and $l=4$ bottom.}
    \label{tpot}
\end{figure}

\begin{figure}
    \centering
    \includegraphics[scale=0.5]{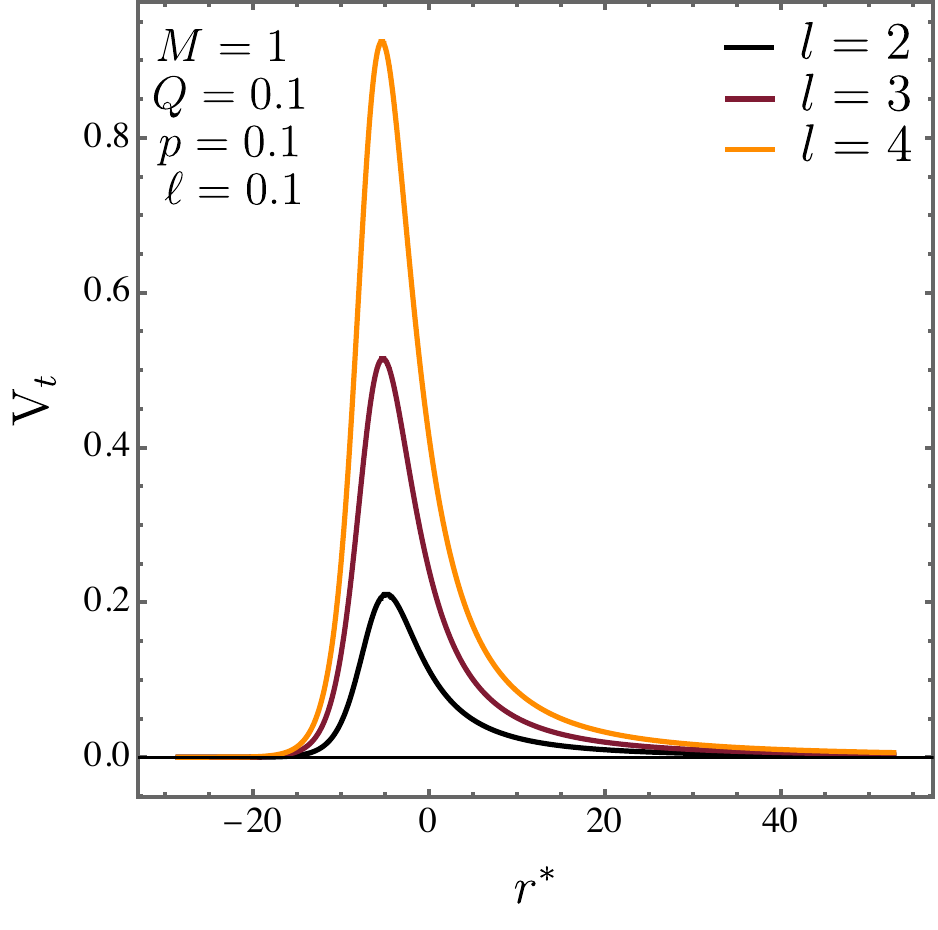}
    \caption{Tensor effective potential $\mathrm{V}_{t}$ written in terms of the tortoise coordinate $r^{*}$ for $M=1$ and fixed parameters $\ell=p=Q=0.1$. The curves correspond to the multipole sectors $l=2,3,4$.}
    \label{ttortoise}
\end{figure}

Figure~\ref{tpot} presents the tensor effective potential $\mathrm{V}_{t}$ for different choices of $l$ and $\ell$. The barrier becomes higher as the Lorentz--violating parameter increases, showing that $\ell$ strengthens the radial confinement of the tensor modes, in agreement with the behavior observed in the scalar and vector sectors. Figure~\ref{ttortoise} shows the same potential written in terms of the tortoise coordinate $r^{*}$. For each multipole number, the profile contains only one maximum, so no secondary barrier appears in this sector. The resulting smooth barrier structure also supports the use of the WKB approximation in the calculation of the tensor quasinormal frequencies.

The tensor spectra reported in Tables~\ref{qnmstensorl2}, \ref{qnmstensorl3}, and \ref{qnmstensorl4} correspond to the multipoles $l=2$, $l=3$, and $l=4$, respectively. These results make it possible to distinguish two sources of modification in the axial sector: the dyonic charges and the Lorentz--violating parameter.

For the configurations in which $\ell=0.1$ remains fixed and the charges are varied according to $Q=p$, the changes in the spectrum are relatively small. As the common charge value increases, the real part of $\omega_n$ becomes larger, while the magnitude of the imaginary part also grows. Thus, the dyonic sector shifts the tensor modes toward slightly higher oscillation frequencies and produces a modest increase in the damping rate. This behavior occurs for all multipoles and overtones listed in the tables.

The role of $\ell$ is considerably more pronounced. When the charges are kept fixed at $Q=p=0.1$, increasing the Lorentz--violating parameter produces a clear upward shift in both $\mathrm{Re}(\omega_n)$ and $|\mathrm{Im}(\omega_n)|$. Hence, the axial perturbations do not become longer lived as $\ell$ grows. Instead, the ringdown becomes faster in two senses: the oscillation frequency increases, and the decay time decreases. Notice again that this trend agrees with the potential analysis, since larger values of $\ell$ raise the height of the effective barrier and make its profile steeper.

The multipole dependence follows the usual ordering of black hole perturbations. At fixed $M$, $Q$, $p$, and $\ell$, the real part of the frequency increases from $l=2$ to $l=4$, meaning that higher tensor multipoles oscillate more rapidly. The damping rate also changes with $l$, but less strongly than the oscillation frequency.  {Within each multipole sector, the fundamental mode $\omega_0$ is the least damped tabulated quasinormal mode. It dominates the ringdown after the overtones $\omega_1$ and $\omega_2$ have decayed, while the subsequent late time response is governed by the tail.}

When compared with the scalar and vector cases, the tensor sector generally yields smaller real frequencies for the same parameter choices and the same multipole number. The corresponding low--lying tensor modes also tend to have slightly weaker damping. Nevertheless, the overall parameter dependence remains the same across the three sectors: the charges $Q$ and $p$ generate mild shifts in the spectrum, whereas the Lorentz--violating parameter $\ell$ gives the dominant increase in both the oscillation frequency and the damping rate.

\begin{table}[!h]
\begin{center}
\caption{\label{qnmstensorl2} Tensor quasinormal spectrum in the $l=2$ sector for $M=1$. The frequencies $\omega_n$ are obtained with the 6th--order WKB approximation for different choices of the $\ell$.}
\begin{tabular}{c| c | c | c} 
 \hline\hline\hline 
 \!\!\!\! $M$,  \, $Q=p$, \,\, $\ell$  & $\omega_{0}$ & $\omega_{1}$ & $\omega_{2}$  \\ [0.2ex] 
 \hline 
 \,  1.0, \,  0.01, \, 0.1  & 0.436727 - 0.109106$i$ & 0.400538 - 0.336570$i$ & 0.337497 - 0.590785$i$ \\
 
\,  1.0, \,  0.10, \, 0.1  & 0.438840 - 0.109279$i$ & 0.402905 - 0.337028$i$ & 0.340353 - 0.591289$i$  \\
 
 \, 1.0, \,  0.15, \, 0.1  & 0.441571 - 0.109495$i$ & 0.405969 - 0.337594$i$ & 0.344060 - 0.591898$i$   \\
 
\, 1.0, \,  0.20, \, 0.1  & 0.445517 - 0.109790$i$ & 0.410405 - 0.338367$i$ & 0.349444 - 0.592705$i$ \\
 
\, 1.0, \,  0.25, \, 0.1  & 0.450820 - 0.110155$i$ & 0.416378 - 0.339314$i$  & 0.356721 - 0.593652$i$  \\
   [0.2ex] 
 \hline \hline \hline 
 \,  1.0, \,  0.1, \, 0.10  & 0.438840 - 0.109279$i$ & 0.402905 - 0.337028$i$ & 0.340353 - 0.591289$i$ \\
 
\,  1.0, \,  0.1, \, 0.15  & 0.478050 - 0.122177$i$ & 0.436142 - 0.377451$i$ & 0.363403 - 0.664381$i$  \\
 
 \, 1.0, \,  0.1, \, 0.20  & 0.523583 - 0.137560$i$ & 0.474157 - 0.425899$i$ & 0.388742 - 0.752699$i$ \\
 
\, 1.0, \,  0.1, \, 0.21  & 0.533578 - 0.140993$i$ & 0.482419 - 0.436750$i$ & 0.394107 - 0.772594$i$ \\
 
\, 1.0, \,  0.1, \, 0.22  & 0.543904 - 0.144560$i$ & 0.490925 - 0.448037$i$ & 0.399587 - 0.793319$i$  \\
   [0.2ex] 
 \hline \hline \hline 
\end{tabular}
\end{center}
\end{table}

\begin{table}[!h]
\begin{center}
\caption{\label{qnmstensorl3} Tensor quasinormal spectrum for the $l=3$ multipole with $M=1$. The values of $\omega_n$ are extracted through the 6th--order WKB approximation for several choices of $\ell$.
}
\begin{tabular}{c| c | c | c} 
 \hline\hline\hline 
  \!\!\!\! $M$,  \, $Q=p$, \,\, $\ell$  & $\omega_{0}$ & $\omega_{1}$ & $\omega_{2}$  \\ [0.2ex] 
 \hline 
 \,  1.0, \,  0.01, \, 0.1  & 0.701136 - 0.114000$i$ & 0.679274 - 0.346410$i$ & 0.639158 - 0.591190$i$ \\
 
\,  1.0, \,  0.10, \, 0.1  & 0.704435 - 0.114214$i$ & 0.682687 - 0.346942$i$ & 0.642787 - 0.592004$i$  \\
 
 \, 1.0, \,  0.15, \, 0.1  & 0.708699 - 0.114443$i$ & 0.687101 - 0.347602$i$ & 0.647487 - 0.593006$i$  \\
 
\, 1.0, \,  0.20, \, 0.1  & 0.714861 - 0.114756$i$ & 0.693486 - 0.348499$i$ & 0.654296 - 0.594358$i$ \\
 
\, 1.0, \,  0.25, \, 0.1  & 0.72314 - 0.115141$i$ & 0.702077 - 0.349597$i$ & 0.663475 - 0.595987$i$  \\
   [0.2ex] 
 \hline \hline \hline 
 \,  1.0, \,  0.1, \, 0.10  & 0.704435 - 0.114214$i$ & 0.682687 - 0.346942$i$ & 0.642787 - 0.592004$i$ \\
 
\,  1.0, \,  0.1, \, 0.15  & 0.767624 - 0.127817$i$ & 0.742562 - 0.388519$i$ & 0.696765 - 0.663725$i$  \\
 
 \, 1.0, \,  0.1, \, 0.20  & 0.841045 - 0.144018$i$ & 0.811921 - 0.438082$i$ & 0.758933 - 0.749355$i$ \\
 
\, 1.0, \,  0.1, \, 0.21  & 0.857170 - 0.147629$i$ & 0.827125 - 0.449134$i$ & 0.772510 - 0.768471$i$  \\
 
\, 1.0, \,  0.1, \, 0.22  & 0.873829 - 0.151378$i$ & 0.842823 - 0.460612$i$ & 0.786513 - 0.788327$i$  \\
   [0.2ex] 
 \hline \hline \hline 
\end{tabular}
\end{center}
\end{table}

\begin{table}[!h]
\begin{center}
\caption{\label{qnmstensorl4} Tensor quasinormal spectrum associated with the $l=4$ multipole for $M=1$. The tabulated frequencies $\omega_n$ are obtained from the 6th--order WKB scheme for several values of $\ell$.}
\begin{tabular}{c| c | c | c} 
 \hline\hline\hline 
 \!\!\!\! $M$,  \, $Q=p$, \,\, $\ell$  & $\omega_{0}$ & $\omega_{1}$ & $\omega_{2}$  \\ [0.2ex] 
 \hline 
 \,  1.0, \,  0.01, \, 0.1  & 0.946906 - 0.115991$i$ & 0.930585 - 0.350480$i$ & 0.899598 - 0.592320$i$ \\
 
\,  1.0, \,  0.10, \, 0.1  & 0.951324 - 0.116174$i$ & 0.935087 - 0.351022$i$ & 0.904262 - 0.593170$i$  \\
 
 \, 1.0, \,  0.15, \, 0.1  & 0.957031 - 0.116402$i$ & 0.940906 - 0.351688$i$ & 0.910296 - 0.594221$i$  \\
 
\, 1.0, \,  0.20, \, 0.1  & 0.965279 - 0.116712$i$ & 0.949319 - 0.352594$i$ & 0.919027 - 0.595644$i$  \\
 
\, 1.0, \,  0.25, \, 0.1  & 0.976357 - 0.117094$i$ & 0.960628 - 0.353704$i$ & 0.930782 - 0.597370$i$  \\
   [0.2ex] 
 \hline \hline \hline 
 \,  1.0, \,  0.1, \, 0.10  & 0.951324 - 0.116174$i$ & 0.935087 - 0.351022$i$ & 0.904262 - 0.593170$i$ \\
 
\,  1.0, \,  0.1, \, 0.15  & 1.036920 - 0.130113$i$ & 1.018220 - 0.393294$i$ & 0.982801 - 0.665091$i$  \\
 
 \, 1.0, \,  0.1, \, 0.20  & 1.136410 - 0.146732$i$ & 1.114680 - 0.443720$i$ & 1.073660 - 0.750972$i$  \\
 
\, 1.0, \,  0.1, \, 0.21  & 1.158260 - 0.150438$i$ & 1.135840 - 0.454970$i$ & 1.093560 - 0.770144$i$  \\
 
\, 1.0, \,  0.1, \, 0.22  & 1.180840 - 0.154287$i$ & 1.157710 - 0.466655$i$ & 1.114100 - 0.790062$i$  \\
   [0.2ex] 
 \hline \hline \hline 
\end{tabular}
\end{center}
\end{table}


\subsection{Spinor perturbations  }

\begin{figure}
    \centering
    \includegraphics[scale=0.45]{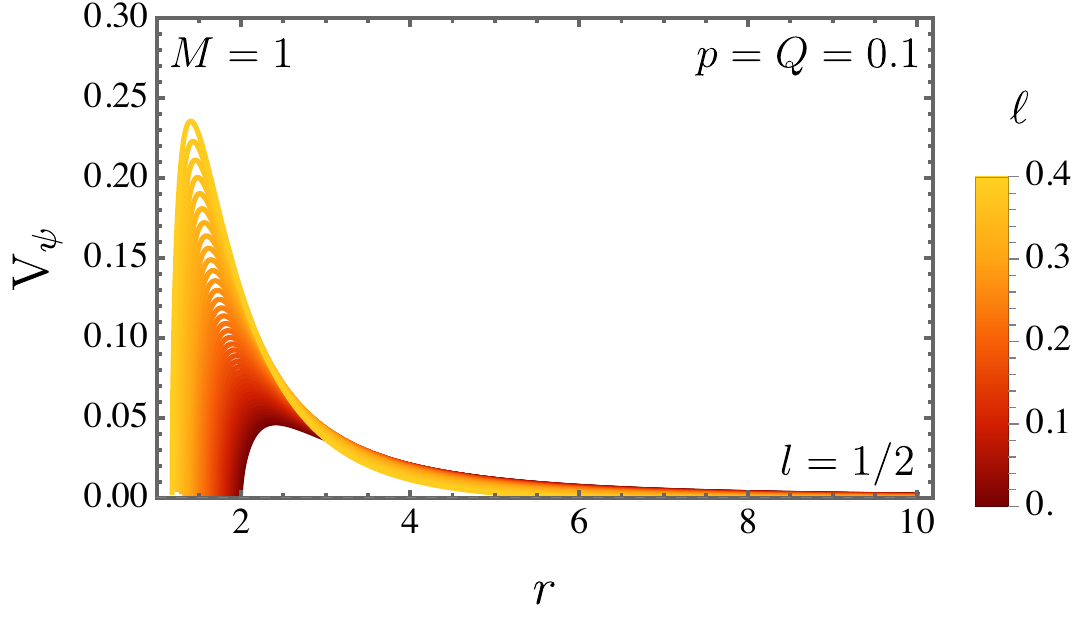}
    \includegraphics[scale=0.45]{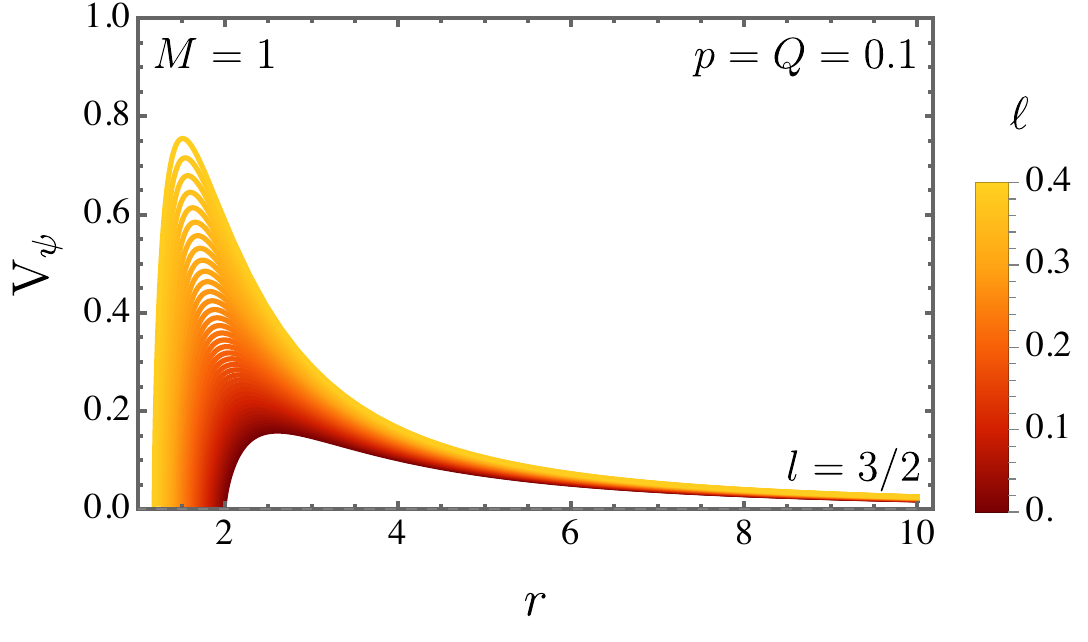}
     \includegraphics[scale=0.45]{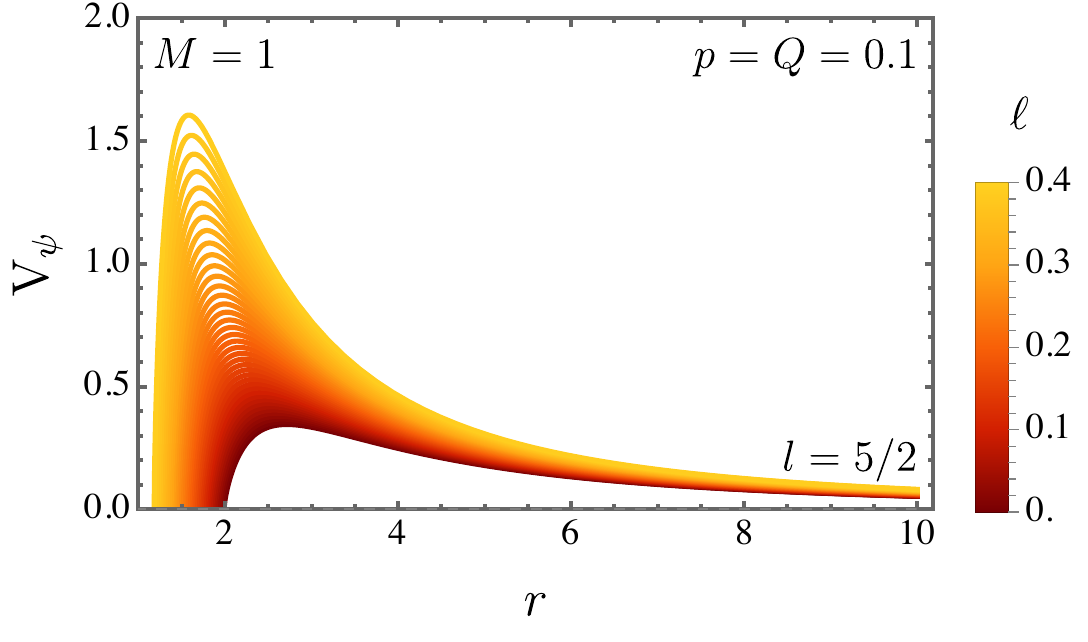}
    \caption{Spinor effective potential $\mathrm{V}_{\psi}$ as a function of the radial coordinate $r$ for $M=1$ and several values of the Lorentz--violating parameter $\ell$. The panels correspond to $l=1/2$ top--left, $l=3/2$ top--right, and $l=5/2$ bottom.}
    \label{psipot}
\end{figure}

After treating the bosonic perturbations, we now extend the discussion to the fermionic sector. The spinor modes are governed by the following effective potential:
\ie
\mathrm{V}_{\psi} = \frac{\left(l+\frac{1}{2}\right) \left(\frac{-2 M r+\frac{p^2}{1-2 \ell }+\frac{Q^2}{(\ell -1)^2}}{r^2}+\frac{1}{1-\ell }\right) \left(2 l-2 \sqrt{\frac{-2 M r+\frac{p^2}{1-2 \ell }+\frac{Q^2}{(\ell -1)^2}}{r^2}+\frac{1}{1-\ell }}+\frac{2 \left(M r+\frac{p^2}{2 \ell -1}-\frac{Q^2}{(\ell -1)^2}\right)}{r^2 \sqrt{\frac{-2 M r+\frac{p^2}{1-2 \ell }+\frac{Q^2}{(\ell -1)^2}}{r^2}+\frac{1}{1-\ell }}}+1\right)}{2 r^2}.
\fe
In the Schwarzschild limit, obtained by setting $\ell=p=Q=0$, the spinor potential reduces to the standard Dirac potential, which shows a direct check of the expression. The behavior of $\mathrm{V}_{\psi}$ is displayed in Fig.~\ref{psipot} for different choices of the angular parameter $l$ and of the Lorentz--violating parameter $\ell$. The curves show that increasing $\ell$ lifts the potential barrier. The Lorentz--violating contribution makes the exterior barrier more pronounced for spinor perturbations, in the same qualitative manner observed for the scalar, vector, and tensor sectors.

\begin{figure}
    \centering
    \includegraphics[scale=0.5]{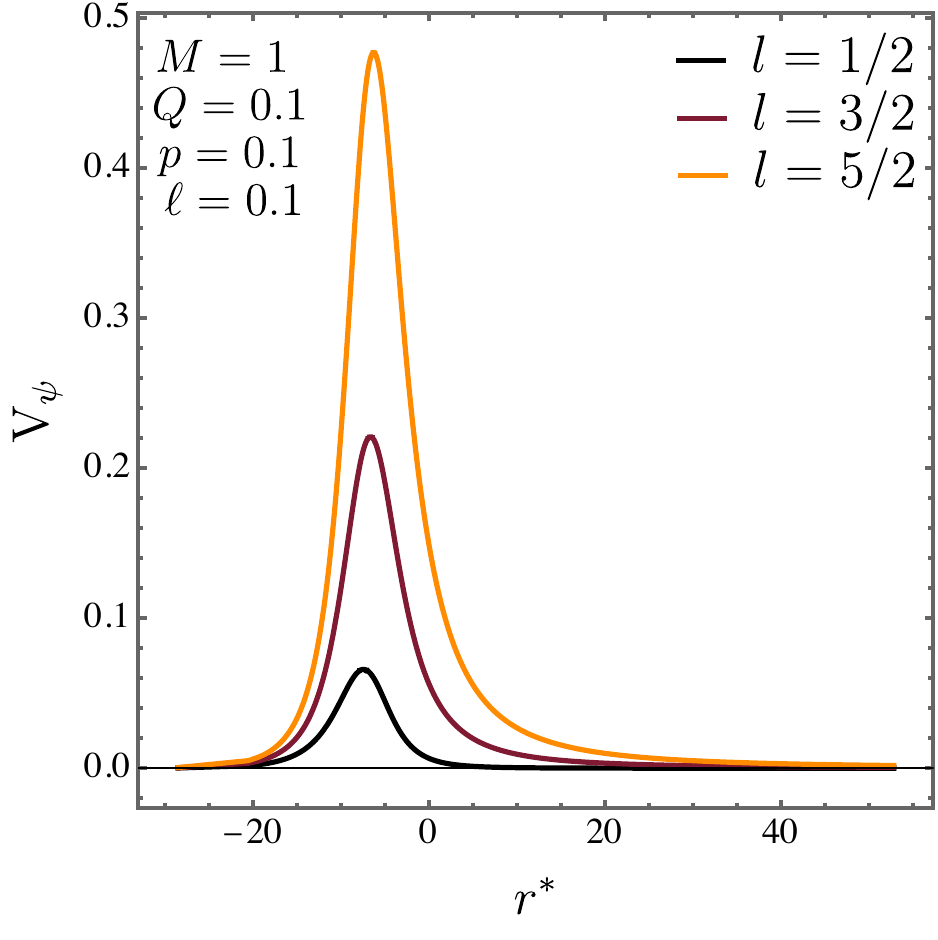}
    \caption{Spinor effective potential $\mathrm{V}_{\psi}$ expressed in terms of the tortoise coordinate $r^{*}$ for $M=1$ and $\ell=0.1$, considering the multipole sectors $l=1/2$, $l=3/2$, and $l=5/2$.}
    \label{spinortortoise}
\end{figure}

\begin{figure}
    \centering
    \includegraphics[scale=0.5]{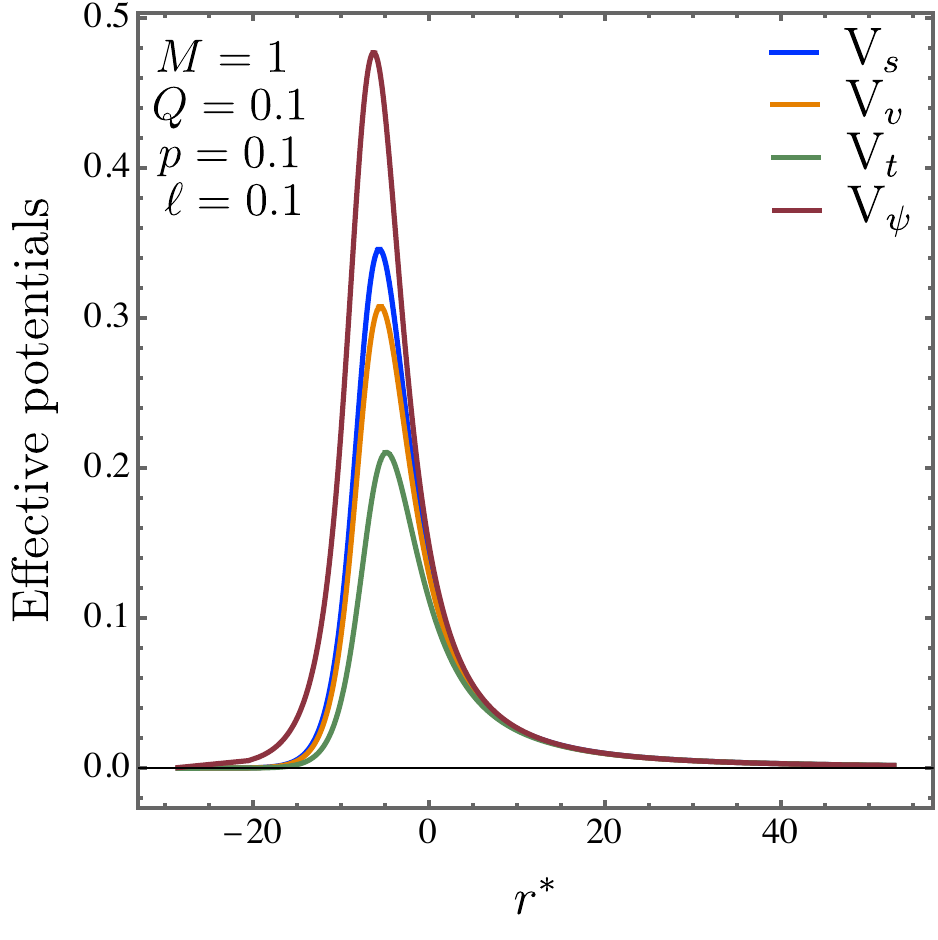}
    \caption{Effective potentials for the different perturbative sectors as functions of the tortoise coordinate $r^{*}$. {The comparison uses $j=l=5/2$ for the spinor field and $l=2$ for the scalar, vector, and effective tensor sectors. The barrier peak heights satisfy $\mathrm{V}_{\psi,\max}>\mathrm{V}_{s,\max}>\mathrm{V}_{v,\max}>\mathrm{V}_{t,\max}$.}}
    \label{comptortoise}
\end{figure}

The representation in terms of the tortoise coordinate is shown in Fig.~\ref{spinortortoise}, where $\mathrm{V}_{\psi}$ is plotted as a function of $r^{*}$ for $M=1$ and $\ell=0.1$. Each value of $l$ leads to a single-barrier profile, with no additional maximum outside the event horizon. A direct comparison with the bosonic potentials is given in Fig.~\ref{comptortoise}, where $\mathrm{V}_{s}$, $\mathrm{V}_{v}$, $\mathrm{V}_{t}$, and $\mathrm{V}_{\psi}$ are plotted in the same $r^{*}$ coordinate. {For the parameters and angular modes considered in Fig.~14, the peak heights satisfy $\mathrm{V}_{\psi,\max}>\mathrm{V}_{s,\max}>\mathrm{V}_{v,\max}>\mathrm{V}_{t,\max}$. Notice that this ordering concerns the barrier maxima and does not imply the same pointwise ordering at all radii.}

{The spinor potential vanishes at the horizon and approaches zero at spatial infinity. For the positive partner used here, its large distance behavior satisfies \[ \lim_{x\to\infty}x^2 \mathrm{V}_\psi^+(x) = (1-\ell)k^2 - k\sqrt{1-\ell}, \qquad x = r_*, \qquad k = |\kappa_D|. \] For $k = 1$ and $0 < \ell < 1/2$, this coefficient is negative, so the potential approaches zero from below and possesses a negative region at sufficiently large radius. As we shall be seeing, the time--domain evolutions examined below show no echo signals.}

Tables~\ref{qnmsspnior12}--\ref{qnmsspnior52} report the spinor quasinormal frequencies associated with the three angular modes $l=1/2$, $l=3/2$, and $l=5/2$, obtained from the 6th--order WKB approximation. A direct comparison among the entries shows that the Lorentz--violating parameter $\ell$ produces the clearest change in the spectrum. When the charges are fixed at $Q=p=0.1$, increasing $\ell$ shifts the modes to larger values of $\mathrm{Re}(\omega_n)$ and also increases the magnitude of $\mathrm{Im}(\omega_n)$. {Then}, the spinor field does not display a slower oscillatory response for stronger Lorentz violation. Instead, the ringdown becomes both faster in frequency and shorter lived in time.

The charge dependence appears in a more restricted form. For the set of rows in which $\ell=0.1$ is kept fixed and $Q=p$ is varied, the fundamental frequency usually moves toward larger real values. In the $l=3/2$ and $l=5/2$ cases, the imaginary part of the fundamental mode also becomes slightly more negative as the dyonic charge increases. This effect is not equally clear for all overtones. Some higher modes change only weakly, while others depart from a simple monotonic pattern, mainly in the $l=1/2$ sector. The charge parameters therefore perturb the spectrum, but they do not drive its main behavior.

The comparison among different spinor multipoles gives another relevant feature of the data. At fixed $M$, $Q=p$, and $\ell$, the real part of the fundamental mode grows as the angular number is increased. Thus, modes with larger $l$ oscillate at higher frequencies. The imaginary part, however, does not follow an equally direct hierarchy. The $l=1/2$ fundamental mode is strongly damped when compared with its oscillation frequency, which makes this sector more delicate from the numerical point of view. This is particularly important because WKB calculations are known to be less robust for low multipoles and higher overtone numbers.

{The tabulated overtones decay more rapidly than the fundamental mode during the ringdown stage. For $M=1$ and $Q=p=\ell=0.1$, the spinor mode with $|\kappa_{\mathrm D}|=1$ and $n=2$ is independently checked using the asymptotic iteration method in Section~VII.E. Moreover, Table~\ref{tab:aim_convergence} compares the corrected WKB frequency with the AIM result and reports their relative differences.}

Taken together, the spinor results reinforce the same physical tendency obtained for the scalar, vector, and tensor perturbations. The parameter $\ell$ increases the height and sharpness of the effective barrier, which is reflected in larger oscillation frequencies and stronger damping rates. The dyonic sector introduces smaller corrections, whereas the Lorentz--violating contribution remains the main source of spectral modification.

\begin{table}[!h]
\centering
\caption{Spinor quasinormal spectrum for the $l=1/2$ sector. The frequencies $\omega_n$ are evaluated using the sixth order WKB approximation for the listed values of $M$, the equal charges $Q=p$, and the Lorentz symmetry breaking parameter $\ell$.}
\label{qnmsspnior12}
\begin{tabular}{c|c|c|c}
\hline\hline
$M,\; Q=p,\; \ell$ & $\omega_0$ & $\omega_1$ & $\omega_2$ \\[0.2ex]
\hline
$1.0,\; 0.01,\; 0.10$ & $0.212280 - 0.117163\,i$ & $0.164826 - 0.390892\,i$ & $0.144628 - 0.707659\,i$ \\
$1.0,\; 0.10,\; 0.10$ & $0.213341 - 0.117383\,i$ & $0.166129 - 0.391293\,i$ & $0.145990 - 0.707902\,i$ \\
$1.0,\; 0.15,\; 0.10$ & $0.214712 - 0.117655\,i$ & $0.167819 - 0.391771\,i$ & $0.147758 - 0.708143\,i$ \\
$1.0,\; 0.20,\; 0.10$ & $0.216698 - 0.118025\,i$ & $0.170275 - 0.392380\,i$ & $0.150326 - 0.708348\,i$ \\
$1.0,\; 0.25,\; 0.10$ & $0.219372 - 0.118476\,i$ & $0.173598 - 0.393044\,i$ & $0.153797 - 0.708345\,i$ \\
\hline\hline
$1.0,\; 0.10,\; 0.10$ & $0.213341 - 0.117383\,i$ & $0.166129 - 0.391293\,i$ & $0.145990 - 0.707902\,i$ \\
$1.0,\; 0.10,\; 0.15$ & $0.231575 - 0.131675\,i$ & $0.178392 - 0.440565\,i$ & $0.158309 - 0.797193\,i$ \\
$1.0,\; 0.10,\; 0.20$ & $0.252573 - 0.148805\,i$ & $0.192332 - 0.499701\,i$ & $0.172660 - 0.904398\,i$ \\
\hline\hline
\end{tabular}
\end{table}

\begin{table}[!h]
\centering
\caption{Spinor quasinormal spectrum for the $l=3/2$ sector. The frequencies $\omega_n$ are evaluated using the sixth order WKB approximation for the listed values of $M$, the equal charges $Q=p$, and the Lorentz symmetry breaking parameter $\ell$.}
\label{qnmsspnior32}
\begin{tabular}{c|c|c|c}
\hline\hline
$M,\; Q=p,\; \ell$ & $\omega_0$ & $\omega_1$ & $\omega_2$ \\[0.2ex]
\hline
$1.0,\; 0.01,\; 0.10$ & $0.444559 - 0.118961\,i$ & $0.413398 - 0.367990\,i$ & $0.367398 - 0.644119\,i$ \\
$1.0,\; 0.10,\; 0.10$ & $0.446653 - 0.119137\,i$ & $0.415655 - 0.368453\,i$ & $0.369889 - 0.644693\,i$ \\
$1.0,\; 0.15,\; 0.10$ & $0.449359 - 0.119355\,i$ & $0.418576 - 0.369021\,i$ & $0.373118 - 0.645381\,i$ \\
$1.0,\; 0.20,\; 0.10$ & $0.453269 - 0.119650\,i$ & $0.422806 - 0.369782\,i$ & $0.377806 - 0.646264\,i$ \\
$1.0,\; 0.25,\; 0.10$ & $0.458521 - 0.120011\,i$ & $0.428504 - 0.370687\,i$ & $0.384140 - 0.647239\,i$ \\
\hline\hline
$1.0,\; 0.10,\; 0.10$ & $0.446653 - 0.119137\,i$ & $0.415655 - 0.368453\,i$ & $0.369889 - 0.644693\,i$ \\
$1.0,\; 0.10,\; 0.15$ & $0.486720 - 0.133592\,i$ & $0.451224 - 0.413798\,i$ & $0.399823 - 0.725333\,i$ \\
$1.0,\; 0.10,\; 0.20$ & $0.533267 - 0.150849\,i$ & $0.492307 - 0.468055\,i$ & $0.434288 - 0.821994\,i$ \\
$1.0,\; 0.10,\; 0.21$ & $0.543489 - 0.154700\,i$ & $0.501297 - 0.480181\,i$ & $0.441819 - 0.843621\,i$ \\
$1.0,\; 0.10,\; 0.22$ & $0.554048 - 0.158702\,i$ & $0.510574 - 0.492786\,i$ & $0.449586 - 0.866108\,i$ \\
\hline\hline
\end{tabular}
\end{table}

\begin{table}[!h]
\centering
\caption{Spinor quasinormal spectrum for the $l=5/2$ sector. The frequencies $\omega_n$ are evaluated using the sixth order WKB approximation for the listed values of $M$, the equal charges $Q=p$, and the Lorentz symmetry breaking parameter $\ell$.}
\label{qnmsspnior52}
\begin{tabular}{c|c|c|c}
\hline\hline
$M,\; Q=p,\; \ell$ & $\omega_0$ & $\omega_1$ & $\omega_2$ \\[0.2ex]
\hline
$1.0,\; 0.01,\; 0.10$ & $0.671996 - 0.118911\,i$ & $0.649883 - 0.361923\,i$ & $0.610943 - 0.619268\,i$ \\
$1.0,\; 0.10,\; 0.10$ & $0.675121 - 0.119089\,i$ & $0.653125 - 0.362427\,i$ & $0.614394 - 0.620009\,i$ \\
$1.0,\; 0.15,\; 0.10$ & $0.679159 - 0.119309\,i$ & $0.657318 - 0.363049\,i$ & $0.618860 - 0.620916\,i$ \\
$1.0,\; 0.20,\; 0.10$ & $0.684992 - 0.119609\,i$ & $0.663382 - 0.363889\,i$ & $0.625329 - 0.622124\,i$ \\
$1.0,\; 0.25,\; 0.10$ & $0.692826 - 0.119975\,i$ & $0.671537 - 0.364906\,i$ & $0.634048 - 0.623550\,i$ \\
\hline\hline
$1.0,\; 0.10,\; 0.10$ & $0.675121 - 0.119089\,i$ & $0.653125 - 0.362427\,i$ & $0.614394 - 0.620009\,i$ \\
$1.0,\; 0.10,\; 0.15$ & $0.735981 - 0.133558\,i$ & $0.710708 - 0.406781\,i$ & $0.666561 - 0.696760\,i$ \\
$1.0,\; 0.10,\; 0.20$ & $0.806725 - 0.150842\,i$ & $0.777456 - 0.459823\,i$ & $0.726791 - 0.788692\,i$ \\
$1.0,\; 0.10,\; 0.21$ & $0.822266 - 0.154701\,i$ & $0.792094 - 0.471675\,i$ & $0.739968 - 0.809253\,i$ \\
$1.0,\; 0.10,\; 0.22$ & $0.838323 - 0.158711\,i$ & $0.807209 - 0.483992\,i$ & $0.753564 - 0.830629\,i$ \\
\hline\hline
\end{tabular}
\end{table}


{
\subsection{Independent tests for the WKB spectrum}
\label{sec:wkb_validation}

The accuracy of the WKB approximation depends on the angular and overtone numbers, and increasing its order does not necessarily improve a given frequency~\cite{KonoplyaZhidenkoZinhailo2019}. We therefore examine the scalar $l=0$ and spinor $l=1/2$ modes with $n=0,1,2$ through the asymptotic iteration method (AIM)~\cite{Cho2010AIM,Cho2012AIM}. The scalar $l=2$ and spinor $l=5/2$ fundamental modes provide comparisons at larger angular numbers. Throughout this subsection, the numerical results correspond to $M=1$, $Q=p=0.1$, and $\ell=0.1$.

With the time dependence $e^{-i\omega t}$, the radial equation is
\begin{equation}
\frac{\mathrm{d}^{2}\Psi}{\mathrm{d}r_*^{2}}+
\left[\omega^{2}-V(r)\right]\Psi=0,
\qquad
\frac{\mathrm{d}r_*}{\mathrm{d}r}=\frac{1}{F(r)},
\nonumber
\end{equation}
and the quasinormal boundary conditions require
\begin{equation}
\Psi\sim e^{-i\omega r_*}\quad(r_*\to-\infty),
\qquad
\Psi\sim e^{i\omega r_*}\quad(r_*\to+\infty).
\nonumber
\end{equation}
Both calculations use the scalar potential
\begin{equation}
V_{\mathrm{sc}}=F(r)\left[
\frac{l(l+1)}{r^{2}}+\frac{1}{r}\frac{\mathrm{d}F}{\mathrm{d}r}
\right]
\nonumber
\end{equation}
and the spinor potential
\begin{equation}
V_{\mathrm{sp}}=W^{2}+F(r)\frac{\mathrm{d}W}{\mathrm{d}r},
\qquad
W=\left(l+\frac{1}{2}\right)\frac{\sqrt{F(r)}}{r}.
\nonumber
\end{equation}
The spinor angular eigenvalue is $k=l+1/2$, and all frequencies refer to the time coordinate of the background metric. The WKB frequencies are recomputed from these potentials after verifying the exterior maximum of each barrier. Evaluations at 50 and 80 working digits yield the same six decimal places reported below. This check controls the numerical evaluation of the WKB formula; its accuracy is assessed separately through AIM.

For the nonextremal geometry considered here,
\begin{equation}
F(r)=f_\infty-\frac{2M}{r}+\frac{P_\ell^{2}}{r^{2}}
=f_\infty\frac{(r-r_+)(r-r_-)}{r^{2}},
\qquad
f_\infty=\frac{1}{1-\ell},
\label{eq:validation_metric}
\end{equation}
where $P_\ell^{2}=Q^{2}/(1-\ell)^{2}+p^{2}/(1-2\ell)$, $\ell<1/2$, and $r_+>r_-$. The tortoise coordinate behaves as
\begin{equation}
r_*=\frac{1}{\alpha_+}\ln\!\left(\frac{r-r_+}{r_+}\right)+O(1),
\qquad r\to r_+,
\label{eq:validation_asymptotics}
\end{equation}
and
\begin{equation}
r_*=\frac{r}{f_\infty}
+\frac{2M}{f_\infty^{2}}\ln\!\left(\frac{r}{r_+}\right)+O(1),
\qquad r\to\infty,
\label{eq:validation_infinity}
\end{equation}
with $\alpha_+=f_\infty(r_+-r_-)/r_+^{2}$. To describe the compact coordinates used in both sectors, we introduce
\begin{equation}
\xi=\left(1-\frac{r_+}{r}\right)^{1/\sigma},
\qquad
r=\frac{r_+}{1-\xi^{\sigma}},
\qquad 0<\xi<1,
\label{eq:validation_coordinate}
\end{equation}
where $\sigma=1$ for scalar perturbations and $\sigma=2$ for spinor perturbations. For spinors, this choice makes $\sqrt{F}$ proportional to $\xi$ at leading order near the event horizon. Factoring out the boundary behavior gives
\begin{equation}
\Psi(\xi)=\mathcal P(\xi,\omega)\chi(\xi),
\qquad
\mathcal P=
\exp\!\left[\frac{i\omega r_+}{f_\infty(1-\xi^{\sigma})}\right]
\xi^{-\sigma i\omega/\alpha_+}
(1-\xi^{\sigma})^{-2iM\omega/f_\infty^{2}}.
\label{eq:validation_factor}
\end{equation}
Defining
\begin{equation}
\mathcal G(\xi)=
\frac{F[r(\xi)](1-\xi^{\sigma})^{2}}
{\sigma r_+\xi^{\sigma-1}},
\qquad
\mathcal H(\xi,\omega)=
\frac{1}{\mathcal P}\frac{\mathrm{d}\mathcal P}{\mathrm{d}\xi},
\label{eq:validation_auxiliary}
\end{equation}
the radial equation becomes
\begin{equation}
\frac{\mathrm{d}^{2}\chi}{\mathrm{d}\xi^{2}}
=\lambda_0(\xi,\omega)\frac{\mathrm{d}\chi}{\mathrm{d}\xi}
+s_0(\xi,\omega)\chi,
\label{eq:validation_aim_form}
\end{equation}
where
\begin{equation}
\lambda_0=-2\mathcal H
-\frac{1}{\mathcal G}\frac{\mathrm{d}\mathcal G}{\mathrm{d}\xi},
\qquad
s_0=\frac{V-\omega^{2}}{\mathcal G^{2}}
-\frac{\mathrm{d}\mathcal H}{\mathrm{d}\xi}
-\mathcal H^{2}
-\frac{\mathcal H}{\mathcal G}\frac{\mathrm{d}\mathcal G}{\mathrm{d}\xi}.
\label{eq:validation_aim_coefficients}
\end{equation}

The numerical implementation evaluates the AIM condition through an equivalent determinant of Taylor coefficients. Around an interior point $\xi_0$, write
\begin{equation}
\lambda_0=\sum_{j=0}^{\infty}c_j(\xi-\xi_0)^j,
\qquad
s_0=\sum_{j=0}^{\infty}d_j(\xi-\xi_0)^j,
\qquad
\chi=\sum_{j=0}^{\infty}a_j(\xi-\xi_0)^j.
\label{eq:validation_aim_series}
\end{equation}
Substitution into Eq.~\eqref{eq:validation_aim_form} yields
\begin{equation}
a_{j+2}=\frac{1}{(j+1)(j+2)}
\sum_{m=0}^{j}\left[
c_m(j-m+1)a_{j-m+1}+d_m a_{j-m}
\right].
\label{eq:validation_taylor_recurrence}
\end{equation}
Two sequences, denoted by $A$ and $B$, are generated from $(a_0,a_1)=(1,0)$ and $(0,1)$, respectively. At order $N$, the frequency condition is
\begin{equation}
\mathcal D_N(\omega)=
a_{N+2}^{(A)}a_{N+1}^{(B)}
-a_{N+1}^{(A)}a_{N+2}^{(B)}=0.
\label{eq:validation_aim_condition}
\end{equation}
In terms of the functions $\lambda_j$ and $s_j$ obtained by successive differentiation in the usual AIM construction,
\begin{equation}
\mathcal D_N=
\frac{s_N(\xi_0)\lambda_{N-1}(\xi_0)
-s_{N-1}(\xi_0)\lambda_N(\xi_0)}
{(N+1)!(N+2)!}.
\label{eq:validation_determinant_equivalence}
\end{equation}
Eq.~\eqref{eq:validation_aim_condition} has the standard AIM zeros. The iteration order $N$ is distinct from the physical overtone number $n$.

Each mode is first checked in the Schwarzschild limit against the corresponding reference frequency~\cite{Mamani2022Schwarzschild,Konoplya2022Overtones,Jing2005Dirac}. The spinor values reported in terms of $2M\omega$ are converted to $M\omega$. All eight comparisons agree within $3\times10^{-6}$ in each component of $M\omega$, with the largest difference below $1.5\times10^{-6}$. This tolerance accommodates the precision of the published values. The frequencies are then followed continuously as $(Q,p,\ell)$ vary from $(0,0,0)$ to the target configuration.

For the sake of completeness, we also evaluate (and display) the sixth order WKB frequencies at $\ell = Q = p =0$ for the fundamental mode of each spin sector. Table~\ref{tab:schwarzschild_wkb_check} compares these results with the Schwarzschild reference frequencies of Ref.~\cite{Mamani2022Schwarzschild}. The calculations use $M = 1$ and the exterior maximum of each potential. Evaluations at 50 and 80 working digits agree at all six decimal places displayed. The relative differences are defined by
\begin{equation}
\delta_A=100\frac{\left|A\!\left(M\omega_0^{(6)}\right)-A\!\left(M\omega_0^{\mathrm{ref}}\right)\right|}{\left|A\!\left(M\omega_0^{\mathrm{ref}}\right)\right|}, \qquad A\in\{\operatorname{Re},\operatorname{Im}\}.
\end{equation}

\begin{table}[htbp]
\centering
\small
\setlength{\tabcolsep}{4pt}
\caption{Schwarzschild fundamental modes at $\ell=Q=p=0$ and $M=1$. The sixth order WKB results are compared with the numerical reference values of Ref.~\cite{Mamani2022Schwarzschild}. All rows have $n=0$. The quantities $\delta_R$ and $\delta_I$ are percentages evaluated using the unrounded WKB results and the reference values displayed here. The spinor multipole follows the convention $k=l+1/2$.}
\label{tab:schwarzschild_wkb_check}
\begin{tabular}{lccccc}
\hline\hline
Sector & $l$ & $M\omega_0^{(6)}$ & $M\omega_0^{\mathrm{ref}}$ & $\delta_R$ & $\delta_I$ \\
\hline
Scalar & $0$ & $0.110467-0.100816i$ & $0.110455-0.104896i$ & $0.0109$ & $3.8893$ \\
Vector & $1$ & $0.248191-0.092637i$ & $0.248263-0.092488i$ & $0.0288$ & $0.1611$ \\
Axial tensor & $2$ & $0.373619-0.088891i$ & $0.373672-0.088962i$ & $0.0141$ & $0.0798$ \\
Spinor & $1/2$ & $0.182646-0.094935i$ & $0.182963-0.096982i$ & $0.1735$ & $2.1109$ \\
\hline\hline
\end{tabular}
\end{table}

The scalar and vector fundamental frequencies are clearly separated in the Schwarzschild limit. The real components agree with the reference values within $0.18\%$ for all four sectors. The imaginary components differ by approximately $3.89\%$ for the scalar monopole and $2.11\%$ for the lowest spinor multipole, while the vector and axial tensor differences remain below $0.17\%$. These discrepancies quantify the accuracy of the sixth order WKB approximation for the modes tested.

Numerical stability is assessed by varying the iteration order, expansion point, and working precision. We use $N\in\{120,180,240\}$ for $n=0$, $N\in\{320,400,480\}$ for $n=1$, and $N\in\{700,850,1000\}$ for $n=2$, with respective precision pairs $(180,240)$, $(360,440)$, and $(700,850)$ decimal digits. The scalar expansion point is chosen near the potential maximum in its compact coordinate. For spinors, it is the smaller of the peak coordinate and $0.45$. In both sectors, the point is shifted independently by $\pm0.02$. The three orders are evaluated at the central point and higher precision, while the shifted points and lower precision are tested at the largest order.

Let $\omega_{\mathrm{ref}}$ be the result at the largest order, central point, and higher precision, and let $\omega_j$ denote the refinement results. The component spreads are defined by
\begin{equation}
\Delta_A=\max\!\left\{
\max_j\left|A\!\left[M(\omega_j-\omega_{\mathrm{ref}})\right]\right|,
\max_j M\eta_j
\right\},
\qquad A\in\{\operatorname{Re},\operatorname{Im}\},
\label{eq:validation_spreads}
\end{equation}
where $\eta_j$ is the magnitude of the final Newton correction. The real and imaginary components define $\Delta_R$ and $\Delta_I$, respectively. Both the final correction and its estimated arithmetic uncertainty must be below $10^{-12}$ in $M\omega$, and a mode is retained only if $\Delta_R,\Delta_I<10^{-6}$. These spreads characterize observed numerical stability and do not constitute rigorous error bounds. All eight modes satisfy the criterion, as shown in Table~\ref{tab:aim_convergence}.

\begin{table}[!ht]
\centering
\caption{Observed refinement spreads for the independent AIM frequencies at $M=1$, $Q=p=0.1$, and $\ell=0.1$. The quantities $\Delta_R$ and $\Delta_I$ refer to the components of $M\omega$. Values below the internal root tolerance are reported as $<10^{-12}$.}
\label{tab:aim_convergence}
\begin{tabular}{c| c| c| c| c}
\hline\hline
Sector & $l$ & $n$ & $\Delta_R$ & $\Delta_I$ \\
\hline
Scalar & $0$ & $0$ & $1.69\times10^{-8}$ & $1.57\times10^{-9}$ \\
Scalar & $0$ & $1$ & $6.97\times10^{-7}$ & $4.17\times10^{-7}$ \\
Scalar & $0$ & $2$ & $6.98\times10^{-7}$ & $1.37\times10^{-7}$ \\
Scalar & $2$ & $0$ & $<10^{-12}$ & $<10^{-12}$ \\
\hline
Spinor & $1/2$ & $0$ & $4.32\times10^{-10}$ & $1.30\times10^{-9}$ \\
Spinor & $1/2$ & $1$ & $2.06\times10^{-8}$ & $1.67\times10^{-8}$ \\
Spinor & $1/2$ & $2$ & $9.52\times10^{-8}$ & $2.37\times10^{-8}$ \\
Spinor & $5/2$ & $0$ & $<10^{-12}$ & $<10^{-12}$ \\
\hline\hline
\end{tabular}
\end{table}

The relative differences between the corrected sixth order WKB and AIM frequencies are evaluated separately for the real and imaginary components,
\begin{equation}
\epsilon_R=100\frac{\left|\operatorname{Re}\omega_{ln}^{(6)}-
\operatorname{Re}\omega_{ln}^{\mathrm{AIM}}\right|}
{\left|\operatorname{Re}\omega_{ln}^{\mathrm{AIM}}\right|},
\qquad
\epsilon_I=100\frac{\left|\operatorname{Im}\omega_{ln}^{(6)}-
\operatorname{Im}\omega_{ln}^{\mathrm{AIM}}\right|}
{\left|\operatorname{Im}\omega_{ln}^{\mathrm{AIM}}\right|}.
\label{eq:validation_errors}
\end{equation}
Table~\ref{tab:wkb_aim_comparison} uses the retained numerical output of the corrected WKB calculation. The discrepancies are computed before rounding the displayed frequencies to six decimal places.

\begin{table*}[!t]
\centering
\caption{Comparison of the corrected sixth order WKB frequencies with the independently converged AIM frequencies at $M=1$, $Q=p=0.1$, and $\ell=0.1$. Frequencies are displayed to six decimal places, and $\epsilon_R$ and $\epsilon_I$ are percentages calculated from the retained numerical values. The AIM refinement spreads are given in Table~\ref{tab:aim_convergence}.}
\label{tab:wkb_aim_comparison}
\setlength{\tabcolsep}{4pt}
\begin{tabular}{c| c| c| c| c| r| r}
\hline\hline
Sector & $l$ & $n$ & $M\omega^{(6)}$ & $M\omega^{\mathrm{AIM}}$ & $\epsilon_R$ & $\epsilon_I$ \\
\hline
Scalar & $0$ & $0$ & $0.137045-0.124638i$ & $0.137030-0.129643i$ & $0.0112$ & $3.8609$ \\
Scalar & $0$ & $1$ & $0.110637-0.425607i$ & $0.107094-0.429936i$ & $3.3080$ & $1.0069$ \\
Scalar & $0$ & $2$ & $0.236232-0.590051i$ & $0.094347-0.742394i$ & $150.3867$ & $20.5205$ \\
Scalar & $2$ & $0$ & $0.570676-0.119713i$ & $0.570679-0.119701i$ & $5.86\times10^{-4}$ & $0.0101$ \\
\hline
Spinor & $1/2$ & $0$ & $0.213341-0.117383i$ & $0.214186-0.120014i$ & $0.3946$ & $2.1925$ \\
Spinor & $1/2$ & $1$ & $0.166129-0.391293i$ & $0.170665-0.394126i$ & $2.6578$ & $0.7190$ \\
Spinor & $1/2$ & $2$ & $0.145990-0.707902i$ & $0.138664-0.700869i$ & $5.2833$ & $1.0034$ \\
Spinor & $5/2$ & $0$ & $0.675121-0.119089i$ & $0.675121-0.119086i$ & $3.19\times10^{-5}$ & $0.0023$ \\
\hline\hline
\end{tabular}
\end{table*}

The scalar $l=2$ fundamental mode agrees with AIM to approximately $5.86\times10^{-4}\%$ in the real component and $0.0101\%$ in the imaginary component. For the scalar monopole, the corresponding differences are $0.0112\%$ and $3.8609\%$ for $n=0$, and $3.3080\%$ and $1.0069\%$ for $n=1$. The second overtone remains poorly described by the sixth order WKB approximation: its real and imaginary components differ by $150.3867\%$ and $20.5205\%$, respectively. 

The spinor frequencies exhibit much closer agreement. For $l=1/2$, the real component differs by $0.3946\%$, $2.6578\%$, and $5.2833\%$ for $n=0,1,2$, while the imaginary component differs by $2.1925\%$, $0.7190\%$, and $1.0034\%$, respectively. For the $l=5/2$ fundamental mode, these differences decrease to $3.19\times10^{-5}\%$ and $0.0023\%$. 

The sixth order WKB approximation agrees closely with AIM for the scalar $l=2$ and spinor $l=5/2$ fundamental modes, but fails to describe the scalar $l=0,n=2$ mode accurately. Thereby, this agreement cannot be extended to the remaining tabulated modes without further verification.
}


{

\subsection{Time normalization and comparison with a global--monopole geometry}
\label{subsec:normalization_monopole}

To interpret the dependence of the observables on the Lorentz--violating parameter, it is necessary to distinguish the normalization of the coordinates from changes in the effective parameters and from the residual deformation of the geometry. Introducing $a\equiv 1-\ell>0$, the metric function can be written as
\begin{equation}
F(r) = \frac{1}{a}-\frac{2M}{r} + \frac{P_\ell^2}{r^2}, \qquad P_\ell^2 = \frac{Q^2}{a^2} + \frac{p^2}{1-2\ell}.
\end{equation}
The substitution $\bar t=t/a$ gives the equivalent representation
\begin{equation}
\mathrm{d}s^2 = a \left[-f(r)\,\mathrm{d}\bar t^2 + \frac{\mathrm{d}r^2}{f(r)}+\frac{r^2}{a}\,\mathrm{d}\Omega^2\right], \qquad f(r) = a  F(r) = 1 - \frac{2aM}{r} + \frac{aP_\ell^2}{r^2}.
\end{equation}
As we can see, the radial function has the Reissner--Nordstr\"om form with mass coefficient $aM$ and squared charge coefficient $aP_\ell^2$. However, this representation also contains an overall constant factor and a modified angular sector. In other words, identifying these radial coefficients does not establish equivalence with the ordinary Reissner--Nordstr\"om spacetime.

For physical comparisons, it is useful to normalize the timelike Killing vector at infinity and simultaneously introduce a radial coordinate with an asymptotically unit radial metric coefficient. Defining
\begin{equation}
\widehat t = \frac{t}{\sqrt a}, \qquad \rho = \sqrt a\,r, \qquad \mu = a^{3/2}M, \qquad q^2 = a^2P_\ell^2 = Q^2 + \frac{a^2p^2}{1 - 2\ell},
\end{equation}
the line element becomes
\begin{equation}
\mathrm{d}s^2 = - h(\rho)\,\mathrm{d}\widehat t^2 + \frac{\mathrm{d}\rho^2}{h(\rho)} + \frac{\rho^2}{a}\,\mathrm{d}\Omega^2, \qquad h(\rho) = 1 - \frac{2\mu}{\rho} + \frac{q^2}{\rho^2}.
\label{eq:normalized_dyonic_geometry}
\end{equation}
Here, $\widehat t$ agrees asymptotically with the proper time of static observers, while $\mu$ and $q^2$ denote the coefficients of the normalized radial function. In particular, $\mu$ is used below as a characteristic length scale; its introduction does not assume an asymptotically flat ADM mass interpretation.

The angular modification cannot be eliminated by a change of units. The area of a sphere at fixed $\rho$ and its asymptotic ratio to the squared proper radial distance are
\begin{equation}
\mathcal A(\rho) = \frac{4\pi\rho^2}{a}, \qquad \lim_{\rho\rightarrow\infty} \frac{\mathcal A(\rho)}{s_{\mathrm{rad}}^2(\rho)} = \frac{4\pi}{a},
\end{equation}
where $s_{\mathrm{rad}}(\rho)$ is the proper radial distance measured from a fixed exterior reference sphere. Positive $\ell$ produces a solid angle excess, whereas negative $\ell$ produces a deficit. The distinction from ordinary Reissner--Nordstr\"om geometry is also apparent in the Ricci scalar,
\begin{equation}
R = \frac{2(a-1)}{\rho^2} =  - \frac{2\ell}{\rho^2} = - \frac{2\ell}{(1-\ell)r^2},
\end{equation}
which does not vanish for $\ell\neq0$.

This structure is closely related to the exterior global-monopole geometry of Barriola and Vilenkin \cite{BarriolaVilenkin1989}. In its normalized form, the angular sector contains a factor $\beta^2 = 1 - 8\pi G\eta^2$, where $\eta$ is the symmetry breaking scale. The corresponding geometric identification is
\begin{equation}
\beta^2 = \frac{1}{a}, \qquad 8\pi G\eta^2 = 1 - \frac{1}{a} = - \frac{\ell}{1 - \ell}.
\end{equation}
For the positive values of $\ell$ considered here, this identification would require a negative value of $\eta^2$ and does not correspond to the conventional global monopole with a real symmetry breaking scale. The comparison is instead with a solid angle excess counterpart of the same metric structure. For negative $\ell$, the angular factor has the conventional deficit sign.

The normalization of time gives an exact conversion for the quasinormal frequencies. Since the temporal dependence used in the preceding analysis is $\exp(-i\omega t)$, the frequency associated with $\widehat t$ satisfies
\begin{equation}
e^{-i\omega t} = e^{-i\widehat\omega\widehat t}, \qquad \widehat\omega = \sqrt a\,\omega, \qquad \mu\widehat\omega = a^2M\omega.
\label{eq:normalized_qnm_frequency}
\end{equation}
Both the real and imaginary parts acquire the same factor $\sqrt a$. Hence, part of the variation of the tabulated frequencies with $\ell$ reflects the normalization of the original time coordinate. In particular,
\begin{equation}
\frac{q^2}{\mu^2} = \frac{P_\ell^2}{aM^2} = \frac{Q^2}{a^3M^2} + \frac{p^2}{a(1 - 2\ell)M^2}.
\end{equation}
A comparison that isolates the angular deformation should examine $\mu\widehat\omega$ at fixed $q/\mu$, multipole number, and overtone number.

The remaining dependence can be displayed directly in the radial wave equations. Introducing the normalized tortoise coordinate and effective potentials gives
\begin{equation}
\mathrm{d}\rho_* = \frac{\mathrm{d}\rho}{h(\rho)}, \qquad \rho_* = \frac{r_*}{\sqrt a}, \qquad \widehat V(\rho) = aV\!\left(\frac{\rho}{\sqrt a}\right), \qquad \frac{\mathrm{d}^2\Psi}{\mathrm{d}\rho_*^2} + \left[\widehat\omega^2 - \widehat V(\rho)\right]\Psi = 0.
\end{equation}
For the minimally coupled massless scalar and vector sectors used in this work, the normalized potentials are
\begin{equation}
\widehat V_s=h(\rho)\left[\frac{a\,l(l+1)}{\rho^2}+\frac{h'(\rho)}{\rho}\right], \qquad \widehat V_v=h(\rho)\frac{a\,l(l+1)}{\rho^2},
\end{equation}
where the prime denotes differentiation with respect to $\rho$. For the massless spinor sector, the partner potentials can be expressed as
\begin{equation}
\widehat V_\psi^\pm = \widehat W^2\pm\frac{\mathrm{d}\widehat W}{\mathrm{d}\rho_*}, \qquad \widehat W = \sqrt a\left(l + \frac{1}{2}\right)\frac{\sqrt{h(\rho)}}{\rho}.
\end{equation}
Within the effective axial source closure adopted in Appendix \ref{dervitensor}, the tensor potential becomes
\begin{equation}
\widehat V_t = h(\rho)\left[\frac{a\bigl(l(l + 1) - 2\bigr) + 2h(\rho)}{\rho^2} - \frac{h'(\rho)}{\rho}\right].
\end{equation}
These expressions retain an explicit dependence on $a$ even when $\mu$ and $q$ are held fixed. This dependence represents a genuine geometric modification of the angular and curvature contributions. An instructive exception is the scalar monopole, $l = 0$, whose normalized potential coincides with that of Reissner--Nordstr\"om at fixed $\mu$ and $q$.

Analogously, the time delay calculation requires a similar distinction between coordinate normalization and the reference background. Denote the coordinate travel time in Section V by $T$ and retain its definition of $T_0$,
\begin{equation}
T_0 = \sqrt{r_S^2 - r_0^2} + \sqrt{r_O^2 - r_0^2}.
\end{equation}
When $M=Q=p=0$ at fixed $\ell$, the metric function is $F=1/a$, and the coordinate travel time for the same prescribed radial endpoints and turning radius is
\begin{equation}
T_{\mathrm{bg}} = a T_0.
\end{equation}
In this manner, the quantity $\Delta T = T - T_0$ defined in Section V contains a contribution from the mismatch between the asymptotically conical background and the flat space reference. A delay relative to the background with the same angular deformation, expressed in normalized asymptotic time, is instead
\begin{equation}
\Delta\widehat T_{\mathrm{cone}}=\frac{T-T_{\mathrm{bg}}}{\sqrt a}=\frac{\Delta T+\ell T_0}{\sqrt a}.
\label{eq:normalized_conical_delay}
\end{equation}
This quantity vanishes when the mass and charges vanish at fixed $\ell$.

The same result follows directly from the normalized geometry. For a ray with turning radius $\rho_0=\sqrt a\,r_0$, its normalized travel time is
\begin{equation}
\widehat T=\sum_{i=S,O}\int_{\rho_0}^{\rho_i}\frac{\mathrm{d}\rho}{h(\rho)\sqrt{1-\dfrac{\rho_0^2h(\rho)}{\rho^2h(\rho_0)}}}, \qquad \rho_i=\sqrt a\,r_i.
\end{equation}
In this view, the corresponding conical background reference and subtracted delay are
\begin{equation}
\widehat T_{\mathrm{bg}} = \sum_{i = S,O}\sqrt{\rho_i^2-\rho_0^2}, \qquad \Delta\widehat T_{\mathrm{cone}} = \widehat T-\widehat T_{\mathrm{bg}}.
\end{equation}
For this prescription, in which the radial endpoints and turning radius are specified, the integral has exactly the Reissner--Nordstr\"om form at fixed $\mu$, $q$, $\rho_S$, $\rho_O$, and $\rho_0$. There is no additional explicit dependence on the angular factor. Equivalently, $\Delta\widehat T_{\mathrm{cone}}/\mu$ depends only on $q/\mu$ and the dimensionless radial positions. If the angular positions of the source and observer are fixed instead, the angular deformation changes the boundary value problem and generally changes the turning radius. Moreover, conversion of a travel time interval to the proper time of a static observer at $\rho_O$ involves
\begin{equation}
\mathrm{d}\tau_O = \sqrt{h(\rho_O)}\,\mathrm{d}\widehat t.
\end{equation}

For the shifted curves in Fig.~\ref{delayyy}, the subtraction between two charge configurations at the same $\ell$ and the same prescribed radial positions already cancels the common background term. Writing $Q_{\mathrm{ref}}$ for the reference electric charge used in that figure, the normalized difference satisfies
\begin{equation}
\Delta\widehat T_{\mathrm{cone}}(Q) -\Delta\widehat T_{\mathrm{cone}}(Q_{\mathrm{ref}}) = \frac{\Delta T(Q) - \Delta T(Q_{\mathrm{ref}})}{\sqrt a},
\end{equation}
with an analogous relation for the magnetic charge. The sign of the charge induced shift is preserved by the normalization. However, it is worth empathizing that the comparison between curves with different $\ell$ still combines changes in the effective radial parameters, changes in the dimensionless endpoint positions, and the normalization of time.

}


\section{Time--domain solution }

To study the time evolution of scalar, vector, and tensor perturbations, we must go beyond a purely spectral description. A frequency--domain treatment identifies the quasinormal spectrum, but it does not by itself display how the perturbing field propagates, decays, and scatters during the evolution. The time--domain approach therefore provides the appropriate setting for tracking the waveform directly and for determining how the quasinormal ringing emerges from the initial disturbance. Because the effective potentials associated with these perturbations generally possess a nontrivial radial profile, the evolution requires a stable and accurate numerical scheme. We therefore employ the characteristic integration method introduced by Gundlach et al.~\cite{Gundlach:1993tp}.

Following the numerical strategies used in Refs.~\cite{Skvortsova:2024wly,Yang:2024rms,Bolokhov:2024ixe,Guo:2023nkd,Baruah:2023rhd,Gundlach:1993tp,Shao:2023qlt,Santos:2025xbk,Lutfuoglu:2025kqp,price1972nonspherical,ching1995wave}, we rewrite the perturbation equation in the double--null variables $u=t-r^{*}$ and $v=t+r^{*}$. This choice places the evolution problem on a null grid and leads to a form that is particularly suitable for numerical propagation. In these coordinates, the master equation becomes:
\ie
\left(4 \frac{\partial^{2}}{\partial u \, \partial v} + V(u,v)\right) \Tilde{\psi} (u,v) = 0.
\fe

The continuous problem is then replaced by its lattice counterpart. The $(u,v)$ plane is sampled by a set of discrete points, and the differential equation is approximated through finite--difference relations on this null mesh. With this discretization, the value of the perturbation field at each new grid point is obtained from the neighboring points already known, so that the waveform is advanced step by step throughout the numerical domain 
\ie
\Tilde{\psi}(N) = -\Tilde{\psi}(S) + \Tilde{\psi}(W) + \Tilde{\psi}(E) - \frac{h^{2}}{8}V(S)\Big[\Tilde{\psi}(W) + \Tilde{\psi}(E)\Big] + \mathcal{O}(h^{4}).
\fe

The numerical evolution is initialized on a uniform lattice in the $(u,v)$ plane, whose spacing is fixed by the step size $h$. For each null cell, four grid points are assigned: the lower corner $S=(u,v)$, the two neighboring points $W=(u+h,v)$ and $E=(u,v+h)$, and the future vertex $N=(u+h,v+h)$, where the field value is obtained from the finite--difference update rule. The integration begins once the perturbation data are specified on the two initial null segments, $u=u_{0}$ and $v=v_{0}$. On the boundary $u=u_{0}$, the initial disturbance is prescribed as a Gaussian pulse, with center $v_{c}$ and width $\sigma$. This pulse supplies the initial signal that is subsequently propagated across the full computational grid
\ie
\Tilde{\psi}(\Tilde{u} = u_{0},v) = A e^{-(v-v_{0})^{2}}/2\sigma^{2}, \,\,\,\,\,\, \Tilde{\psi}(u,v_{0}) = \Tilde{\psi}_{0}.
\fe

The evolution is started by fixing the second null boundary at $v=v_{0}$, where the condition $\tilde{\psi}(u,v_{0})=0$ is imposed. This choice provides a simple initial segment and avoids introducing spurious contributions from that side of the grid. Once the boundary data have been specified, the finite--difference algorithm advances the field across the null lattice, determining $\tilde{\psi}$ at each new point from the neighboring values already computed. The update therefore follows the causal ordering naturally encoded in the $(u,v)$ discretization.

In the present analysis, we restrict attention to massless perturbations and set the black hole mass to $M=1$. The initial signal is introduced as a Gaussian packet located at $v=0$ with width $\sigma=1$, while the complementary boundary contribution is taken to vanish. The numerical domain covers the interval $0\leq u,v\leq 1000$ and is divided with the uniform step size $h=0.1$. This resolution is sufficient to resolve the quasinormal ringing and the subsequent decay of the perturbation signal.

To make the time--domain treatment consistent with the previous frequency--domain analysis, the numerical evolution is carried out for the different spin sectors considered in this work. The corresponding results are presented (with their comparison as well) in the following subsections.


\subsection{Scalar perturbations }

\begin{figure}
    \centering
    \includegraphics[scale=0.55]{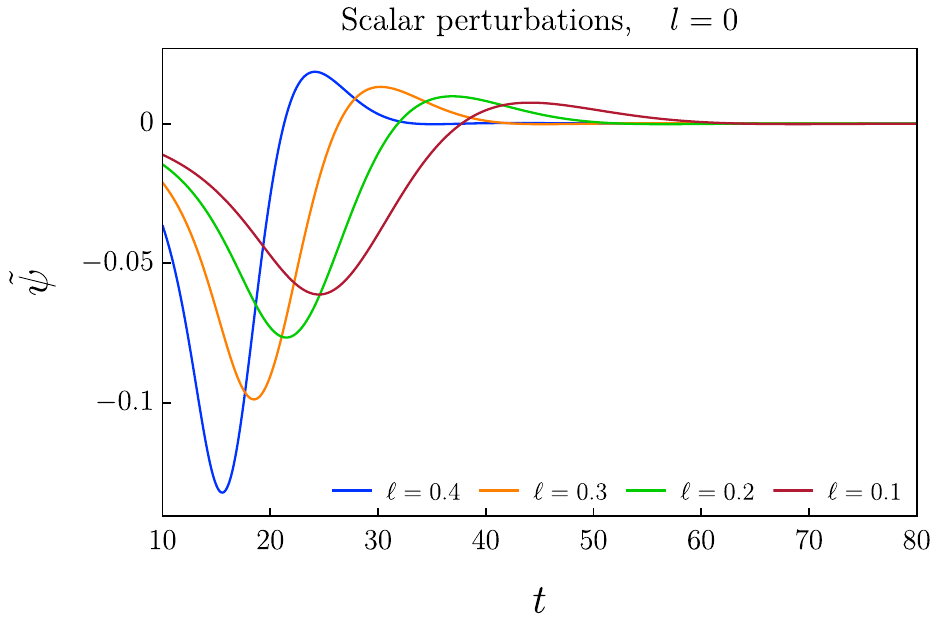}
    \includegraphics[scale=0.55]{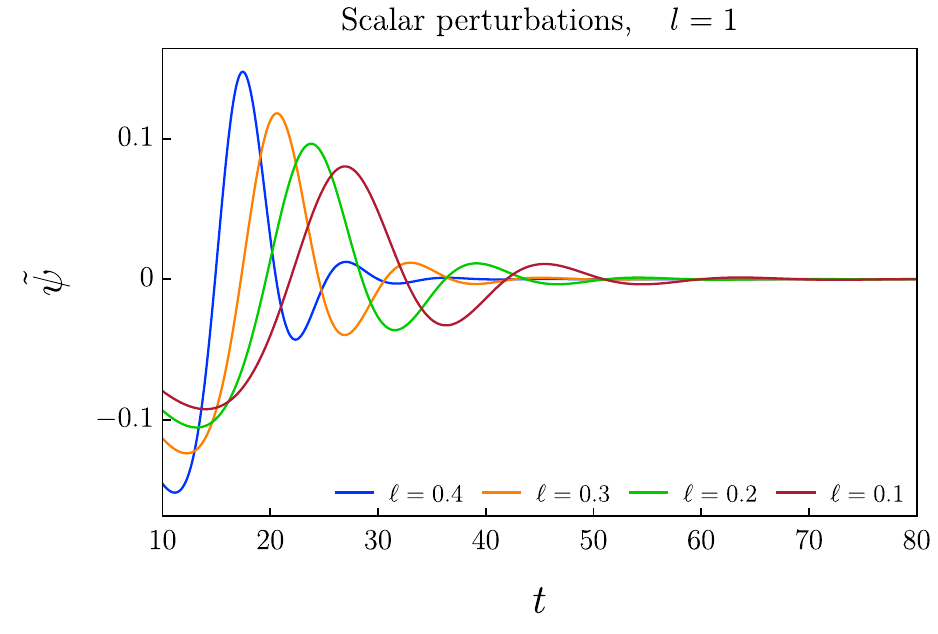}
     \includegraphics[scale=0.55]{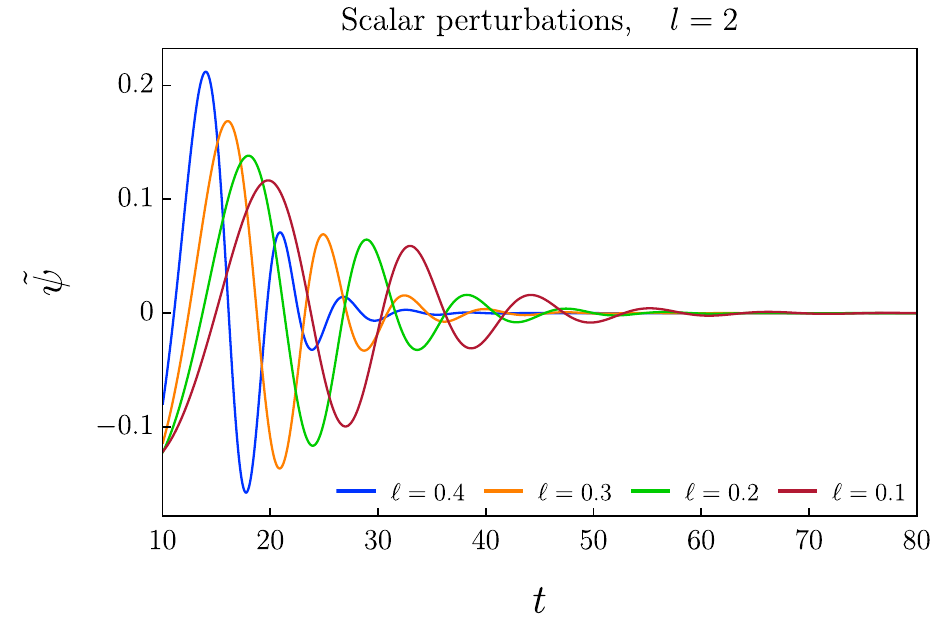}
    \caption{Scalar time--domain response $\tilde{\psi}$ for a black hole with $M=1$ under four choices of the Lorentz--violating parameter, $\ell=0.1,0.2,0.3,0.4$. The upper--left, upper--right, and lower panels correspond to $l=0$, $l=1$, and $l=2$, respectively}
    \label{timedomainscalar0}
\end{figure}

We begin with the scalar channel and evolve the corresponding master variable on the background given by Eq.~(\ref{metric_ansatz_KR}). The resulting profiles are displayed in Fig.~\ref{timedomainscalar0} for the parameter choice $M=1$ and $Q=p=0.1$, while the Lorentz--violating parameter is varied as $\ell=0.1,0.2,0.3,0.4$. The upper--left, upper--right, and lower panels refer, respectively, to the modes $l=0$, $l=1$, and $l=2$. After the initial pulse crosses the numerical domain, the field enters a ringdown stage described by oscillations with a decreasing amplitude. This behavior is the time--domain counterpart of the quasinormal spectrum. A direct comparison among the curves shows that increasing $\ell$ shifts the scalar profile toward faster oscillations and stronger attenuation. {In this manner, increasing $\ell$ shortens the scalar ringdown damping time in the original time coordinate, with the increase in $|\operatorname{Im}\omega_0|$.}

\begin{figure}
    \centering
    \includegraphics[scale=0.55]{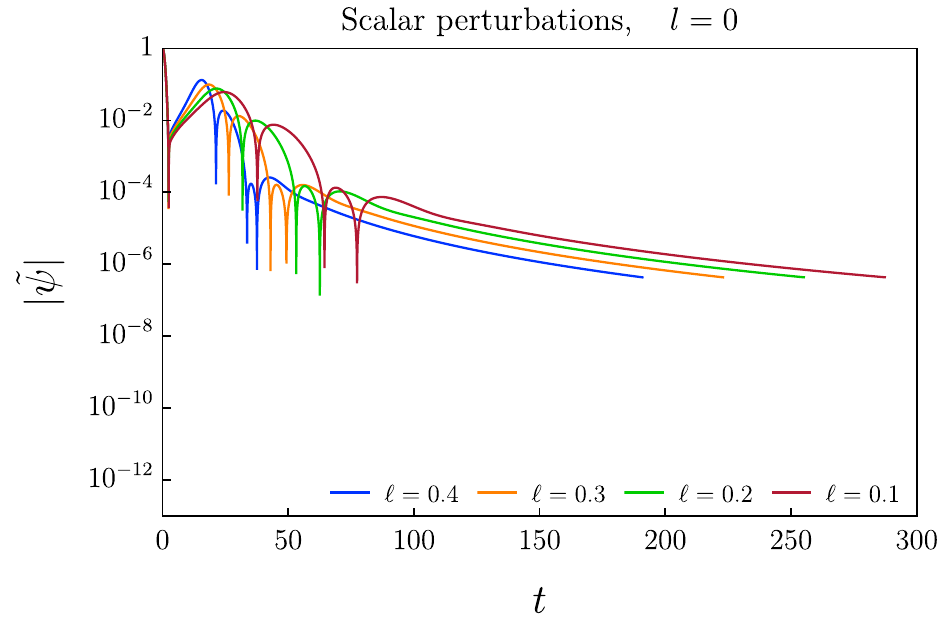}
    \includegraphics[scale=0.55]{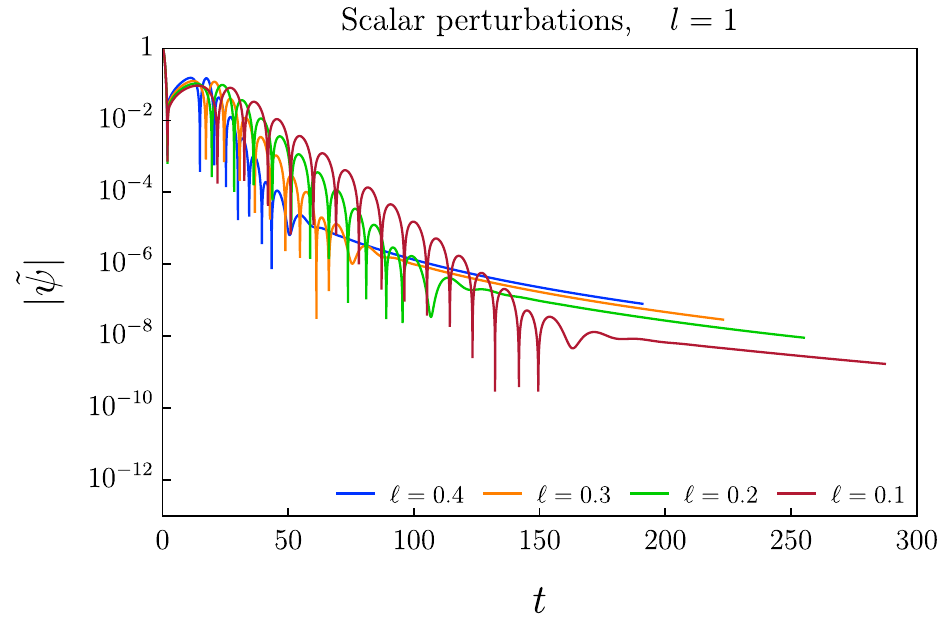}
     \includegraphics[scale=0.55]{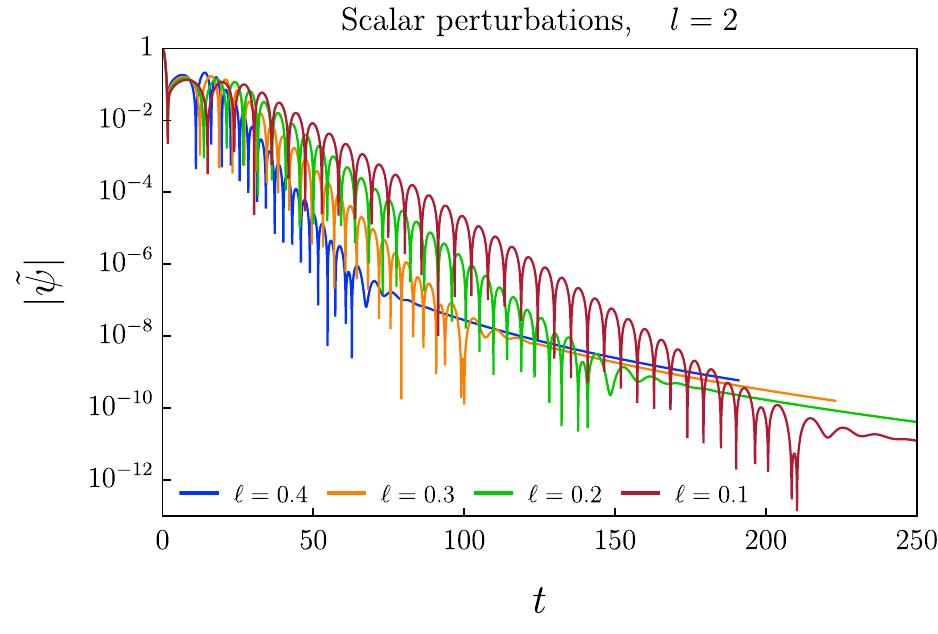}
    \caption{Logarithmic time--domain profile $|\tilde{\psi}|$ of the scalar perturbation for $M=1$ and $\ell=0.1,0.2,0.3,0.4$. The upper--left, upper--right, and lower panels display the modes $l=0$, $l=1$, and $l=2$, respectively. {The horizontal axis is linear, while the vertical axis is logarithmic.}}
    \label{timedomainscalar1}
\end{figure}

Figure~\ref{timedomainscalar1} presents the same evolutions through a {logarithmic scale of} $|\tilde{\psi}|$, keeping the same values of $l$ and $\ell$. This representation makes the decay pattern easier to identify. During the ringdown interval, the logarithmic amplitude follows an almost linear trend, as expected for an exponentially damped signal. At later times, the curves depart from this linear behavior, indicating the gradual dominance of the tail contribution. {During ringdown, larger values of $\ell$ produce a more rapidly decaying oscillation envelope. Notice that this statement concerns exponential quasinormal damping; also, it does not determine the dependence of the late--time power law exponent on $\ell$.}

\begin{figure}
    \centering
    \includegraphics[scale=0.55]{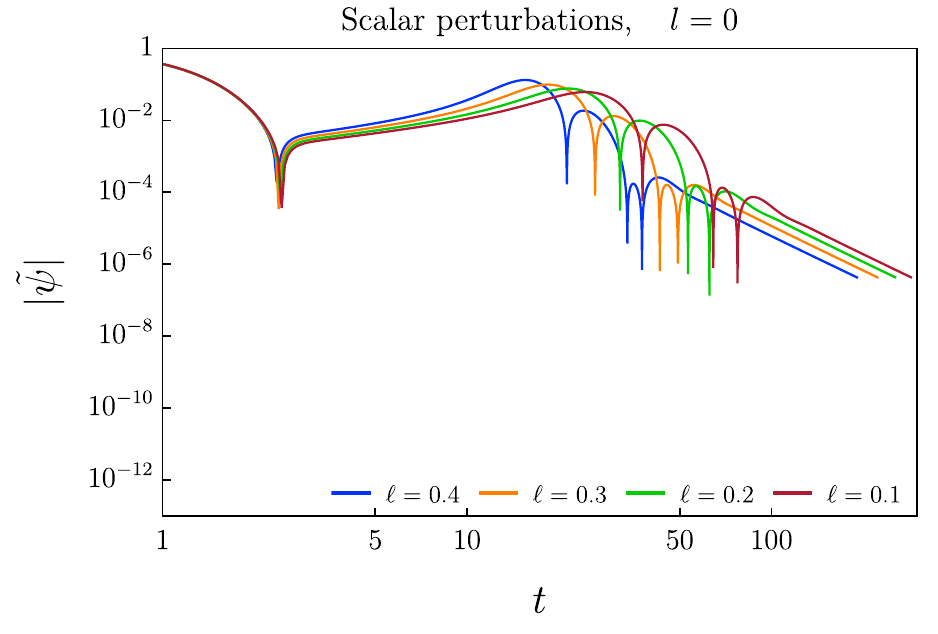}
    \includegraphics[scale=0.55]{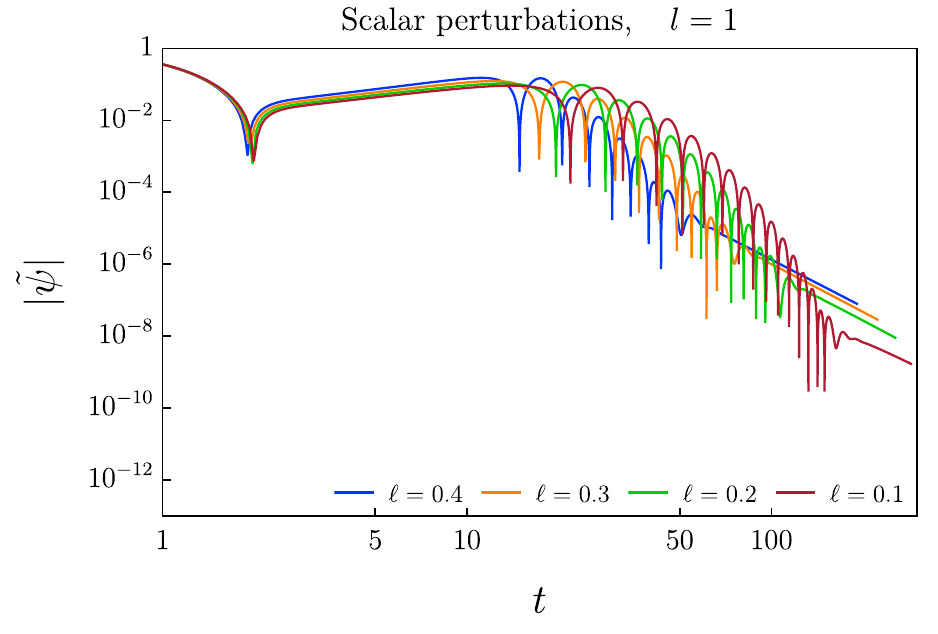}
     \includegraphics[scale=0.55]{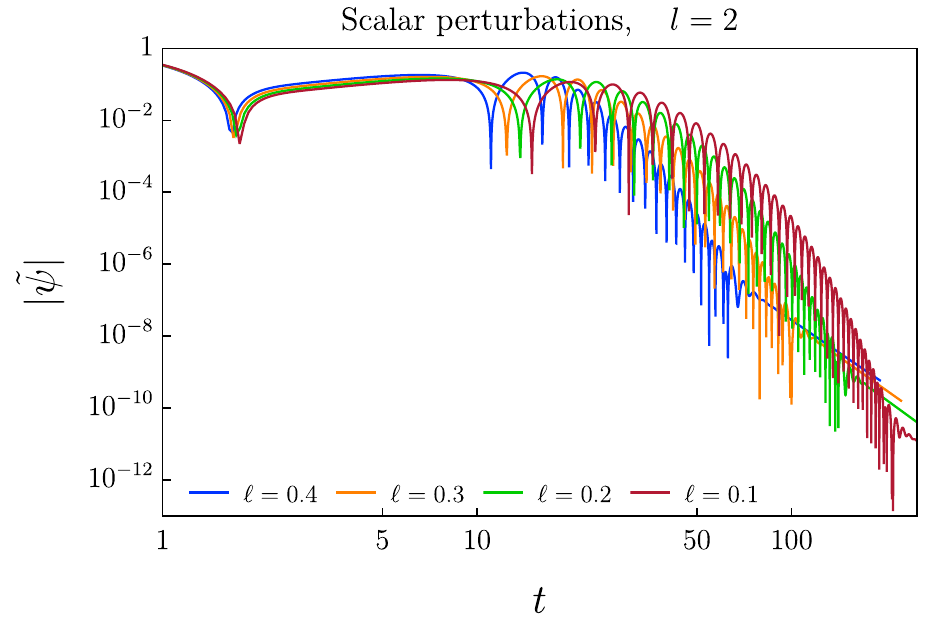}
    \caption{Late--time scalar response {shown with both axes on logarithmic scales}, with $|\tilde{\psi}|$ displayed as a function of $ t$ for $M=1$, $Q=p=0.1$, and $\ell=0.1,0.2,0.3,0.4$. The upper--left, upper--right, and lower panels refer to the modes $l=0$, $l=1$, and $l=2$, respectively.}
    \label{timedomainscalar2}
\end{figure}

The final stage of the scalar evolution is emphasized in Fig.~\ref{timedomainscalar2}, where the signal is represented on a log--log scale. The same organization of panels is adopted in order to compare the three angular sectors directly. In this form, the late--time regime appears more clearly. {
Section~\ref{sec:tail_analysis} presents a separate quantitative tail analysis for the scalar mode $l=2$ at $\ell=0.1$, together with selected modes from the other perturbative sectors. Table~\ref{tab:tail_exponents} reports the fitting interval and estimated uncertainties for those specific calculations.}

\subsection{Vector perturbations  }

\begin{figure}
    \centering
    \includegraphics[scale=0.55]{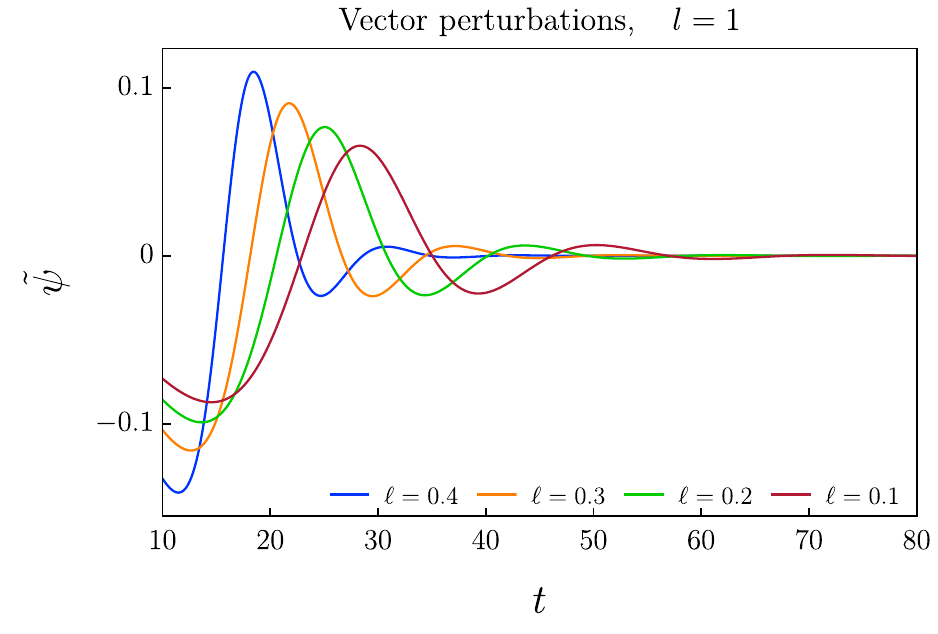}
    \includegraphics[scale=0.55]{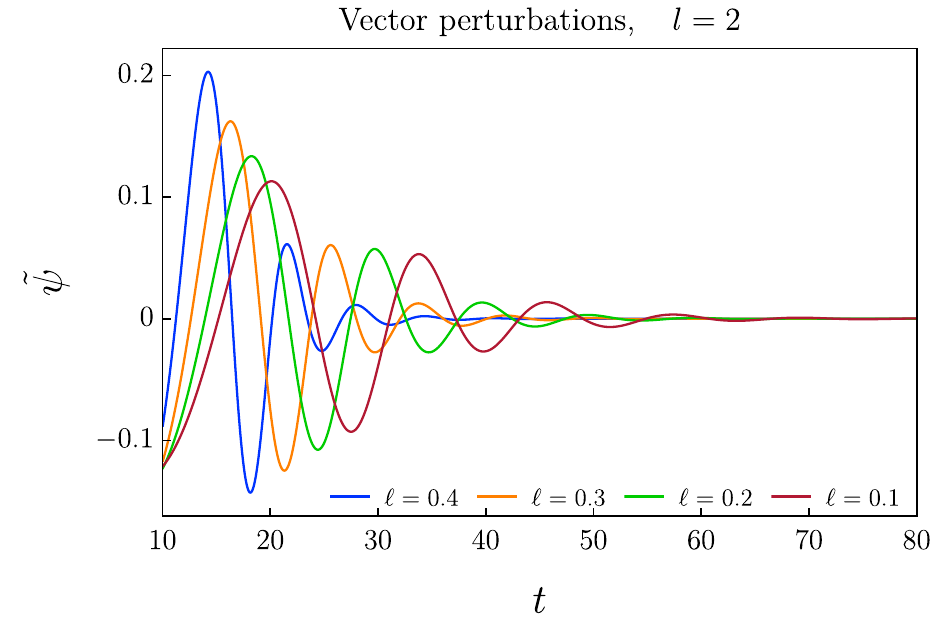}
     \includegraphics[scale=0.55]{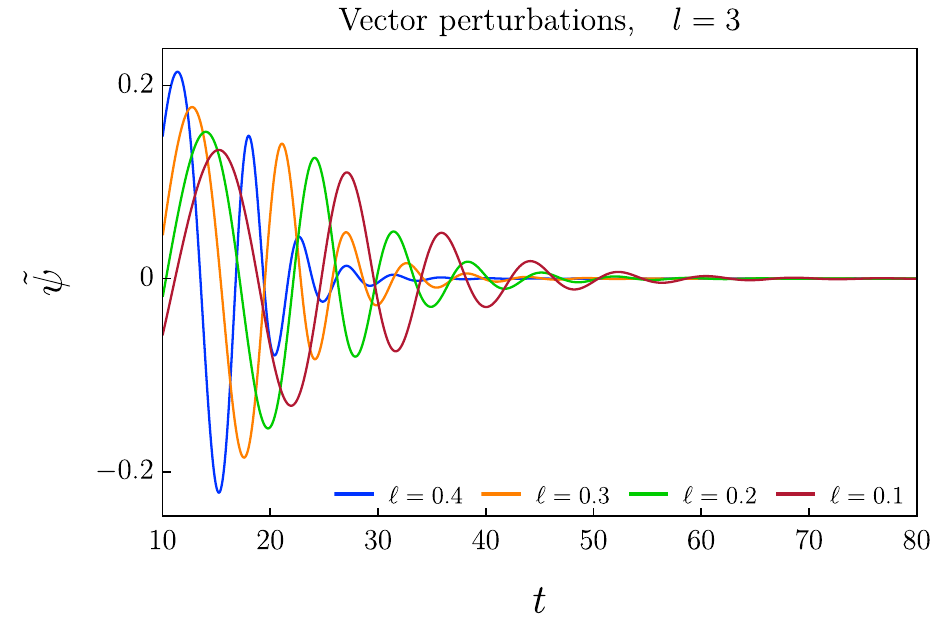}
    \caption{Vector perturbation profiles $\tilde{\psi}$ obtained in the time domain for $M=1$, $Q=p=0.1$, and $\ell=0.1,0.2,0.3,0.4$. The panels show the multipoles $l=1$, $l=2$, and $l=3$ in the upper--left, upper--right, and lower positions, respectively.}
    \label{timedomainsvector1}
\end{figure}

We now consider the vector sector associated with the background metric in Eq.~(\ref{metric_ansatz_KR}). The numerical profiles of $\tilde{\psi}$ are reported in Fig.~\ref{timedomainsvector1} for $M=1$, $Q=p=0.1$, and $\ell=0.1,0.2,0.3,0.4$. The three panels correspond to the multipoles $l=1$, $l=2$, and $l=3$, arranged in the upper--left, upper--right, and lower panels, respectively. After the initial pulse, the field enters the usual ringdown stage, with oscillations whose amplitudes decrease in time. {For the common multipoles $l=1,2$ and the same background parameters, the tabulated vector fundamental modes have lower oscillation frequencies and smaller damping rates than the corresponding scalar modes. Their ringdown damping times are therefore longer, i.e., with the comparison in Section~VII.B.} {For fixed background mass and charges, the tabulated vector frequencies show that increasing $\ell$ increases the damping rate and decreases the fundamental mode ringdown damping time.}

\begin{figure}
    \centering
    \includegraphics[scale=0.55]{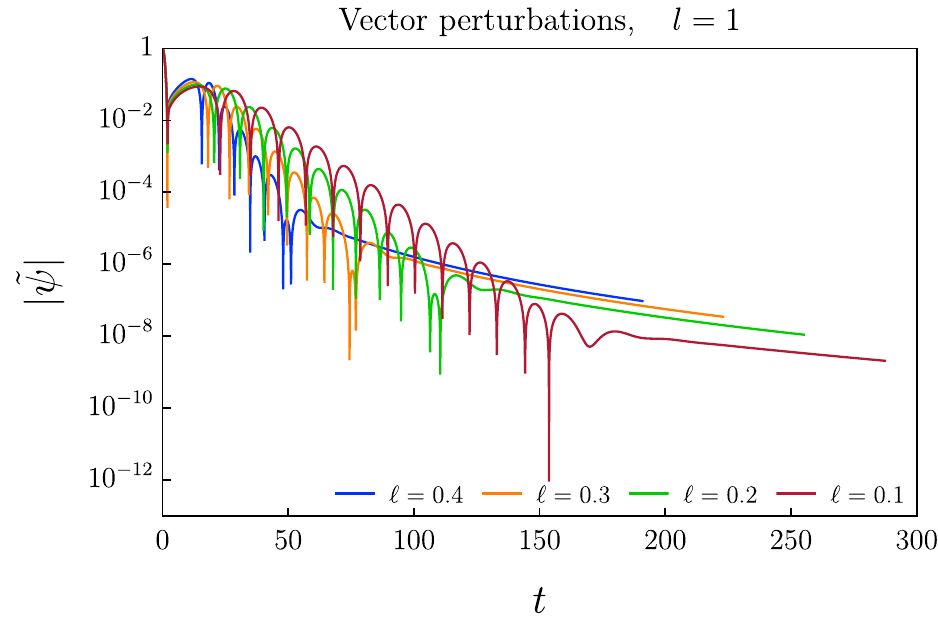}
    \includegraphics[scale=0.55]{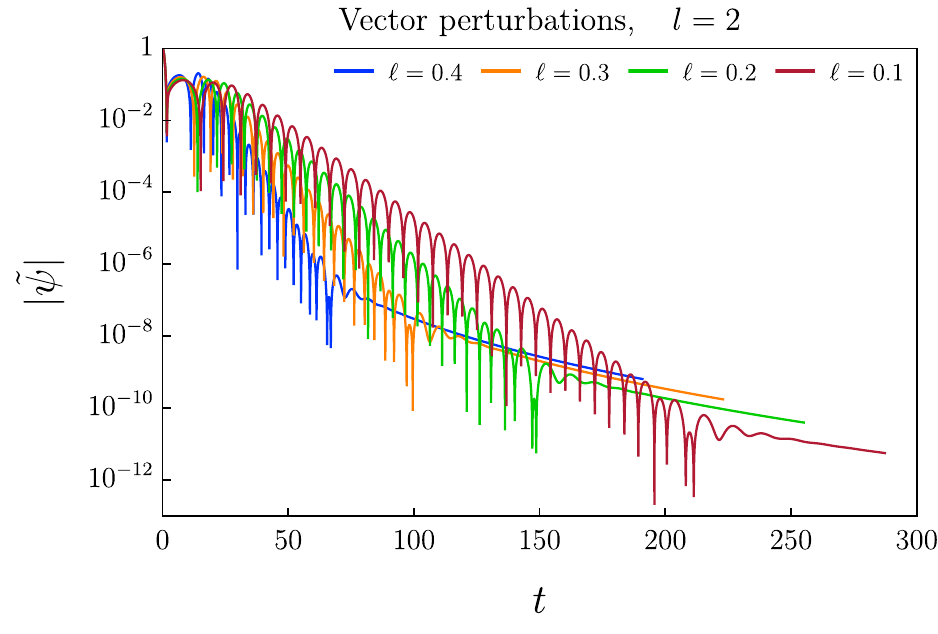}
     \includegraphics[scale=0.55]{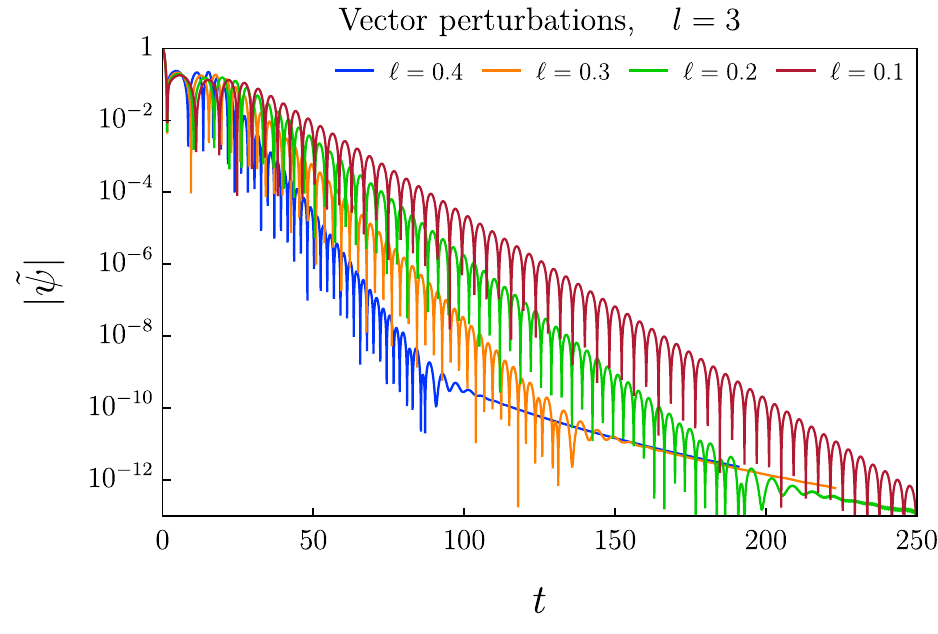}
    \caption{Logarithmic time--domain evolution of the vector perturbation, represented by $|\tilde{\psi}|$, for $M=1$, $Q=p=0.1$, and $\ell=0.1,0.2,0.3,0.4$. The upper--left, upper--right, and lower panels correspond to $l=1$, $l=2$, and $l=3$. {The horizontal axis is linear, while the vertical axis is logarithmic.} }
    \label{timedomainsvector2}
\end{figure}

The decay pattern is displayed more directly in Fig.~\ref{timedomainsvector2}, where the quantity $|\tilde{\psi}|$ is plotted for the same set of parameters {in a logarithmic scale}. In this representation, the quasinormal stage appears through nearly linear portions of the curves, reflecting the exponential suppression of the amplitude. At later times, the profiles depart from this behavior and approach the tail regime. This time--domain behavior agrees with the frequency--domain analysis, since the modes with larger $\ell$ have higher damping rates. {For the time dependence $e^{-i\omega t}$, notice that the damping time of the dominant quasinormal mode is $\tau=1/|\operatorname{Im}\omega_0|$. The smaller imaginary magnitudes of the vector fundamental modes imply slower exponential decay than in the corresponding scalar cases.}

\begin{figure}
    \centering
    \includegraphics[scale=0.55]{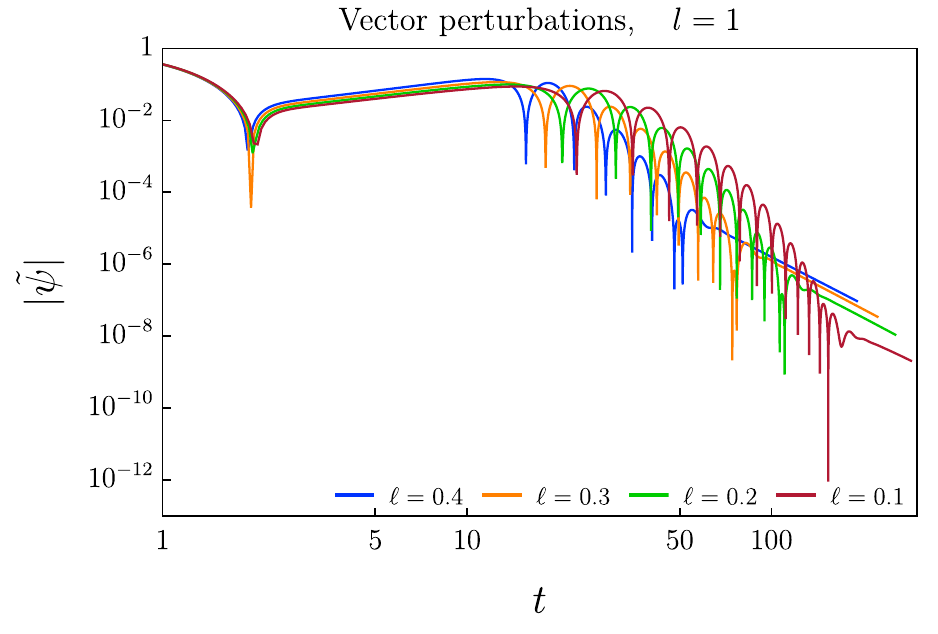}
    \includegraphics[scale=0.55]{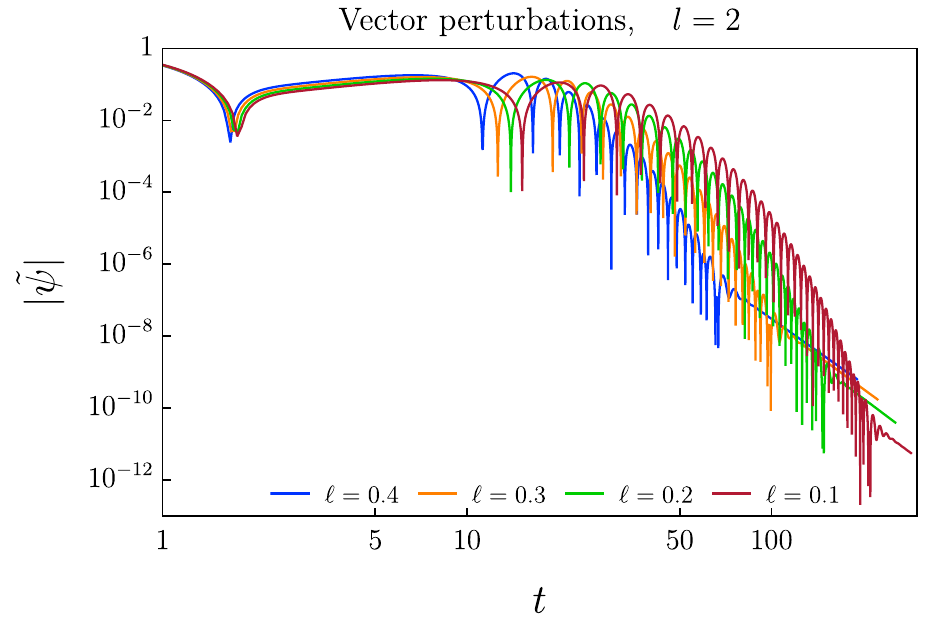}
     \includegraphics[scale=0.55]{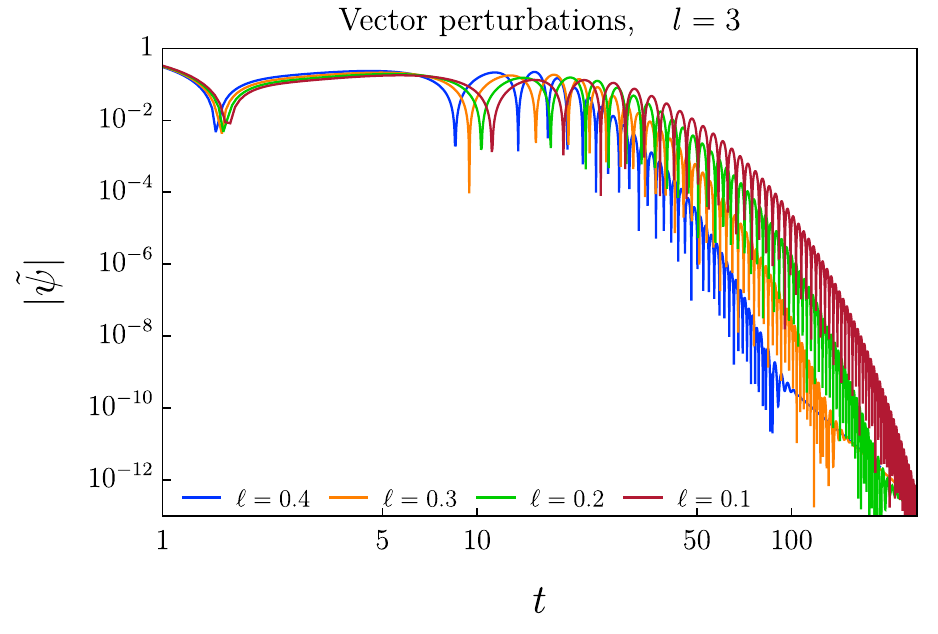}
    \caption{Asymptotic vector response displayed in a log--log representation, with $|\tilde{\psi}|$ plotted versus $ t$ for $M=1$, $Q=p=0.1$, and $\ell=0.1,0.2,0.3,0.4$. The panels show the modes $l=1$, $l=2$, and $l=3$ in the upper--left, upper--right, and lower positions, respectively. }
    \label{timedomainsvector3}
\end{figure}

The asymptotic part of the vector evolution is shown in Fig.~\ref{timedomainsvector3} by using a double--logarithmic representation. The same panel structure is kept in order to compare the three multipoles. In this scale, the end of the exponentially damped ringdown and the subsequent power--law falloff become clearer.

\subsection{Tensor perturbations }

{We consider the effective axial tensor model defined by the potential in Eq.~(\ref{tensdas}) and the source closure in Appendix~\ref{dervitensor}. Fig.~\ref{timedomaintensor2} shows its time--domain evolution for $M=1$, $Q=p=0.1$, and $\ell=0.1,0.2,0.3,0.4$, with multipoles $l=2,3,4$. For the common multipole $l=2$ and the same background parameters, the tabulated tensor fundamental mode has a lower oscillation frequency and a longer damping time than the scalar and vector modes.

Fig.~\ref{timedomaintensor3} displays the absolute waveform on a logarithmic vertical scale. During ringdown, increasing $\ell$ produces stronger exponential damping, consistently with the larger values of $|\operatorname{Im}\omega_0|$ in the tensor tables. The comparison of damping times refers specifically to the quasinormal stage.

Fig.~\ref{timedomaintensor4} presents the same response on logarithmic axes, showing the transition toward late--time decay. It is important to mention that the ordering of ringdown damping times does not determine the ordering of power law tail exponents. These exponents are estimated separately for the selected modes and parameters in Section~VIII~E.}

\begin{figure}
    \centering
    \includegraphics[scale=0.55]{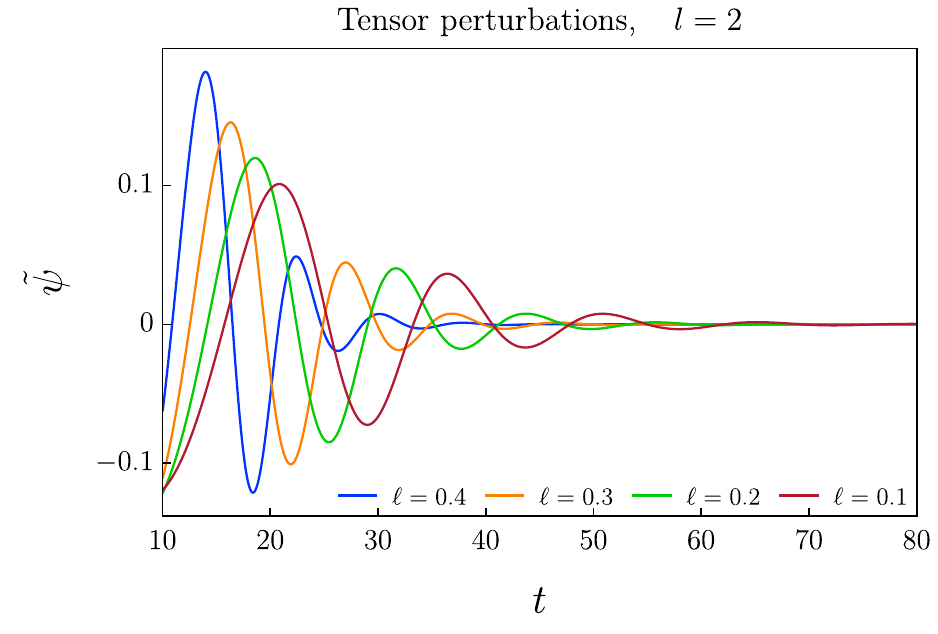}
    \includegraphics[scale=0.55]{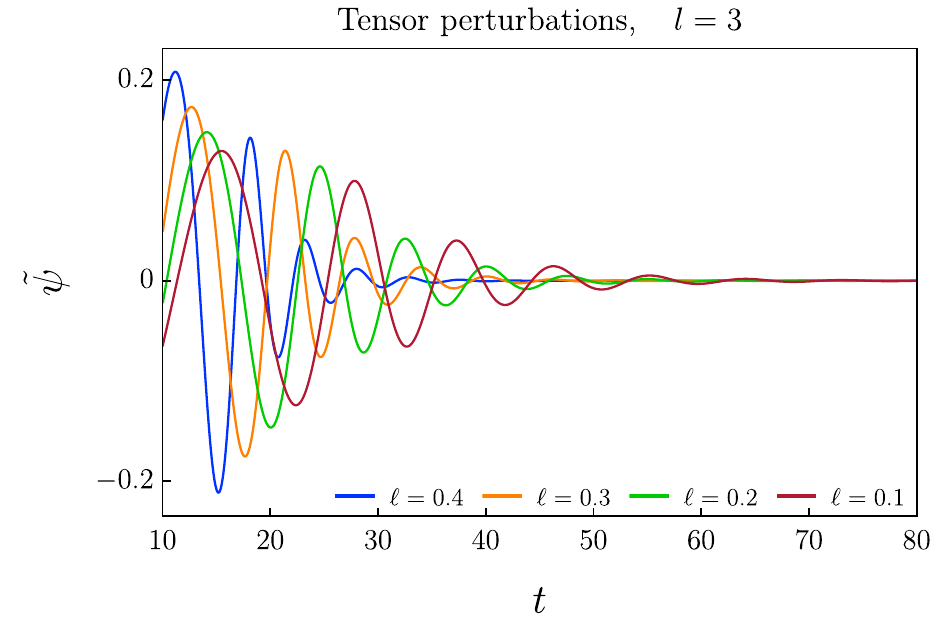}
     \includegraphics[scale=0.55]{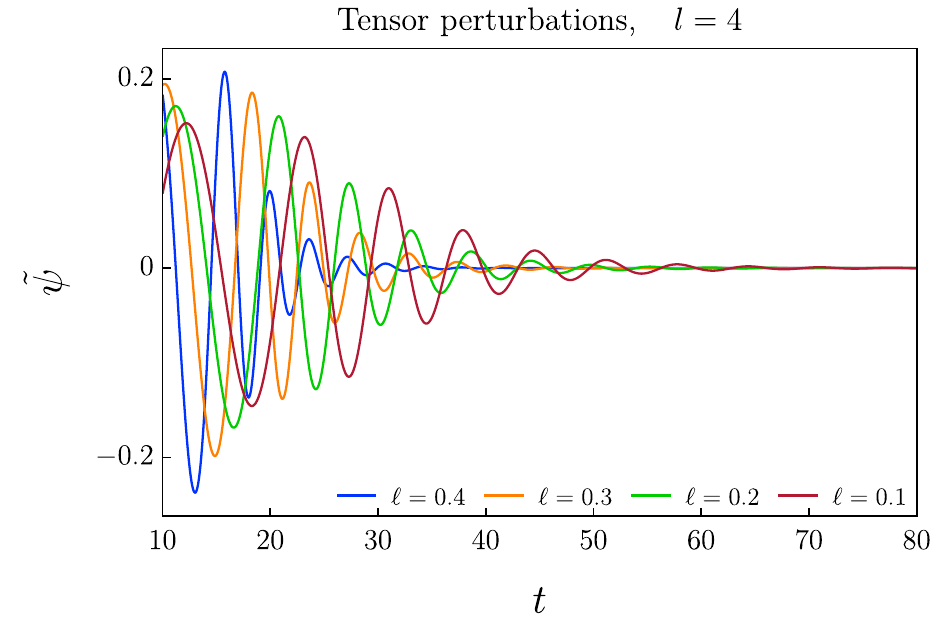}
    \caption{Time--domain tensor waveforms $\tilde{\psi}$ for $M=1$, $Q=p=0.1$, and $\ell=0.1,0.2,0.3,0.4$. The upper--left, upper--right, and lower panels correspond to the multipoles $l=2$, $l=3$, and $l=4$, respectively. }
    \label{timedomaintensor2}
\end{figure}

\begin{figure}
    \centering
    \includegraphics[scale=0.55]{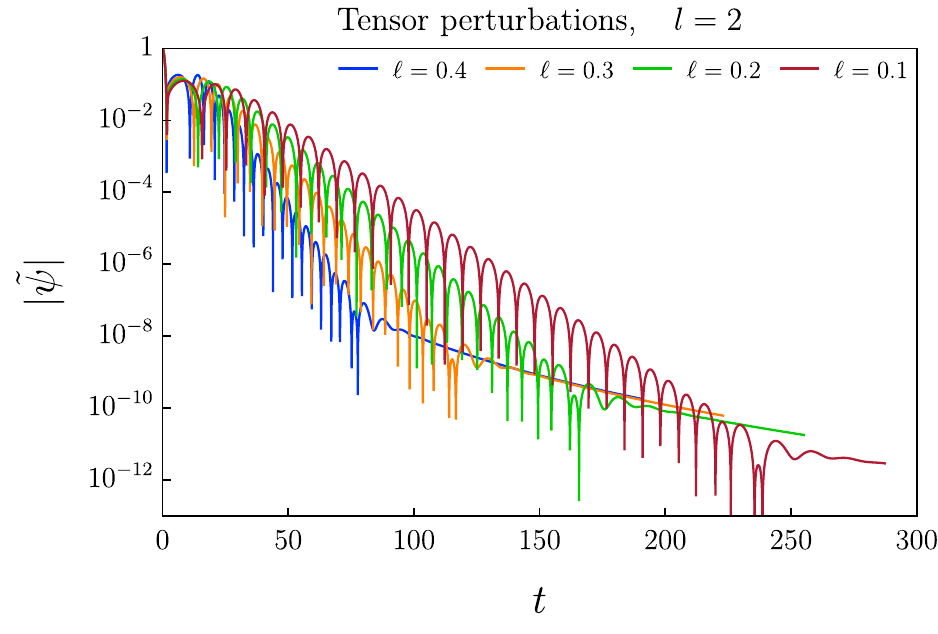}
    \includegraphics[scale=0.55]{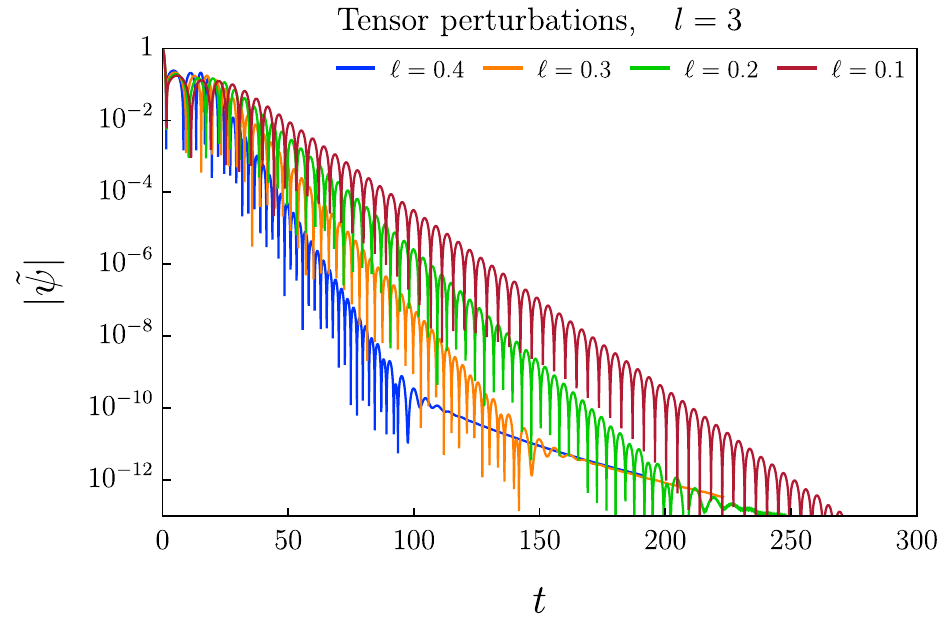}
     \includegraphics[scale=0.55]{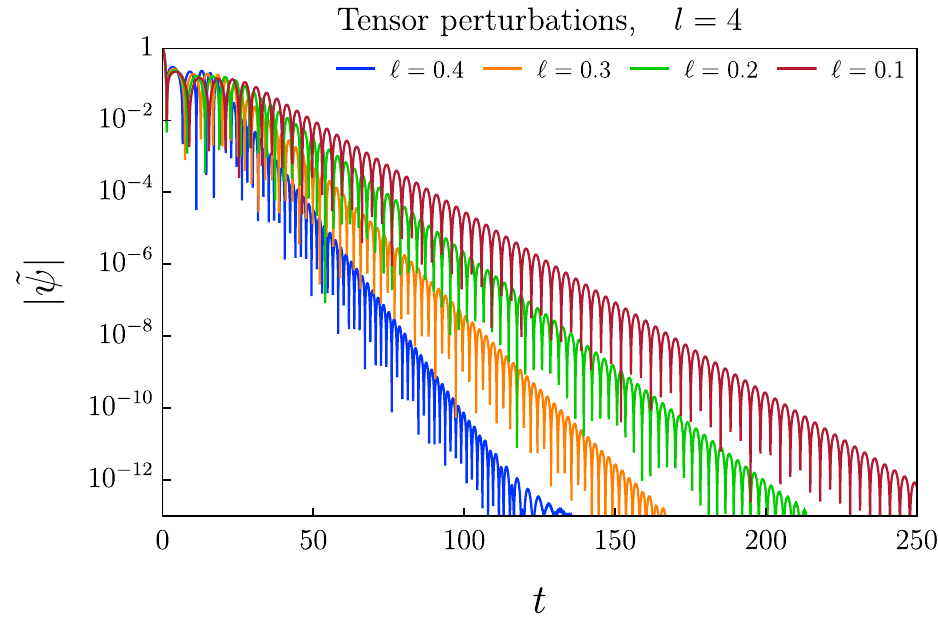}
    \caption{Logarithmic time--domain profiles of tensor perturbations, given by $|\tilde{\psi}|$, for $M=1$, $Q=p=0.1$, and $\ell=0.1,0.2,0.3,0.4$. The panels correspond to the modes $l=2$, $l=3$, and $l=4$ in the upper--left, upper--right, and lower positions, respectively. {The horizontal axis is linear, while the vertical axis is logarithmic.} }
    \label{timedomaintensor3}
\end{figure}

\begin{figure}
    \centering
    \includegraphics[scale=0.55]{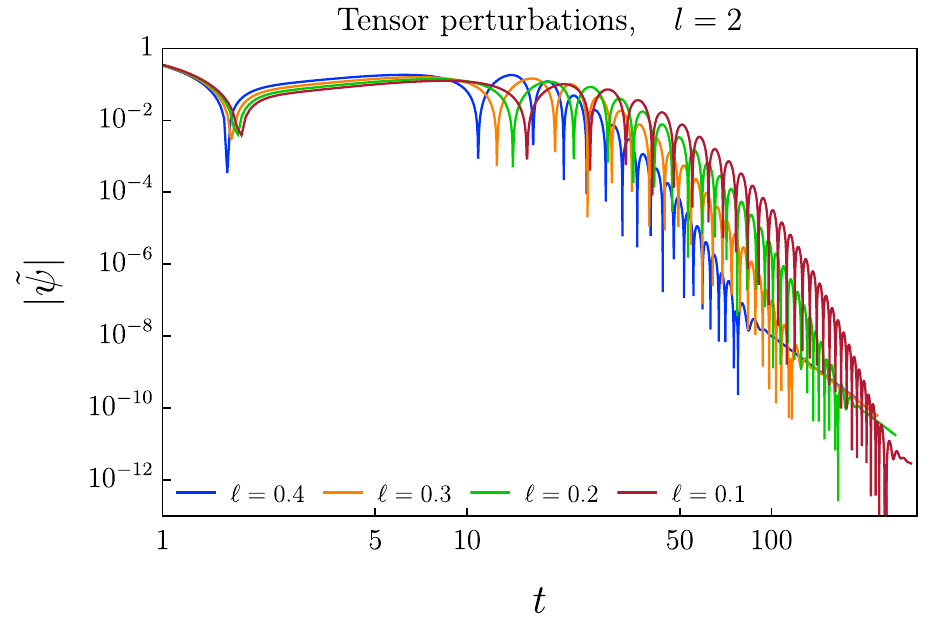}
    \includegraphics[scale=0.55]{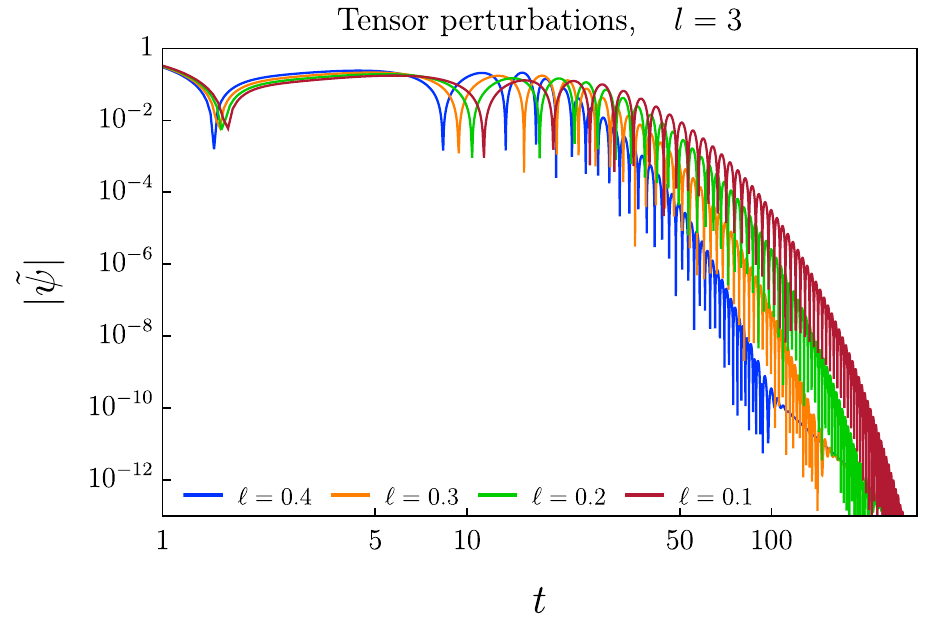}
     \includegraphics[scale=0.55]{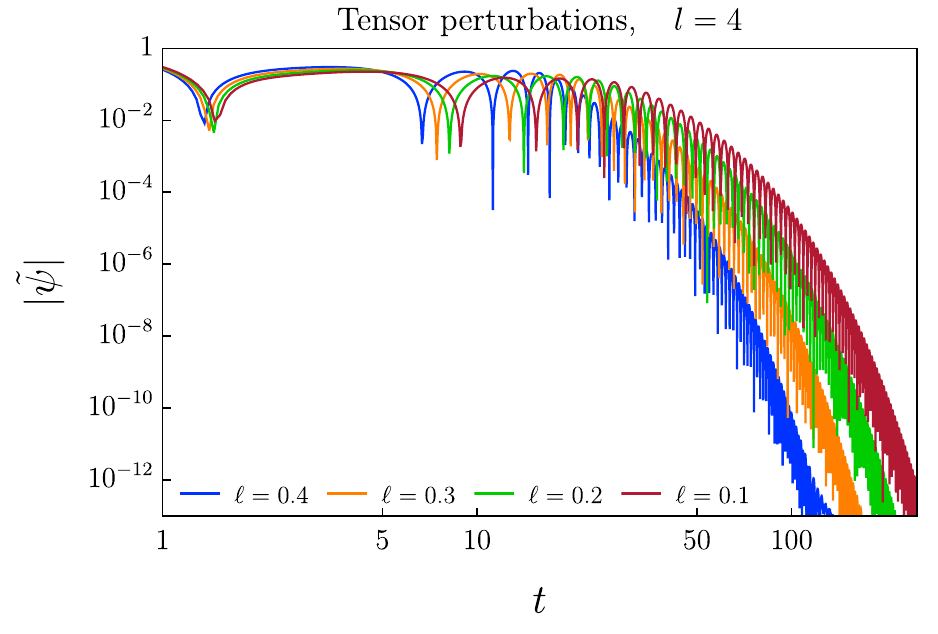}
    \caption{Late--time tensor signal {with both axes on logarithmic scales}, with $|\tilde{\psi}|$ plotted as a function of $ t$ for $M=1$, $Q=p=0.1$, and $\ell=0.1,0.2,0.3,0.4$. The upper--left, upper--right, and lower panels display the modes $l=2$, $l=3$, and $l=4$, respectively. }
    \label{timedomaintensor4}
\end{figure}


\subsection{Spinor perturbations  }

\begin{figure}
    \centering
    \includegraphics[scale=0.55]{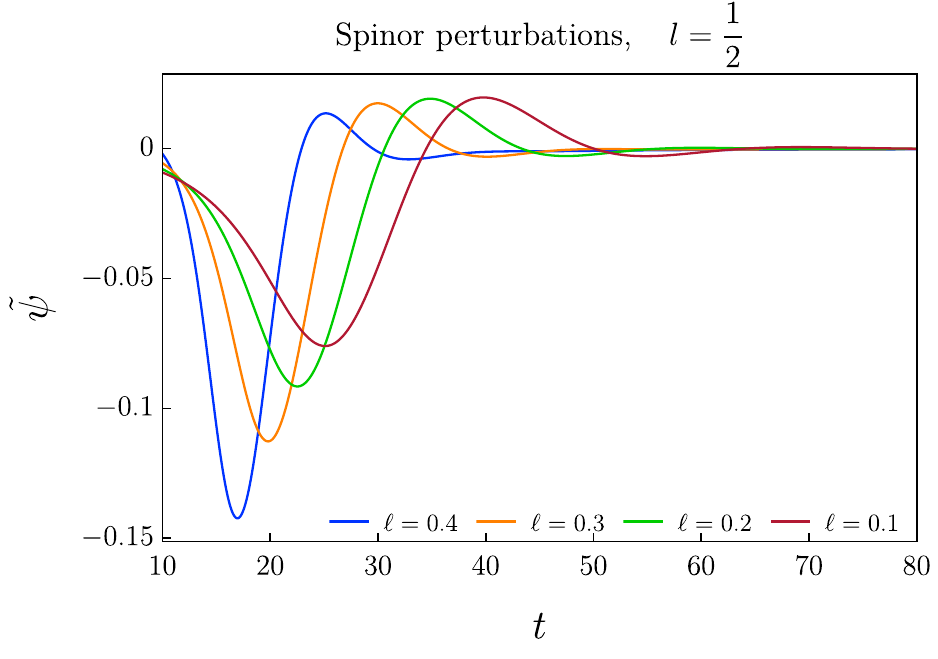}
    \includegraphics[scale=0.55]{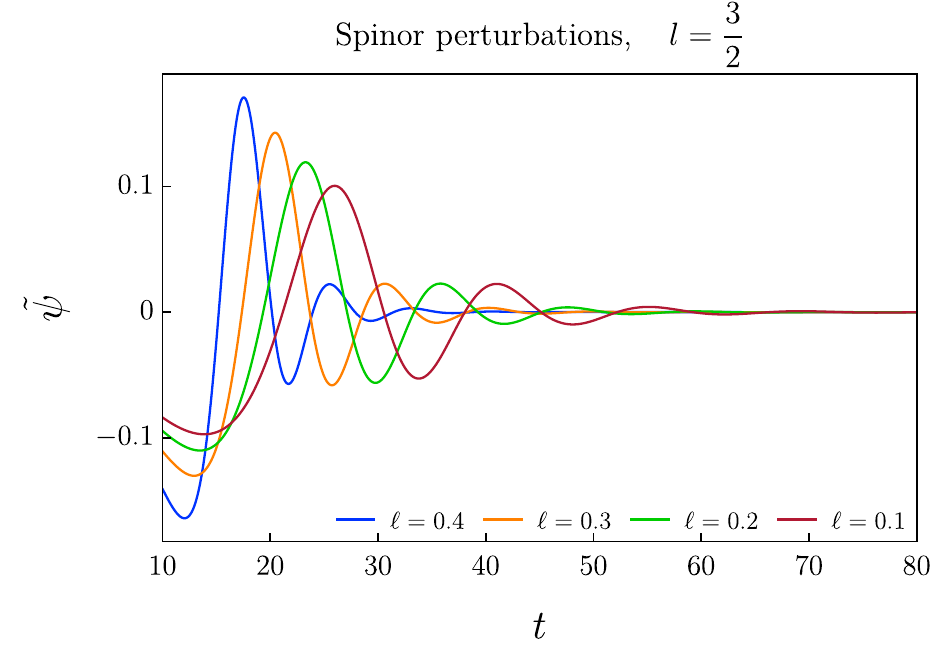}
     \includegraphics[scale=0.55]{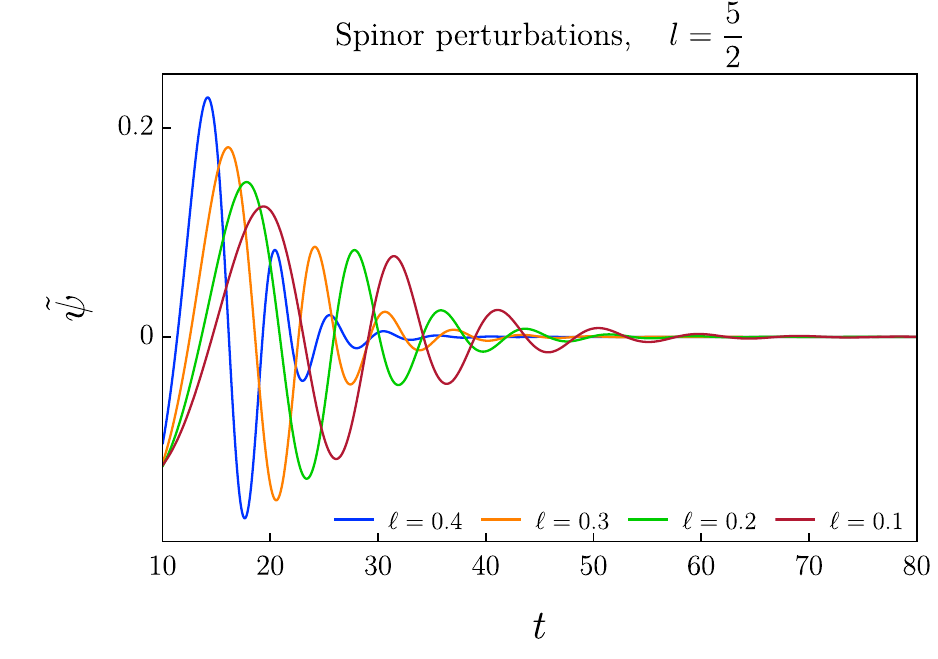}
    \caption{Time--domain spinor waveforms $\tilde{\psi}$ for $M=1$, $Q=p=0.1$, and $\ell=0.1,0.2,0.3,0.4$. The upper--left, upper--right, and lower panels display the modes $l=1/2$, $l=3/2$, and $l=5/2$, respectively. }
    \label{timedomainspinor12}
\end{figure}

Spinor perturbations provide the final part of the time--domain analysis for the spacetime introduced in Eq.~(\ref{metric_ansatz_KR}). The numerical signal $\tilde{\psi}$ is shown in Fig.~\ref{timedomainspinor12} for $M=1$, $Q=p=0.1$, and $\ell=0.1,0.2,0.3,0.4$. The three panels display the half integer modes $l=1/2$, $l=3/2$, and $l=5/2$, respectively. After the initial Gaussian pulse evolves through the grid, the waveform enters a ringdown interval, where the oscillations are progressively suppressed. {The comparison between sectors depends on the angular modes chosen. For the modes and parameters compared in Fig.~\ref{comtimeall}, the spinor mode with $j=5/2$ has a higher fundamental oscillation frequency than the bosonic modes with $l=2$. Within each tabulated spinor multipole, increasing $\ell$ at fixed mass and charges increases the damping rate and shortens the ringdown damping time.}

\begin{figure}
    \centering
    \includegraphics[scale=0.55]{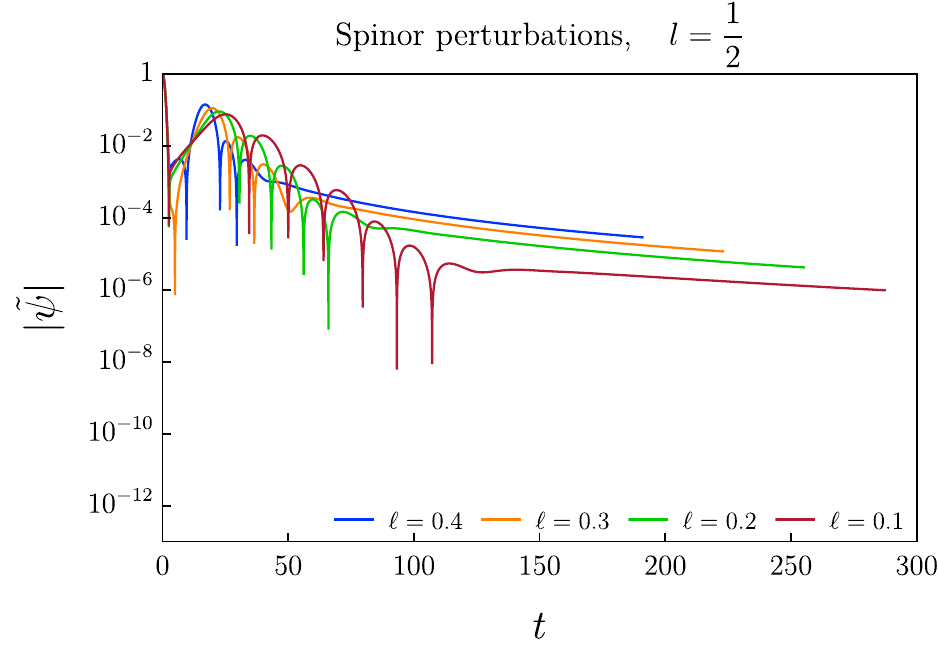}
    \includegraphics[scale=0.55]{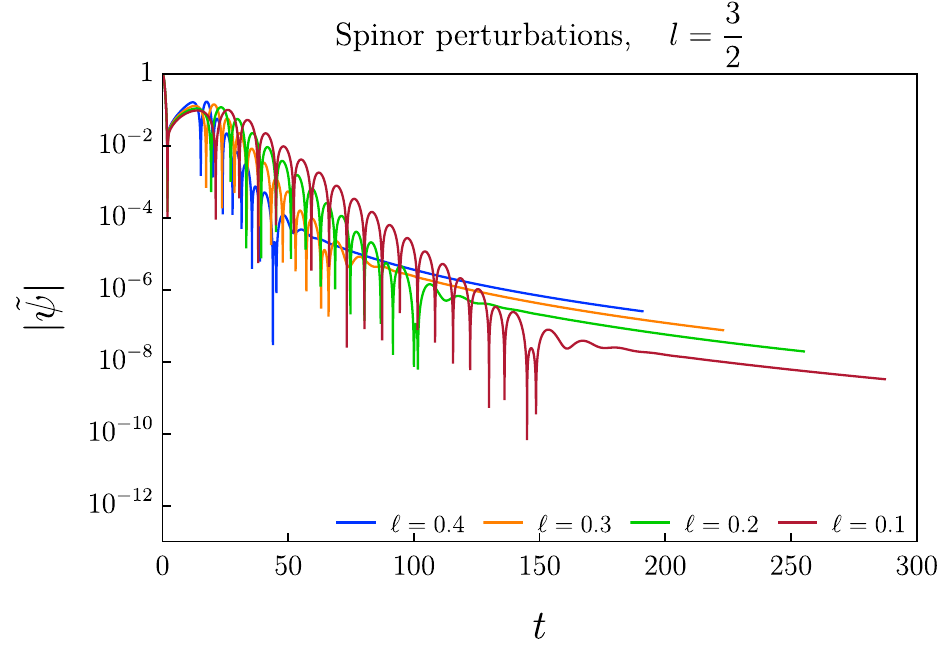}
     \includegraphics[scale=0.55]{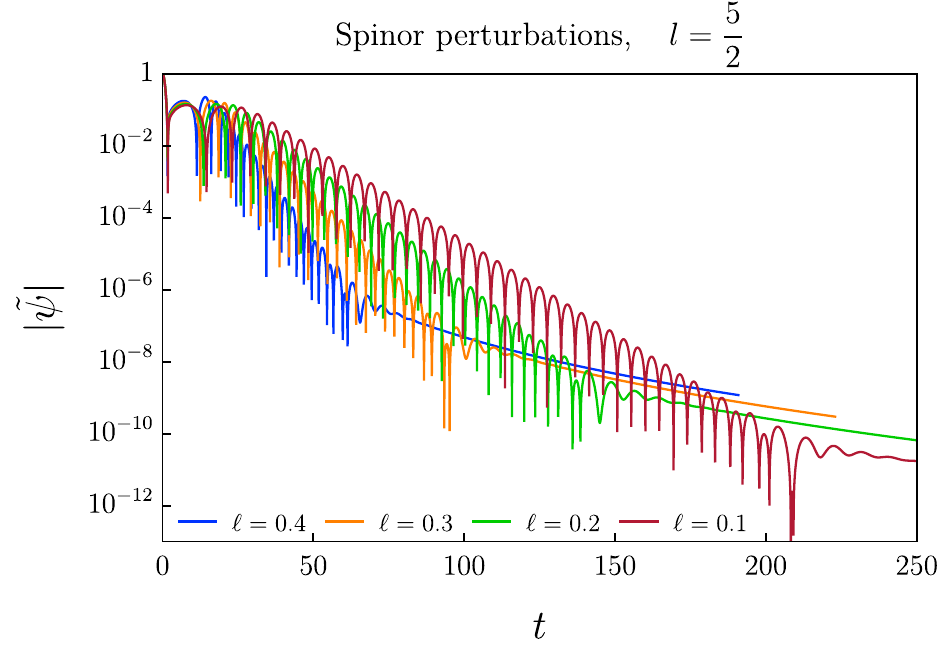}
    \caption{Logarithmic time--domain profiles of the spinor perturbation, represented by $|\tilde{\psi}|$, for $M=1$, $Q=p=0.1$, and $\ell=0.1,0.2,0.3,0.4$. The upper--left, upper--right, and lower panels correspond to the modes $l=1/2$, $l=3/2$, and $l=5/2$, respectively. {The horizontal axis is linear, while the vertical axis is logarithmic.}}
    \label{timedomainspinor32}
\end{figure}

{The decay rate is examined in Fig. \ref{timedomainspinor32} by plotting $|\tilde{\psi}|$ as a function of $t$ with a logarithmic vertical axis.} In this plot, the quasinormal regime appears as an approximately linear portion of each curve, while the subsequent bending indicates the departure from exponential damping and the onset of the tail. {During the quasinormal stage, the oscillation envelope decays more rapidly as $\ell$ increases.}

\begin{figure}
    \centering
    \includegraphics[scale=0.55]{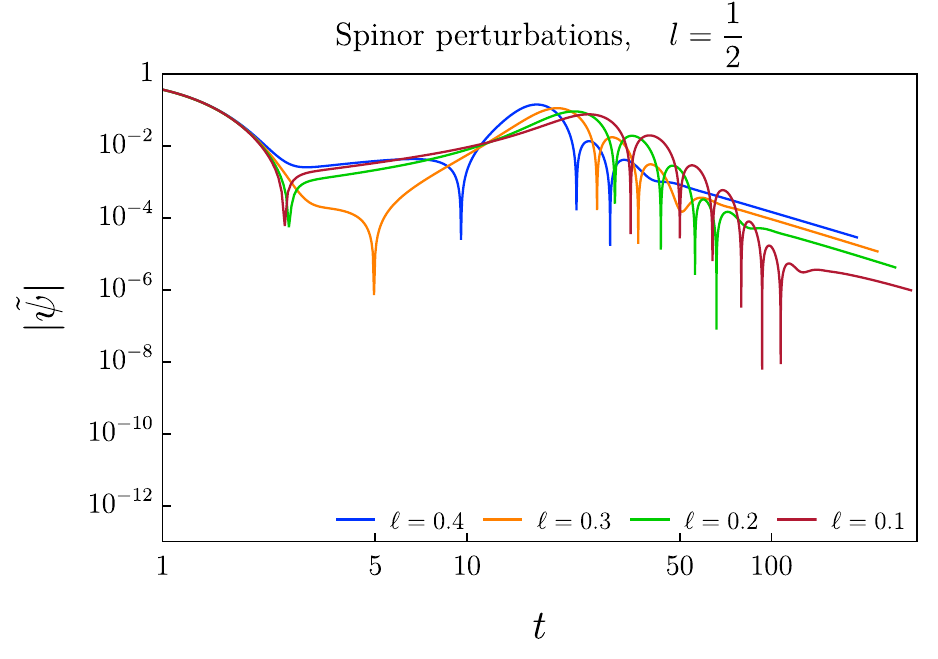}
    \includegraphics[scale=0.55]{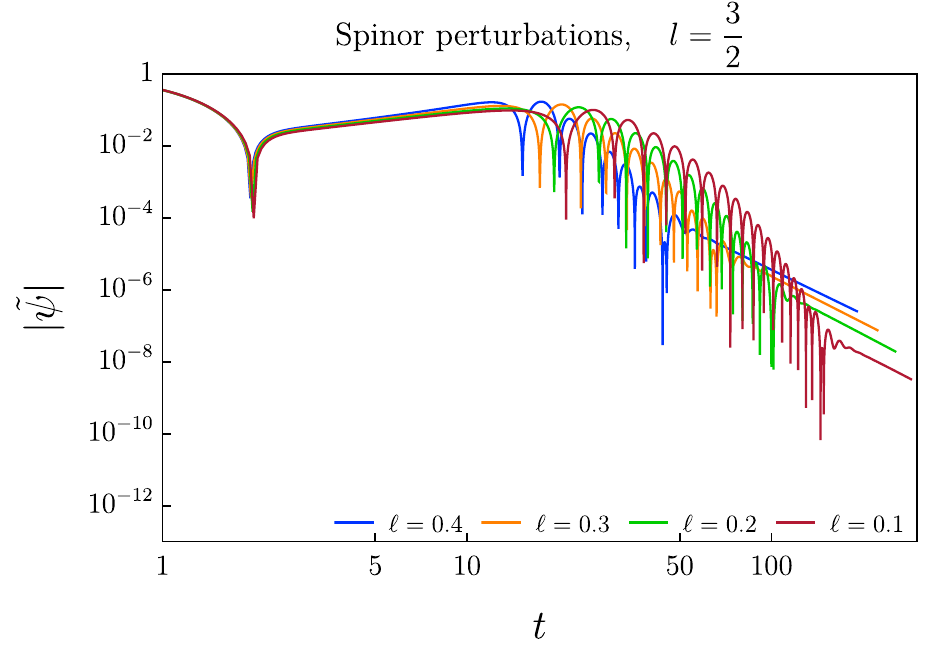}
     \includegraphics[scale=0.55]{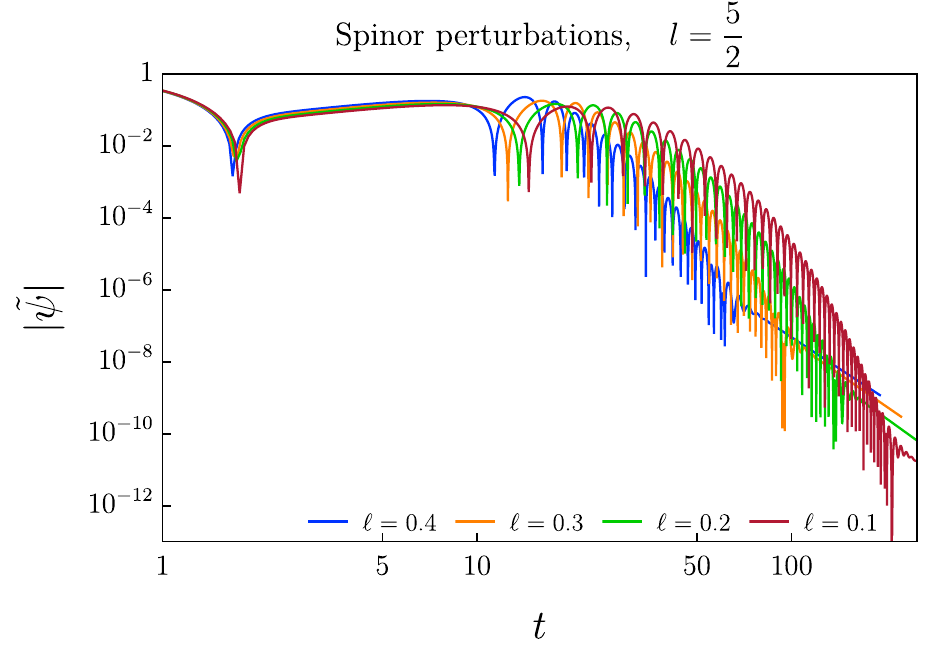}
    \caption{Asymptotic spinor signal shown in a log--log representation, where $|\tilde{\psi}|$ is plotted as a function of $ t$ for $M=1$, $Q=p=0.1$, and $\ell=0.1,0.2,0.3,0.4$. The upper--left, upper--right, and lower panels correspond to the modes $l=1/2$, $l=3/2$, and $l=5/2$, respectively. }
    \label{timedomainspinor52}
\end{figure}

\begin{figure}
    \centering
    \includegraphics[scale=0.55]{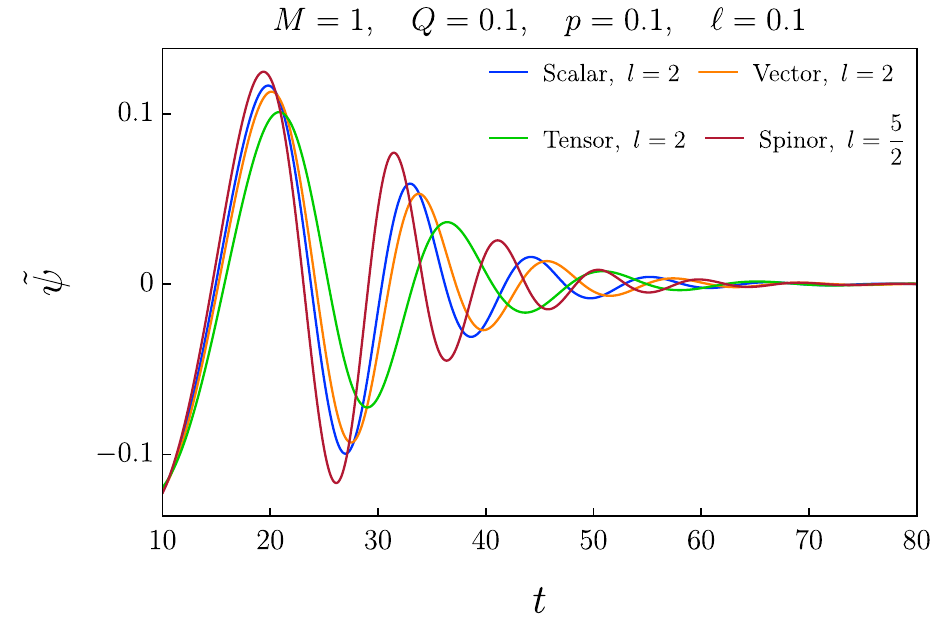}
    \includegraphics[scale=0.55]{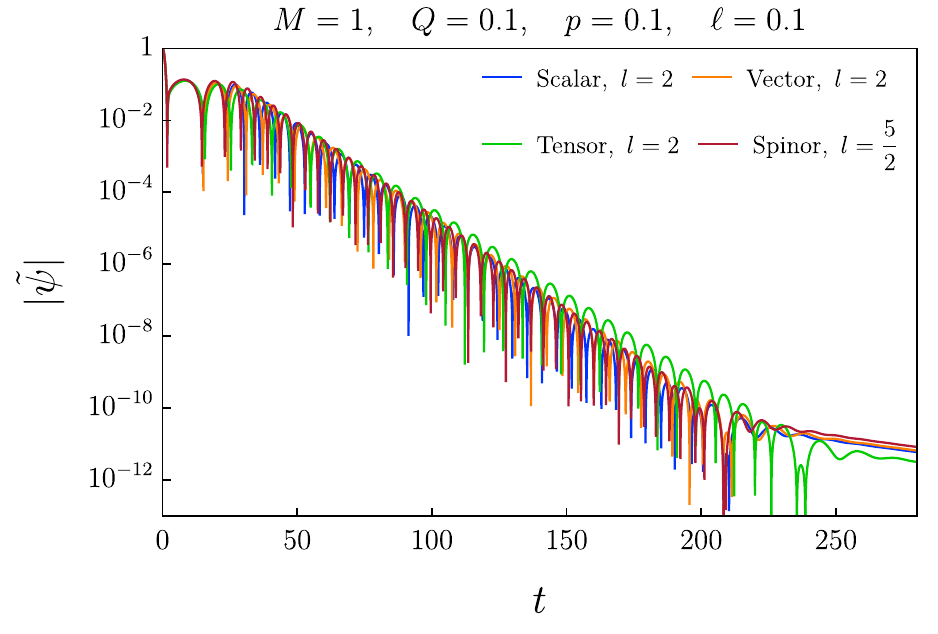}
     \includegraphics[scale=0.55]{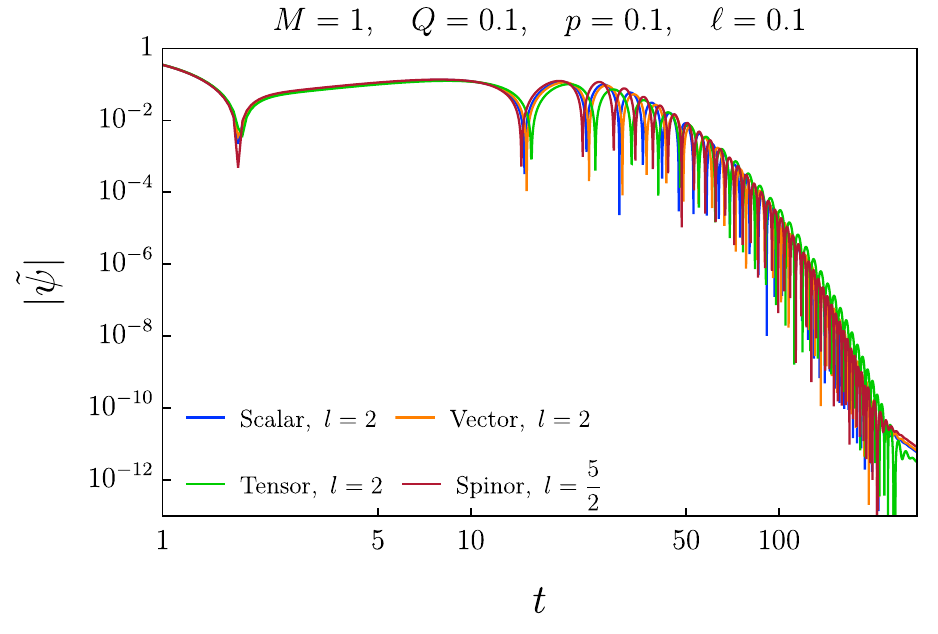}
    \caption{Time--domain comparison among the scalar, vector, tensor, and spinor sectors for $\ell=0.1$, $M=1$, and $Q=p=0.1$.  {{ The upper--left panel shows $\tilde{\psi}$ as a function of $t$ on linear axes. The upper--right and lower panels show $|\tilde{\psi}|$ as a function of $t$, with a logarithmic vertical axis and logarithmic scales on both axes, respectively.}}}
    \label{comtimeall}
\end{figure}

Fig.~\ref{timedomainspinor52} isolates the asymptotic regime by representing the signal on a log--log scale. This form makes the algebraic tail visible after the quasinormal ringing has faded. Although the same general pattern found for bosonic perturbations is recovered, the spinor sector follows its own attenuation scale. 

{Fig.~\ref{comtimeall} compares the four perturbative sectors for $M=1$ and $Q=p=\ell=0.1$, with $l=2$ for the scalar, vector, and tensor fields and $l=j=5/2$, corresponding to $|\kappa_{\mathrm D}|=3$, for the spinor field. The fundamental frequencies in Tables~\ref{qnmsscalarl2}, \ref{qnmsvectorl2}, \ref{qnmstensorl2}, and \ref{qnmsspnior52} give the damping times $\tau=1/|\operatorname{Im}\omega_0|$, with $\tau_{\mathrm t}\simeq9.151$, $\tau_{\mathrm v}\simeq8.524$, $\tau_{\psi}\simeq8.397$, and $\tau_{\mathrm s}\simeq8.353$. Thereby, $\tau_{\mathrm t}>\tau_{\mathrm v}>\tau_{\psi}>\tau_{\mathrm s}$ for these modes during the fundamental mode ringdown. At the latest times displayed in Fig.~\ref{comtimeall}, the spinor curve has the largest amplitude and the tensor curve the smallest. The comparison of late--time amplitudes therefore differs from the ordering of the ringdown damping times.}

{No echo signals were observed in the time--domain evolutions performed for the scalar, vector, effective axial tensor, and spinor sectors. Within the simulated intervals, the waveforms exhibit ringdown followed by late--time decay, without repeated delayed pulses.}


{

\subsection{Late--time tail analysis}
\label{sec:tail_analysis}

For a quantitative tail analysis, we perform a separate evolution with momentarily stationary initial data for the four angular modes compared in Fig.~\ref{comtimeall}, keeping $M=1$ and $Q = p = \ell = 0.1$. The initial data are $\widetilde{\psi}(0,x)  =  \exp[-(x/M)^2]$ and $\partial_t\widetilde{\psi}(0,x) = 0$, where $x=r_*$ is fixed by $x(10M) = 0$. The waveform is extracted at $x=0$, corresponding to $r=10M$. The tensor sector uses the effective axial potential adopted in the preceding sections. This separate evolution is carried out up to $t/M=2000$ using approximately 31--digit arithmetic. Calculations with $\Delta x/M = 0.16,0.08,0.04$ and $\Delta t/\Delta x = 0.4$ give a spatial resolution check. Additional runs halve the Courant factor and move the boundaries from $x/M=\pm1050$ to $x/M = \pm 1250$.

Fig.~\ref{fig:tail_exponents}(a) displays the absolute waveforms on logarithmic axes. Their nearly linear profiles indicate an approximate power law decay over the displayed interval. To measure the remaining variation of the slope, we define the local decay exponent as
\begin{equation} \alpha_{\mathrm{eff}}(t)=-\frac{\mathrm{d}\ln|\widetilde{\psi}(t,0)|}{\mathrm{d}\ln(t/M)}. \label{eq:local_tail_exponent} \end{equation}
Fig.~\ref{fig:tail_exponents}(b) shows that the local exponents continue to increase slowly throughout the fitting interval. We account for this finite time behavior by fitting
\begin{equation} \ln|\widetilde{\psi}(t,0)|=c_0-\alpha\ln u+\frac{b_1}{u}+\frac{b_2}{u^2}+\frac{b_3}{u^3}, \qquad u=\frac{t}{1000M}. \label{eq:corrected_tail_fit} \end{equation}
The coefficient $\alpha$ estimates the asymptotic power law exponent. For this fitted model, the local exponent is $\alpha_{\mathrm{eff}}(t)=\alpha+b_1/u+2b_2/u^2+3b_3/u^3$, which approaches $\alpha$ as the inverse time corrections decrease. The reference fit uses $1000\leq t/M\leq2000$ and 501 samples distributed uniformly in $\ln t$. All waveform samples in this interval have the same sign. The open circles in Fig.~\ref{fig:tail_exponents}(a) show the fitted profiles, and the shaded region marks the fitting interval. The fits closely follow the numerical waveforms in all four sectors. Fitting a pure power law over the same interval gives the finite window slopes $6.689$, $6.689$, $6.730$, and $6.643$ for the scalar, vector, tensor, and spinor sectors, respectively.

The sensitivity to extrapolation is assessed using lower fitting endpoints $t_1/M = 700,1000,1200$ and upper endpoints $t_2/M = 1800,2000$. Besides the cubic inverse time correction in Eq.~\eqref{eq:corrected_tail_fit}, we test a quadratic correction and alternatives containing $u^{-1}\ln u$ and $u^{-2}\ln u$. The uncertainty estimate combines the largest change across these models and intervals with the changes under spatial refinement, time step refinement, and boundary displacement. This envelope is enlarged by $25\%$ and rounded upward to $10^{-3}$. The quoted uncertainties quantify numerical and extrapolation sensitivity. The largest change in the fitted exponent under spatial refinement is below $1.5\times10^{-7}$; the correction model and fitting interval dominate the uncertainty. Table~\ref{tab:tail_exponents} lists the extrapolated exponents and their estimated uncertainties.

\begin{table}[ht]
\centering
\caption{Extrapolated late--time tail exponents for $M=1$ and $Q=p=\ell=0.1$, extracted at $r=10M$ from the initially stationary Gaussian pulse. Fits include the finite time corrections in Eq.~\eqref{eq:corrected_tail_fit}. Uncertainties account for variations of the fitting interval, correction model, spatial resolution, time step, and boundary location.}
\label{tab:tail_exponents}
\begin{tabular}{lccccc}
\hline
Sector & Angular mode & $\ell$ & $[t_1/M,t_2/M]$ & $\alpha$ & Uncertainty \\
\hline
Scalar & $l=2$ & $0.1$ & $[1000,2000]$ & $6.754$ & $0.003$ \\
Vector & $l=2$ & $0.1$ & $[1000,2000]$ & $6.754$ & $0.003$ \\
Tensor & $l=2$ & $0.1$ & $[1000,2000]$ & $6.838$ & $0.011$ \\
Spinor & $j=5/2$, $|\kappa_{\mathrm D}|=3$ & $0.1$ & $[1000,2000]$ & $6.692$ & $0.002$ \\
\hline
\end{tabular}
\end{table}

\begin{figure}[ht]
\centering
\includegraphics[scale=0.46]{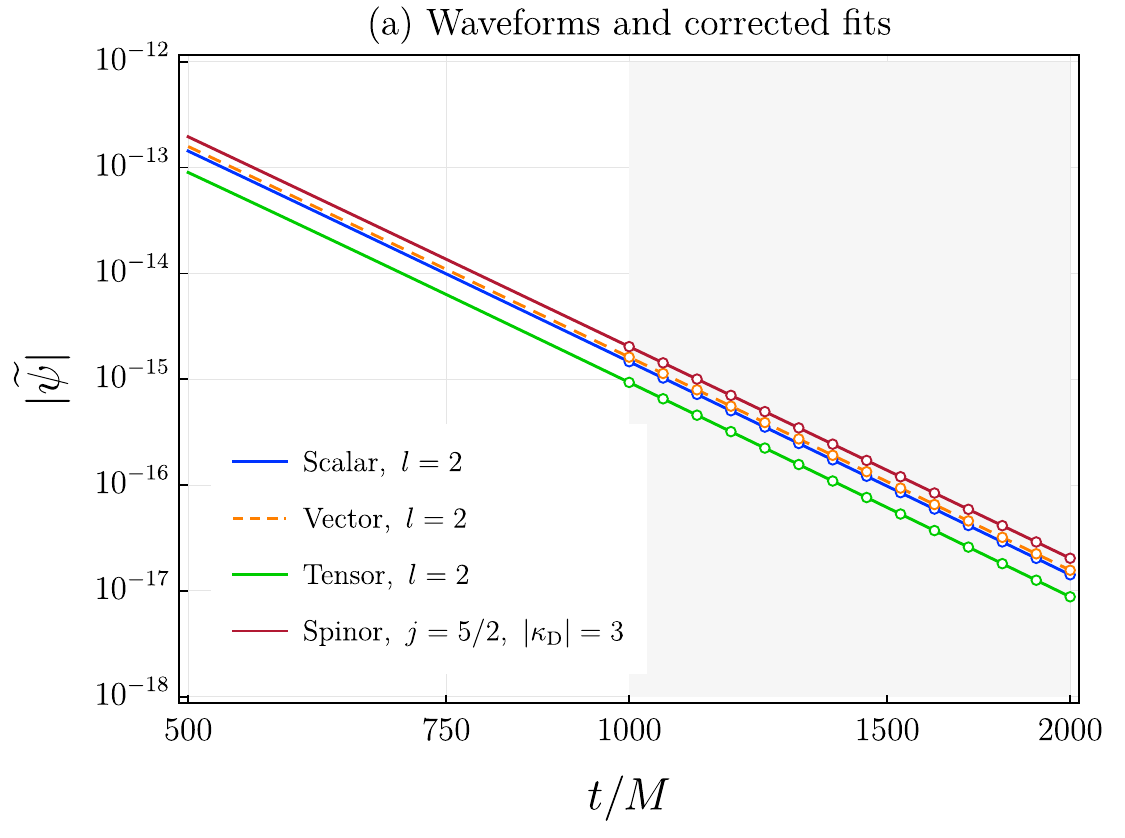}
\includegraphics[scale=0.46]{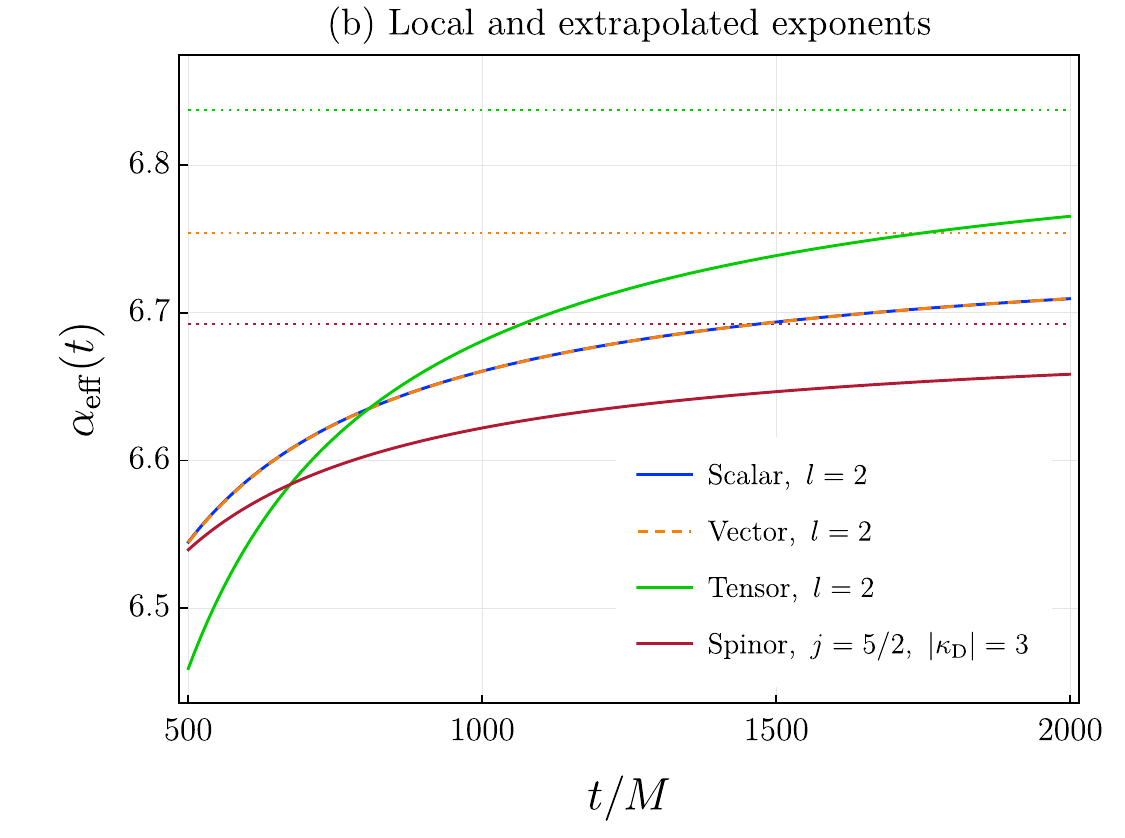}
\caption{Late--time waveforms and decay exponents for $M=1$ and $Q = p = \ell = 0.1$, extracted at $r = 10M$, with $l = 2$ in the scalar, vector, and effective tensor sectors and $j=5/2$, $|\kappa_{\mathrm D}|=3$ in the spinor sector. (a) Absolute numerical waveforms on logarithmic axes. The open circles show the fitted profiles from Eq.~\eqref{eq:corrected_tail_fit}, and the shaded region marks $1000\leq t/M\leq2000$. (b) Local decay exponents defined by Eq.~\eqref{eq:local_tail_exponent}. The horizontal dotted lines indicate the extrapolated exponents listed in Table~\ref{tab:tail_exponents}. The scalar and vector local exponents nearly overlap.}
\label{fig:tail_exponents}
\end{figure}

The horizontal dotted lines in Fig.~\ref{fig:tail_exponents}(b) show the extrapolated values of $\alpha$. The local exponents remain below these values at $t/M = 2000$, with the largest separation occurring in the tensor sector. This separation accounts for the difference between the pure power law slopes and the corrected estimates in Table~\ref{tab:tail_exponents}. Over the fitting interval, the tensor local exponent is the largest and the spinor local exponent is the smallest. The extrapolated exponents preserve this relation: the tensor tail has the steepest asymptotic decay, while the spinor tail has the shallowest decay for the parameters and angular modes considered here.

The scalar and vector local exponents nearly coincide throughout Fig.~\ref{fig:tail_exponents}(b), and their extrapolated values agree within the estimated uncertainties. Their leading large distance potentials have the same coefficient,
\begin{equation} \lim_{x\to\infty}x^2V_{\mathrm s}(x) = \lim_{x\to\infty}x^2V_{\mathrm v}(x) = (1 - \ell)l(l+1). \label{eq:scalar_vector_tail_coefficient} \end{equation}

{ The fitted exponents admit a closed form interpretation. Defining $c=\lim_{x\to\infty}x^2V(x)=\nu(\nu+1)$, with $\nu=(\sqrt{1+4c}-1)/2$, the inverse square contribution to the retarded Green function behaves as $G_{\mathrm{tail}}\propto t^{-(2\nu+2)}$ for noninteger $\nu$ \cite{KRChing1995}. For the localized, momentarily stationary initial data used here, the evolution involves $\partial_tG_{\mathrm{tail}}$, giving, at fixed $x$, \[ |\tilde{\psi}(t,x)|\propto (t/M)^{-\alpha}, \qquad \alpha=2\nu+3=2+\sqrt{1+4c}. \] 

Besides the common scalar and vector coefficient in Eq.~(\ref{eq:scalar_vector_tail_coefficient}), the positive spinor partner and the effective axial tensor potential give \[ c_\psi=(1-\ell)k^2-k\sqrt{1-\ell}, \qquad c_t=(1-\ell)l(l+1)+2\ell. \] For $\ell=0.1$, $l=2$, and $k=3$, these coefficients yield $\alpha_s=\alpha_v=6.753946$, $\alpha_\psi=6.692100$, and $\alpha_t=6.837355$, in agreement with Table~XVI within the quoted uncertainties. The common scalar and vector exponent follows from their identical leading coefficient, while the larger tensor coefficient accounts for the steepest decay among the modes considered.

In the Schwarzschild limit $\ell=Q=p=0$, the scalar index becomes $\nu=l$, and the inverse square tail contribution vanishes because its amplitude is proportional to $\sin(\pi\nu)$. The subleading curvature terms then give $t^{-(2l+4)}$ for momentarily stationary scalar data \cite{KRPriceBurko2004}, consistent with the numerical value $\alpha=8.000037$ obtained for $l=2$. }

}


\section{Conclusion }\label{Sec:Conclusion}

In this work, we investigated the perturbative dynamics, tidal effects, time delay, and relativistic frequency shifts associated with a dyonic Kalb--Ramond black hole. The background geometry was controlled by the mass $M$, the electric charge $Q$, the magnetic charge $p$, and the Lorentz--violating parameter $\ell$, while the electric and magnetic sectors entered the main physical quantities through the effective combination $P_{\ell}^{2}=\frac{Q^{2}}{(1-\ell)^{2}}+\frac{p^{2}}{1-2\ell}$.
This structure showed that the Lorentz--violating parameter weighted the electric and magnetic charges differently, thereby modifying the horizon structure, the free-fall dynamics, the tidal response, and the perturbative spectrum. In the limit $\ell\to0$, the dyonic Reissner--Nordstr\"om result was recovered, whereas the purely electric and uncharged Kalb--Ramond configurations were obtained through the corresponding charge reductions.

We first analyzed the gravitational Doppler effect for radial signal exchange between freely falling and static observers. For a source released from the asymptotic region, the exterior velocity approached the speed of light at the event horizon, and the frequency ratio measured by a static observer tended to zero for signals emitted outward by the infalling source. Conversely, signals emitted by the static observer and received by the freely falling observer yielded a finite horizon value equal to $1/2$. {Inside the two horizon region, the frequency ratio attained its minimum at $r_m=\mathcal{P}_\ell^2/M$ and then increased along the inward trajectory, corresponding to a weakening redshift while remaining below unity.} The numerical analysis showed that increasing the dyonic charge shifted the frequency ratio upward and weakened the corresponding gravitational Doppler redshift.

We then computed the tidal forces experienced by a neutral body in radial free fall. The radial component contained the usual Schwarzschild stretching term together with a dyonic contribution that acted in the opposite direction near the inner region. As a result, the radial tidal force changed sign at $R_{0}^{\rm rad}=\frac{3P_{\ell}^{2}}{2M}$,
whereas its extremum occurred at $R_{\rm max}^{\rm rad}=\frac{2P_{\ell}^{2}}{M}$.
The angular tidal force also changed sign, with its zero located at $R_{0}^{\rm ang}=\frac{P_{\ell}^{2}}{M}$, and its minimum located at $R_{\rm min}^{\rm ang}=\frac{4P_{\ell}^{2}}{3M}$.
Together with the turning radius $R_{\rm stop}=P_{\ell}^{2}/(2M)$, these quantities satisfied the ordering $R_{\rm stop}<R_{0}^{\rm ang}<R_{\rm min}^{\rm ang}<R_{0}^{\rm rad}<R_{\rm max}^{\rm rad}$. This hierarchy indicated that the radial tidal component reversed its sign before the angular one during inward motion. The charge-sector plots showed that increasing either $Q/M$ or $p/M$ shifted all characteristic tidal radii toward larger values, while the outer and inner horizons approached each other until the extremal limit was reached.

The gravitational time delay was also evaluated for null trajectories in the same geometry. After subtracting the flat-space contribution, the resulting delay depended on the source position, the observer position, the closest-approach radius, and the dyonic parameters. The shifted time delay showed that the electric and magnetic charges reduced the delay relative to the reference configuration used in the numerical analysis. The Lorentz--violating parameter modified this behavior through the same effective dyonic combination that had appeared in the horizon, Doppler, and tidal sectors.

{The perturbative analysis treated massless scalar, vector, and spinor test fields and an effective axial metric model. The axial results depend on the source closure specified in Appendix~A. Over the tabulated parameter ranges, increasing $\ell$ at fixed mass and charges increased both the oscillation frequencies and damping rates, shortening the ringdown damping times in the original time coordinate. These features combined changes in time normalization, effective parameters, and the residual angular deformation as well.

Independent AIM calculations showed close agreement with WKB for the tested higher multipole fundamental modes. However, the scalar $l=0$, $n=2$ mode exhibited a substantial discrepancy, which demonstrated that the accuracy established for selected modes cannot be extended to the entire tabulated spectrum.

A separate evolution with initially stationary Gaussian data quantified the late--time tails. For $M=1$, $Q=p=\ell=0.1$, bosonic $l=2$, and spinor $j=5/2$, the extrapolated exponents were $6.754\pm0.003$, $6.754\pm0.003$, $6.838\pm0.011$, and $6.692\pm0.002$ for the scalar, vector, effective tensor, and spinor sectors, respectively. These estimates included finite time corrections and uncertainties from numerical and extrapolation sensitivity. For this comparison, the effective tensor mode had the longest fundamental mode ringdown damping time but the steepest extrapolated tail decay. The Schwarzschild scalar validation recovered the exponent appropriate to initially stationary data. No echo signals were observed in the time--domain evolutions examined.}

As a further extension, we may examine the gravitational lensing sector by applying the Gauss--Bonnet method in the weak--deflection regime and Tsukamoto's formalism in the strong--deflection regime. Additional developments may include the calculation of greybody factors and greybody bounds, together with particle creation through the quantum tunneling method for bosonic and fermionic modes. Neutrino oscillations and matter accretion also provide natural directions for broadening the physical analysis of this spacetime. These studies have already been carried out, and the corresponding manuscripts are now in their final stage; they are expected to appear on arXiv shortly.


\section*{Acknowledgments}
\hspace{0.5cm} A.A.A.F. is supported by Conselho Nacional de Desenvolvimento Cient\'{\i}fico e Tecnol\'{o}gico (CNPq) with project number.


\appendix

{

\section{Derivation of the effective axial perturbation potential}
\label{dervitensor}

We derive the potential in Eq.~(\ref{tententen}) within an effective source description of the axial metric perturbations. The background is represented as a solution of the Einstein equations with an anisotropic source, following the approach employed in Refs.~\cite{AraujoFilho2025NLED,ChenChen2019}. To obtain a closed equation for the metric perturbation, we neglect the independent odd--parity response of this effective source, as specified below. It is important to mention that this prescription is an additional assumption about the perturbations; the background solution of the electromagnetic--Kalb--Ramond theory alone does not imply it.

Consider the static line element
\begin{equation}
\mathrm{d}\bar s^{2} = - A(r)\,\mathrm{d}t^{2} + \frac{\mathrm{d}r^{2}}{B(r)}  + r^{2}\gamma_{ij}\,\mathrm{d}x^{i}\mathrm{d}x^{j},
\label{eq:ax_background}
\end{equation}
where $\gamma_{ij}$ is the metric of the unit two sphere and $A,B > 0$ in the exterior region. We use $a,b\in\{t,r\}$ and $i,j\in\{\theta,\varphi\}$ for the radial--temporal and angular indices, respectively. Writing the background equations as $\bar G_{\mu\nu} = \kappa\bar T^{\mathrm{eff}}_{\mu\nu}$, with $\kappa = 8\pi$ in geometrized units, spherical symmetry gives
\begin{equation}
\bar T^{\mathrm{eff}\,\mu}{}_{\nu}  = \operatorname{diag}(- \rho,p_r,p_\perp,p_\perp).
\label{eq:ax_source}
\end{equation}
The temporal and radial equations determine
\begin{equation}
\kappa\rho  =  \frac{1-B-rB'}{r^{2}},  \qquad  \kappa p_r = \frac{B-1}{r^{2}}+\frac{BA'}{Ar},
\label{eq:ax_background_stresses}
\end{equation}
where a prime denotes differentiation with respect to $r$. These quantities encode the complete background source, including the nonminimal curvature contribution; they are not identified with the electromagnetic stress tensor alone.

For a radiative multipole $l\geq2$, introduce the axial vector harmonic
\begin{equation}
S_i^{lm} = \epsilon_i{}^{j}\widehat D_jY_{lm},  \qquad  \widehat D^{i}S_i^{lm} = 0,  \qquad  \widehat D^{j}\widehat D_jS_i^{lm}  = [1 - l( l + 1)]S_i^{lm},
\label{eq:ax_harmonics}
\end{equation}
where $\widehat D_i$ and $\epsilon_{ij}$ are the covariant derivative and area form associated with $\gamma_{ij}$. Here, let us remember that the multipole index is represented by $l$, while the Lorentz--violating parameter $\ell$. In the Regge--Wheeler gauge~\cite{ReggeWheeler1957,moncrief1974stability}, the nonvanishing metric perturbations can be written as
\begin{equation}
h_{ti} = h_0(r)S_i^{lm}e^{-i\omega t},  \qquad  h_{ri} = h_1(r)S_i^{lm}e^{-i\omega t},  \qquad h_{ab} = h_{ij} = 0.
\label{eq:ax_metric_perturbation}
\end{equation}
The odd--parity source amplitudes are defined by
\begin{equation}
\delta T^{\mathrm{eff}}_{ai} = \tau_a(r)S_i^{lm}e^{-i\omega t},  \qquad  \delta T^{\mathrm{eff}}_{ij}  = \tau_2(r)\widehat D_{(i}S_{j)}^{lm}e^{-i\omega t}.  
\label{eq:ax_matter_perturbation}
\end{equation}
The source free axial closure is
\begin{equation}
\tau_t - p_\perp h_0 = 0,  \qquad  \tau_r-p_\perp h_1 = 0,  \qquad  \tau_2 = 0.
\label{eq:ax_closure}
\end{equation}

The ansatz in Eq.~(\ref{Bmunu_ansatz_KR}) specifies the background Kalb--Ramond field. To distinguish this background from its perturbations, we write
\begin{equation}
g_{\mu\nu}=\bar g_{\mu\nu}+h_{\mu\nu}, \qquad A_\mu=\bar A_\mu+\delta A_\mu, \qquad B_{\mu\nu}=\bar B_{\mu\nu}+\delta B_{\mu\nu}.
\end{equation}
The background has only a temporal--radial Kalb--Ramond component and satisfies $\bar H_{\mu\nu\rho}=0$. These properties do not impose $\delta B_{\mu\nu}=0$. In particular, the perturbed field strength is
\begin{equation}
\delta H_{\mu\nu\rho}=\partial_\mu\delta B_{\nu\rho}+\partial_\nu\delta B_{\rho\mu}+\partial_\rho\delta B_{\mu\nu},
\end{equation}
which can be nonzero. Preserving the background norm would impose the scalar condition $\delta(B_{\mu\nu}B^{\mu\nu})=0$, without eliminating all two form perturbations.

In other words, the prescription in Eq.~(\ref{eq:ax_closure}) is imposed directly on the effective axial source. We do not solve the independent equations for $\delta A_\mu$ and $\delta B_{\mu\nu}$ in this calculation. Setting these field perturbations to zero would require a separate check in the linearized electromagnetic and Kalb--Ramond equations. In particular, the variation of the nonminimal curvature coupling contains terms involving $\bar B_{\mu\nu}$ and the perturbed curvature even when $\delta B_{\mu\nu}=0$. Eq.~(\ref{Bmunu_ansatz_KR}) supplies no dynamical justification for suppressing the two-form perturbations.

Furthermore, the combinations in Eq.~\eqref{eq:ax_closure} are the independent axial source variables in this gauge~\cite{GerlachSengupta1979,Karlovini2002}. In other words, the closure retains the algebraic variation $\delta T^{\mathrm{eff}}_{ai}=p_\perp h_{ai}$ required by the perturbed geometry. Setting all coordinate components of $\delta T^{\mathrm{eff}}_{\mu\nu}$ to zero would be a different prescription.

To evaluate the linearized equations, we use
\begin{align}
\delta\Gamma^{\alpha}_{\mu\nu}  & = \frac{1}{2}\bar g^{\alpha\beta}  \left(\bar\nabla_\mu h_{\beta\nu}  + \bar\nabla_\nu h_{\beta\mu}  - \bar\nabla_\beta h_{\mu\nu}\right),  \label{eq:ax_connection_variation}\\  
\delta R_{\mu\nu}  & = \bar\nabla_\alpha\delta\Gamma^\alpha_{\mu\nu}   - \bar\nabla_\nu\delta\Gamma^\alpha_{\mu\alpha}.
\label{eq:ax_ricci_variation}
\end{align}
The axial perturbation has vanishing trace and $\delta R=0$. Using the background angular equation, the mixed components simply reduce to
\begin{equation}
\delta G_{ai} - \kappa p_\perp h_{ai}  = \delta R_{ai}  -  \left(\frac{\bar R}{2}+\kappa p_\perp\right)h_{ai}  = \delta R_{ai} - \bar R^{\theta}{}_{\theta}h_{ai}.
\label{eq:ax_mixed_equation}
\end{equation}
Substituting Eq.~\eqref{eq:ax_metric_perturbation}, imposing Eq.~\eqref{eq:ax_closure}, and extracting the coefficients of $\widehat D_{(i}S_{j)}^{lm}$ and $S_i^{lm}$ in the $ij$ and $ri$ equations, respectively, gives
\begin{align}
i\omega h_0 + C(Ch_1)'& = 0,  \label{eq:ax_constraint}\\  
i\omega\left(h_0' - \frac{2h_0}{r}\right)  +\left(\frac{\lambda_l A}{r^2} - \omega^2\right)h_1&=0,  
\label{eq:ax_dynamical}
\end{align}
where $C(r)=\sqrt{A(r)B(r)}$ and $\lambda_l=(l-1)(l+2)$. For the nonstationary modes considered here, the remaining $ti$ equation follows from these relations and the linearized conservation identity.

Let us define the master amplitude and tortoise coordinate by
\begin{equation}
\Psi(r) = \frac{C(r)}{r} h_1(r), \qquad  \frac{\mathrm{d}r_*}{\mathrm{d}r} = \frac{1}{C(r)}.
\label{eq:ax_master_definition}
\end{equation}
For $\omega\neq0$, Eq.~\eqref{eq:ax_constraint} gives
\begin{equation}
h_0  =  \frac{iC}{\omega}(r\Psi)'.
\label{eq:ax_h0_elimination}
\end{equation}
Substituting this expression and $h_1=r\Psi/C$ into Eq.~\eqref{eq:ax_dynamical} gives
\begin{equation}
C^{2}\Psi'' + CC'\Psi' +\left[\omega^{2} -\frac{A\lambda_l}{r^{2}} - \frac{2C^{2}}{r^{2}}+\frac{CC'}{r}\right]\Psi = 0.
\label{eq:ax_radial_master}
\end{equation}
Since $\mathrm{d}^{2}\Psi/\mathrm{d}r_*^{2} = C^{2}\Psi''+CC'\Psi'$, the radial equation takes the form
\begin{equation}
\frac{\mathrm{d}^{2}\Psi}{\mathrm{d}r_*^{2}} + 
\left[\omega^{2}-V_t(r)\right]\Psi = 0,
\label{eq:ax_schrodinger}
\end{equation}
with
\begin{align}
V_t(r)  & = \frac{A}{r^{2}}\left[l(l+1) + 2(B-1)\right]  - \frac{(AB)'}{2r}  \label{eq:ax_potential_compact}\\  & = A\left[  \frac{l(l+1)}{r^{2}} + \frac{2(B-1)}{r^{2}}  -\frac{1}{r\sqrt{AB^{-1}}}\frac{\mathrm{d}\sqrt{AB}}{\mathrm{d}r}  \right].
\label{eq:ax_potential_general}
\end{align}
This is Eq.~(\ref{tententen}). Its dependence on the background source can also be displayed by introducing $m(r) = r[1-B(r)]/2$ and using Eq.~\eqref{eq:ax_background_stresses}:
\begin{equation}
V_t(r) = A(r) \left[  \frac{l(l+1)}{r^{2}}-\frac{6m(r)}{r^{3}}  + \frac{\kappa}{2}(\rho-p_r)  \right],
\label{eq:ax_potential_source_form}
\end{equation}
in agreement with the general axial equation on a spherical background~\cite{Karlovini2002}.

For the dyonic Kalb--Ramond geometry, $A= B = F$, with
\begin{equation}
F(r) = \frac{1}{1-\ell}-\frac{2M}{r}+\frac{P_\ell^{2}}{r^{2}},  \qquad  P_\ell^{2} = \frac{Q^{2}}{(1-\ell)^{2}} + \frac{p^{2}}{1-2\ell}.
\label{eq:ax_dyonic_background}
\end{equation}
In this manner,
\begin{equation}
\begin{aligned}
V_t(r)& = F(r)\left[  \frac{l(l+1)}{r^{2}}+\frac{2[F(r)-1]}{r^{2}}  -\frac{F'(r)}{r}\right]\\  & = F(r)\left[  \frac{l(l+1)}{r^{2}} - \frac{6M}{r^{3}}  +\frac{4P_\ell^{2}}{r^{4}}  +  \frac{2\ell}{(1-\ell)r^{2}}\right],
\end{aligned}
\label{eq:ax_dyonic_potential}
\end{equation}
which reproduces Eq.~(\ref{tensdas}). The limit $\ell = Q = p = 0$ gives the Schwarzschild Regge--Wheeler potential~\cite{ReggeWheeler1957}.

The scope of Eq.~\eqref{eq:ax_dyonic_potential} is fixed by the closure in Eq.~\eqref{eq:ax_closure}. In particular, setting the independent electromagnetic and Kalb--Ramond field perturbations to zero would not, by itself, establish this closure: the nonminimal curvature coupling also varies with the metric. A complete perturbation analysis of the action must retain the coupled field equations or demonstrate a decoupling. Even in the Einstein--Maxwell charged limit, the physical gravitational and electromagnetic perturbations form a coupled system~\cite{Moncrief1974}; the limit $\ell\to0$ of the present expression is understood within the same effective source prescription. It is important to highlight that the frequencies obtained from this potential characterize that effective axial metric equation and do not establish the mode spectrum of the full electromagnetic--Kalb--Ramond theory.

}


\section*{Data Availability Statement}

Data Availability Statement: No datasets were generated or analyzed in this study. The code used in this work is available from the author upon reasonable request.

\bibliographystyle{ieeetr}
\bibliography{main}

\end{document}